\documentclass[PAPER, cernpreprint, atlasdraft=false, texlive=2023, USenglish, texmf, orcidlogo]{atlasdoc}
 
\usepackage[subfig]{atlaspackage}
\usepackage{etoolbox} 
\robustify{\subref} 
\usepackage{atlasbiblatex}
 
\usepackage{atlasphysics}
 
\usepackage{amsmath}
\usepackage{colortbl}
\usepackage{hhline}
\usepackage{multirow}
 
\graphicspath{{logos/}{figures/}}
 
\usepackage{ANA-TOPQ-2019-22-PAPER-defs}

\AtlasTitle{
Measurement of jet substructure in boosted $t\bar{t}$ events with the
ATLAS detector using 140 fb$^{-1}$ of
13 TeV $pp$ collisions
}
 
\PreprintIdNumber{CERN-EP-2023-239}
\AtlasVersion{3.2}
\AtlasAbstract{
Measurements of the substructure of top-quark jets are presented, using 140 fb$^{-1}$ of 13 TeV $pp$ collision data recorded with the ATLAS detector at the LHC. Top-quark jets reconstructed with the anti-$k_{t}$ algorithm with a radius parameter $R=1.0$ are selected in top-quark pair ($t\bar{t}$) events where one top quark decays  semileptonically and the other hadronically, or where both top quarks decay hadronically. The top-quark jets are required to have transverse momentum $p_\mathrm{T} > 350$ GeV, yielding large samples of data events with jet  $p_\mathrm{T}$ values between 350 and  600 GeV. One- and two-dimensional differential cross-sections for eight substructure variables, defined using only the charged components of the jets, are measured in a particle-level phase space by correcting for the smearing and acceptance effects induced by the detector. The  differential cross-sections are compared with the predictions of several Monte Carlo  simulations in which top-quark pair-production quantum chromodynamic matrix-element calculations at next-to-leading-order precision in the strong coupling constant $\alpha_\mathrm{S}$ are passed to leading-order parton shower and hadronization generators. The Monte Carlo predictions for measures of the broadness, and also the two-body structure, of the top-quark jets are found to be in good agreement with the measurements, while variables sensitive to the three-body structure of the top-quark jets exhibit some tension with the measured distributions.
}
 
\author{The ATLAS Collaboration}
 
\AtlasRefCode{TOPQ-2019-22}
 
\AtlasJournalRef{Phys. Rev. D 109 (2024) 112016}
\AtlasDOI{10.1103/PhysRevD.109.112016}

\hypersetup{pdftitle={ATLAS document},pdfauthor={The ATLAS Collaboration}}
 
\begin{document}
 
\maketitle
 
\tableofcontents
\newpage
 
\section{Introduction}
\label{sec:intro}


Observation of the production and decay of top quarks in high-energy proton--proton (\pp)
collisions has provided information about the production of massive particles through the strong force.
The Large Hadron Collider (LHC) has enabled the collection of large samples of top-quark events and the measurement of top-quark production processes with excellent kinematic resolution~\cite{ATLAS:2023gsl,TOPQ-2019-23,CMS-TOP-20-001,CMS-TOP-18-004}. These results have been compared with quantum chromodynamics (QCD) calculations at lowest order in the strong coupling constant \alphas, as well as next-to-leading-order (NLO) and next-to-next-to-leading-order (NNLO) results.
The top-quark decay process and the resulting parton-showering and hadronization effects
have been measured with less precision.
This paper reports a measurement of the substructure observed in high transverse-momentum (\pT) top-quark jets produced in \pp\ collisions at a center-of-mass energy of 13~\TeV\ using 140~$\mathrm{fb}^{-1}$ of data recorded by the ATLAS detector.
 
The large top-quark pair-production (\ttbar)\ cross-section at the LHC allows detailed studies of
the substructure of jets arising from light quarks, $b$-quarks, and gluons, as well as
large-radius jets containing the decay products of boosted top quarks.
It also allows tests of the Standard Model (SM) at both the top-quark mass scale relevant for
substructure measurements and the \TeV\ scales relevant for jet production
measurements~\cite{Kogler_2019,Larkoski_2020}.
Furthermore, effects beyond the SM can appear as deviations of
top-quark jet substructure from the SM predictions~\cite{
EXOT-2015-04,
SUSY-2016-10,
EXOT-2016-25,
SUSY-2017-01,
HDBS-2018-31}\
that may not be detectable with inclusive cross-section measurements.
A precise measurement of top-quark differential cross-sections as a function
of several complementary substructure variables therefore
tests SM predictions and increases the sensitivity
to possible effects beyond the SM. Top-quark jet substructure variables are also employed in the development of sophisticated taggers, used to separate top-quark jets from energetic jets arising from
lighter quarks or gluons~\cite{JETM-2018-03,CMS-JME-18-002,Kasieczka:2019dbj}. These taggers are used in measurements and searches targeting final states including highly energetic top quarks~\cite{TOPQ-2018-11,ATLAS:2022znu,TOPQ-2019-28,EXOT-2018-48,CMS-B2G-16-015,CMS-B2G-17-017}.
 
The SM predicts that top quarks decay into a \Wboson~boson and a $b$-quark
almost 100\%\ of the time.
In the cases where the \Wboson~boson decays hadronically ($\Wboson\rightarrow q\bar{q^\prime}$),
the final state consists of an all-hadronic top-quark decay with three charged decay products.
The resulting quarks and gluons shower and hadronize to form semi-stable objects that can be studied as a single, collimated system when the
top quark is produced at high \pT, above approximately 300~\GeV.
This top-quark jet remains connected to the rest of the event through color connections, specifically to additional radiation and the other top quark's decay products.
Measurements of the substructure of these top-quark jets
test the models used to describe this decay, the parton shower, and the hadronization process.
 
Interest in the properties of jets, both experimental and theoretical, extends back several decades and remains an active area of study~\cite{Salam:2010nqg}.
At the LHC, measurements of jet substructure have been performed in proton--proton collisions by the ATLAS, CMS, ALICE, and LHCb collaborations in several topologies, as dijet and inclusive jet production~\cite{STDM-2011-19,STDM-2014-17,STDM-2017-33,CMS-QCD-10-029,CMS-SMP-15-003,CMS-SMP-16-010,CMS-SMP-20-010,CMS-HIN-16-006,CMS-SMP-12-019,CERN-EP-2019-087,EPS-HEP2021-ALICE,Vertesi:2020gqr,ALICE:2021aqk,LHCb-PAPER-2016-064,STDM-2017-04,STDM-2011-38,STDM-2018-57}, electroweak boson production and top-quark production~\cite{PhysRevD.91.032006,CMS-SMP-20-010,STDM-2011-38,TOPQ-2014-09,TOPQ-2017-13,STDM-2017-34,CMS-SMP-12-019,CMS-TOP-17-013,LHCb-PAPER-2019-012,CMS-TOP-19-005,CMS-TOP-15-015}.
The substructure studies examined variables based on energy flow within the jet, such as angularity, planar flow, and jet mass, and also variables that are sensitive to the three-prong structure of the top-quark jet, such as $N$-subjettiness and energy correlation variables. These measurements indicated possible disagreement with the SM predictions for
some substructure variables, although most of the measurements were limited by large statistical and systematic uncertainties.
In addition, the measurements focusing on jets from top-quark decays typically involved a lower average jet \pT\ than employed in the analysis reported here. At higher \pT, the three-body structure of high-\pT\ top-quark jets is less well defined because the three partons are more collimated, making it more difficult to separate out their showers.
 
The measurements reported here are performed in two independent channels, using two classes of \ttbar\ final states, the lepton+jets (\ellplusjets) channel where one top quark has decayed semileptonically and the other hadronically, and the all-hadronic final state where both top quarks have decayed hadronically. Even though the top-quark jet substructure in the two channels is described by the same underlying theory, the different final states require different approaches to their selection and reconstruction to separate the signal events from the background, giving rise to differing effects on the resolution, uncertainty and shape of the measured observables. In addition, the two final states differ in the amount of final-state radiation.
 
In both channels, reconstructed top-quark jets with $\pT > 350$~\GeV\ are used to measure the substructure variables by \enquote{unfolding} detector effects to stable-particle level in order to obtain differential cross-section measurements.
These unfolded variables are defined in specific kinematic regions, or fiducial phase-spaces,
in order to avoid significant extrapolation of results from the kinematic region in which the
measurements are made to the region in which they are compared with predictions.
The \ellplusjets\ fiducial phase-space is characterized by the requirement of a single lepton,
a top-quark jet with $\pT > 350$~\GeV\ and an additional jet arising from the $b$-quark hadronization.
The all-hadronic fiducial phase-space is defined by requiring two top-quark jets,
each containing a $b$-hadron, with one jet having $\pT > 500$ \GeV\ and a second jet having $\pT > 350$~\GeV, and the absence of isolated leptons.
Only the charged components of the top-quark jets are considered when defining the substructure,
as the substructure measurements using this approach have higher resolution than those using a combination of charged and neutral components.
The unfolded differential cross-sections are then compared with the predictions of different Monte Carlo (MC) generators.
The top-quark pair-production matrix element is calculated at NLO precision in QCD.
The modeling of radiation beyond the first emission, the top-quark decay, and the showering
is at leading-order (LO) precision, and the hadronization is described by a phenomenological model.
These predictions are referred to later in the text as NLOME+PS.
 
The measured substructure variables are
the $N$-subjettiness variable $\tau_3$, as well as the ratios
$\tau_{32}\equiv{\tau_3}/{\tau_2}$ and
$\tau_{21}\equiv{\tau_2}/{\tau_1}$~\cite{Thaler_2011,Thaler_2012},
the normalized energy-correlation function \ECF, as well as the associated variables $D_2$ and
$C_3$~\cite{Larkoski_2013,Larkoski:2014gra}, and
the Les Houches angularity \LHA\ and the scaled \pT dispersion \ptdispersion,
which are two generalized angularities~\cite{Larkoski_2014_ga}. This list of variables has been selected to provide complete and condensed information about the jet substructure while taking into consideration multiple factors: sensitivity to the differences between NLOME+PS predictions, resolution, minimal correlation among the measured variables, and their relevance in tagging algorithms.
 
The results are intended to complement the recent measurements of top-quark jet substructure performed by the
ATLAS~\cite{STDM-2017-34} and CMS~\cite{CMS-TOP-17-013}\ collaborations by including several new features. The measurement focuses on boosted top quarks, and the inclusion of an all-hadronic channel allows the substructure to be studied for top-quark jets with an average \pT\ around 500~\GeV\ for the first time. In addition, using tracks to reconstruct the substructure improves the resolution, with the diagonal elements of the migration matrix increasing by approximately 50$\%$, and also leads to lower uncertainties than in the previous results.
 
The paper is organized as follows.
Section~\ref{sec:detector}\ describes the ATLAS detector, while Section~\ref{sec:datasets} presents the data and simulation samples used in the measurements.
Both the reconstruction of lepton and jet candidates and the event selection is explained in Section~\ref{sec:objects_and_selection}.
The estimation of the backgrounds in the selected \ttbar\ final states is described in Section~\ref{sec:background}.
The measured substructure variables are defined in Section~\ref{sec:variables}\ and the unfolding is described in Section~\ref{sec:unfolding}.
Systematic uncertainties are discussed in Section~\ref{sec:systematics}.
The results are presented in Section~\ref{sec:results}, where the comparison with theoretical predictions is also discussed.
Finally, a summary is presented in Section~\ref{sec:conclusions}.


\section{ATLAS detector}
\label{sec:detector}

 
\newcommand{\AtlasCoordFootnote}{
ATLAS uses a right-handed coordinate system with its origin at the nominal interaction point (IP) in the center of the detector and the \(z\)-axis along the beam pipe. The \(x\)-axis points from the IP to the center of the LHC ring, and the \(y\)-axis points upwards. Cylindrical coordinates \((r,\phi)\) are used in the transverse plane, \(\phi\) being the azimuthal angle around the \(z\)-axis. The pseudorapidity is defined in terms of the polar angle \(\theta\) as \(\eta = -\ln \tan(\theta/2)\). Angular distance is measured in units of \(\Delta R \equiv \sqrt{(\Delta\eta)^{2} + (\Delta\phi)^{2}}\).}
 
The ATLAS detector~\cite{PERF-2007-01} at the LHC covers nearly the entire solid angle around the collision point.\footnote{\AtlasCoordFootnote} It is a multipurpose detector consisting of an inner tracking detector (ID) surrounded by a thin superconducting solenoid, electromagnetic and hadronic sampling calorimeters, and a muon spectrometer incorporating three large superconducting air-core toroidal magnets.
 
The ID is immersed in a \SI{2}{\tesla} axial magnetic field and provides charged-particle tracking in the range \(|\eta| < 2.5\).
The calorimeter system covers the pseudorapidity range \(|\eta| < 4.9\). Within the region \(|\eta|< 3.2\), electromagnetic calorimetry is provided by barrel and endcap high-granularity lead/liquid-argon (LAr) calorimeters. Hadron calorimetry is provided by the steel/scintillator-tile calorimeter, segmented into three barrel structures within \(|\eta| < 1.7\), and two copper/LAr endcap hadron calorimeters.
The muon spectrometer (MS) comprises separate trigger and high-precision tracking chambers measuring the deflection of muons in a magnetic field generated by the superconducting air-core toroidal magnets.

Events of interest are selected by the first-level trigger system implemented in custom hardware, followed by selections made by algorithms implemented in software in the high-level trigger~\cite{TRIG-2016-01}.
 
An extensive software suite~\cite{ATL-SOFT-PUB-2021-001} is used in data simulation, in the reconstruction and analysis of real and simulated data, in detector operations, and in the trigger and data acquisition systems of the experiment.


\section{Data samples, simulation and theoretical predictions}
\label{sec:datasets}

 
\label{sec:MCttbar}
\label{sec:MCbkg}
 
\newcommand{\hdampFootnote}{
The \hdamp parameter controls the matching in \powheg and effectively regulates the high-\pt\ radiation against which the \ttbar\ system recoils.}
 
The data were collected from LHC $pp$~collisions at $\sqrt{s}=13$~\TeV\ during 2015--2018. Data quality requirements are applied to ensure that all detector components were fully operational~\cite{DAPR-2018-01}.
The resulting data sample corresponds to an integrated luminosity of 140.1~\ifb. The uncertainty in the combined 2015--2018 integrated luminosity is 0.83\%~\cite{DAPR-2021-01}, obtained using the LUCID-2 detector \cite{LUCID2} for the primary luminosity measurements, complemented by measurements using the inner detector and calorimeters.
 
The \ttbar\ \ellplusjets\ events were selected with inclusive lepton triggers~\cite{TRIG-2016-01,TRIG-2018-05,TRIG-2018-01}\ that required an event to have an electron or muon candidate with $\pT > 26$~\GeV\ (or $\pT > 24$~\GeV\ and $\pT > 20$~\GeV, respectively, for electrons and muons in 2015 data).
Events in the all-hadronic channel were selected inclusively using a collection of single- and double-\largeR~jet triggers, with the single-jet triggers using both trimmed\footnote{Here the trimming~\cite{Krohn2010} employs a radius parameter of $R=0.2$ and removes subjets with less than 4\% of the \largeR\ jet \pT.} and untrimmed jets.
The required \pt for the single-jet triggers ranged from 360 to 480~\GeV, depending on
the instantaneous luminosity when the data were selected.
For the double-jet triggers, both \largeR~jets were required to have $\pt > 330$~\GeV.
The lower-\pt jet triggers additionally required the \largeR~jets to have a mass above either 35 or 40~\GeV, depending on the trigger.
Although both the single- and double-jet triggers were employed to select the events
for the all-hadronic channel,
the single-jet triggers were highly efficient for events that pass the
preselection described in Section~\ref{sec:selection_allhad}. The same trigger requirements are applied to data and simulated events; this allows corrections for trigger inefficiencies to be made in the cross-section extraction procedure described in Section~\ref{sec:unfolding}.
 
Simulated events were produced using several MC calculations, and both the effects and response of the detector were incorporated using a simulation of the ATLAS detector~\cite{SOFT-2010-01} based on either the \GEANT program~\cite{Agostinelli:2002hh} or a faster approach employing parameterized showers in the calorimeter~\cite{ATL-PHYS-PUB-2010-013}.
The faster approach was only used to simulate a subset of alternative \ttbar\ samples produced for the \ellplusjets\ channel.
The data events and simulated events were processed with the same reconstruction and analysis algorithms.
 
The production of \ttbar\ events was modeled using the \POWHEGBOX[v2]~\cite{Frixione:2007nw,Nason:2004rx,Frixione:2007vw,Alioli:2010xd} generator to provide matrix elements at NLO precision with the \NNPDF[3.0nlo]~\cite{Ball:2014uwa} parton distribution functions~(PDFs) and the \hdamp\ parameter\footnote{\hdampFootnote} set to $1.5\,\mt$~\cite{ATL-PHYS-PUB-2016-020},
where the top-quark mass \mt\  was set to 172.5~\GeV\ unless otherwise noted.
To assess the effect of an uncertainty in \hdamp, an additional $t\bar{t}$ sample with \hdamp\ set to $3.0\,\mt$\ is used.
To assess the effect of a top-quark mass uncertainty, two additional $t\bar{t}$ samples were simulated with \mt\ set to 169~\GeV\ and 176~\GeV.
These samples are used to accurately estimate the dependance on \mt\ of the measured observables, and estimate the uncertainty arising from a $\pm$0.5 GeV  variation of \mt\ from the nominal value.
The functional form of the renormalization and factorization scale was set to the default scale $\sqrt{m^2 + p_{\textrm T}^2}$, where $m$ and $p_{\textrm T}$ represent the mass and transverse momentum of the top quark in each event.
Parton showering and hadronization were modeled with the \PYTHIA[8.230] generator~\cite{Sjostrand:2014zea} using the A14 set of tuned parameters~\cite{ATL-PHYS-PUB-2014-021} and the \NNPDF[2.3lo]\ set of PDFs~\cite{Ball:2012cx}.
The value of \alphaFSR set in the tuning is $\alphaFSR(m_Z) = 0.127$.

The uncertainty due to the parton-shower and hadronization model is estimated by comparing the
results from the nominal generator setup with those from a sample produced with the \POWHEGBOX[v2] generator and the \HERWIG[7.1.3]~\cite{Bahr:2008pv,Bellm:2015jjp} parton-shower and hadronization model.
In this alternative sample, the \HERWIG[7.1] default set of tuned parameters~\cite{Bellm:2015jjp,Bellm:2017jjp} and the \MMHT[lo] PDF set~\cite{Harland-Lang:2014zoa} were used.
 
An uncertainty also arises from the procedure used to match the hard-scatter and parton-shower generators.
The results from the nominal generator were compared with those from a sample produced by the
\MGNLOPY[8] generator~\cite{Alwall:2014hca}. For the calculation of the hard scattering,
the \MGNLO[2.6.0] generator (hereafter called \AMCatNLO) was used with the \NNPDF[3.0nlo] PDF set.
Parton showering and hadronization were modeled with the \PYTHIA[8.230] generator, using the A14 tune and the \NNPDF[2.3lo] set of PDFs.
Top quarks were decayed at LO, using the \MADSPIN~model~\cite{Frixione:2007zp,Artoisenet:2012st} to preserve top-quark spin correlations.
The shower-starting scale had the functional form $\muQ = \HT/2$~\cite{ATL-PHYS-PUB-2017-007}, where \HT is defined as the scalar sum of the \pT of all outgoing partons.
The renormalization and factorization scale choice was the same as for the \POWHEG\ generator
setup.
Some of the \PYTHIA\ settings used in the alternative sample differ from those in the nominal sample,
since they must accommodate a different matching algorithm.
One important difference concerns the matrix element (ME) corrections applied to the radiation modeled by the \PYTHIA\ parton shower,
which cannot be enabled when the \PYTHIA\ parton-shower and hadronization model is matched to
the \AMCatNLO ME calculation.
To remove the effects of ME corrections from the matching uncertainty, an additional \POWPY sample was used in comparison with the \AMCatNLOPY[8]\ sample, differing from the nominal sample only in the \PYTHIA ME corrections, which were disabled.
 
All \ttbar\ samples were normalized to the cross-section prediction at NNLO in QCD including the resummation of next-to-next-to-leading logarithmic (NNLL) soft-gluon terms calculated using the \TOPpp[2.0]~program~\cite{Beneke:2011mq,Cacciari:2011hy,Baernreuther:2012ws,Czakon:2012zr,Czakon:2012pz,Czakon:2013goa,Czakon:2011xx}.
This cross-section corresponds to $\sigma(\ttbar)_{\textrm{NNLO+NNLL}} = \mathrm{832\pm51~pb}$,
where the uncertainty is due to scale variations, \alphas and the PDFs.
Their contributions are calculated using the PDF4LHC prescription~\cite{Butterworth:2015oua} with the
\MSTW[nnlo] (68\% CL)~\cite{Martin:2009iq,Martin:2009bu}, \CT[10nnlo]~\cite{Lai:2010vv,Gao:2013xoa} and \NNPDF[2.3lo] PDF sets.
The samples employed for the signal modeling are filtered according to the number of leptons arising from the \ttbar decay: only events with no leptons are used in the all-hadronic channel and only events with at least one lepton are used in the \ellplusjets\ channel.
 
In addition to the signal, the contributions of several background processes to the selected event sample are estimated using MC predictions.
 
The production of a top quark in association with a $W$ boson ($tW$) was modeled with the \POWHEGBOX[v2]~\cite{Re:2010bp,Nason:2004rx,Frixione:2007vw,Alioli:2010xd} generator at NLO in QCD, using the five-flavor scheme and the \NNPDF[3.0nlo] set of PDFs.
The diagram-removal (DR) scheme~\cite{Frixione:2008yi} was used to remove interference and overlap with \ttbar\ production. The related uncertainty was estimated by comparison with an alternative sample generated using the diagram-subtraction (DS) scheme~\cite{Frixione:2008yi,ATL-PHYS-PUB-2016-020}.
The \PYTHIA[8.230] generator modeled the parton showering and hadronization, using the A14 tune and the \NNPDF[2.3lo] set of PDFs.
 
Single-top-quark $t$-channel production was modeled with the \POWHEGBOX[v2] generator at NLO in QCD, using the four-flavor scheme and the corresponding \NNPDF[3.0nlo] PDFs.
Parton showering and hadronization in these events was modeled with the \PYTHIA[8.230] generator using the A14 tune and the \NNPDF[2.3lo] PDFs.
Single-top $s$-channel production was modeled using the \POWHEGBOX[v2]\ generator at NLO in QCD in the five-flavor scheme with the \NNPDF[3.0nlo]\ PDFs.
Parton showering and hadronization in these events was modeled with the \PYTHIA[8.230] generator using the A14 tune and the \NNPDF[2.3lo] PDFs.
For both $t$-channel and $tW$-channel single-top production, an additional sample, produced with \POWHEGBOX[v2]\ and \HERWIG[7.04] and using the H7UE set of tuned parameters,
is employed to estimate the uncertainty arising from the choice of parton-showering and hadronization models.
 
Simulated $W/Z+$jets events were produced with the \SHERPA[2.2.1]~\cite{Bothmann:2019yzt} generator using NLO matrix elements for up to two partons,
and LO matrix elements for up to four partons, calculated with the Comix~\cite{Gleisberg:2008fv} and \OPENLOOPS~\cite{Buccioni:2019sur,Cascioli:2011va,Denner:2016kdg} libraries.
They were matched with the \SHERPA\ parton-shower model~\cite{Schumann:2007mg} using the MEPS@NLO prescription~\cite{Hoeche:2011fd,Hoeche:2012yf,Catani:2001cc,Hoeche:2009rj} and the set of tuned parameters developed by the \SHERPA authors.
The \NNPDF[3.0nnlo] PDFs were used and the samples were normalized to a NNLO prediction~\cite{Anastasiou:2003ds}.
 
The production of \ttZ\ and \ttW\ events was modeled using the \AMCatNLO[v2.3.3] generator at NLO with the \NNPDF[3.0nlo] PDFs.
Parton showering and hadronization in these events was modeled with  the \PYTHIA[8.210] generator using the A14 tune and the \NNPDF[2.3lo] PDFs.
The production of \ttH\ events was modeled using the \POWHEGBOX[v2]~\cite{Frixione:2007nw,Nason:2004rx,Frixione:2007vw,Alioli:2010xd,Hartanto:2015uka} generator at NLO with the \NNPDF[3.0nlo] PDFs.
Parton showering and hadronization in these events was performed with \PYTHIA[8.230] using the A14 tune and the \NNPDF[2.3lo] PDFs.
 
Samples of diboson final states ($VV$) were modeled with the \SHERPA[2.2.2]generator, including off-shell effects and virtual Higgs boson contributions where appropriate. The samples were generated using NLO matrix elements for up to one additional parton and at LO accuracy for up to three additional parton emissions. The matrix-element calculations were matched and merged with the \SHERPA\ parton shower based on Catani--Seymour dipole factorization using the MEPS@NLO prescription. The virtual QCD correction were provided by the \OPENLOOPS library. The \NNPDF[3.0nnlo] PDFs were used along with the dedicated set of tuned parton-shower parameters developed by the \SHERPA\ authors.
 
The decays of bottom and charm hadrons in processes containing top quarks were simulated using the \EVTGEN\ program~\cite{Lange:2001uf}.
 
In all simulated samples, the effect of multiple interactions in the same bunch
crossing (\pileup) was incorporated by overlaying the hard-scattering interaction with a number
of simulated inelastic $pp$\ events generated with the \PYTHIA[8.186] MC generator using the \NNPDF[2.3lo] PDFs and the A3 tune~\cite{ATL-PHYS-PUB-2016-017}.


\section{Event reconstruction and selection}
\label{sec:objects_and_selection}


\subsection{Detector-level objects reconstruction}
\label{sec:objects_reco}
 
The jet substructure measurements at detector level use
electron candidates, muon candidates, jet candidates, the missing transverse momentum, and charged tracks,
all of which are required to be associated with the primary vertex.
The primary vertex is defined as the reconstructed vertex with the highest $\sum p_{\mathrm{T}}^2$ of the associated tracks and is required to have at least two associated tracks with $\pt > 0.5$~\GeV.
 
Four types of jets are reconstructed independently:
(1) trimmed \largeR\ jets created from calorimeter energy deposits~\cite{PERF-2014-07}\ using the \antikt\ clustering algorithm~\cite{Cacciari:2008gp}\ with radius parameter $R=1.0$ implemented
in the FastJet package~\cite{Fastjet},
(2) \smallR\ jets created from both tracking and calorimeter information using the particle-flow
algorithm~\cite{PERF-2015-09} and the \antikt\ algorithm with radius parameter $R=0.4$,
(3) trimmed \largeR\ reclustered jets (RC \largeR\ jets) created from the \smallR\ jets using the \antikt\ algorithm with radius parameter $R=1.0$~\cite{Nachman:2014kla}, and
(4) \varR\ track-jets created from ID tracks with the variable-$R$\ algorithm~\cite{VRjets}\ implemented in the FastJet package and defining jets with a \pT-dependent variable-radius parameter ranging between $R = 0.02$ and $R = 0.4$ in accord with a constant parameter $\rho = 30$~\GeV\ scaling as $\rho/\pT$~\cite{ATL-PHYS-PUB-2020-019}. The two channels use different combinations of these jets in their selection.
 
In order to suppress jets originating from \pileup, \smallR\ jets with $20 < \pT < 60$~\GeV\ and $\abseta < 2.5$\ must be associated with the primary vertex according to the jet-vertex-tagger (JVT) algorithm~\cite{PERF-2014-03}. The \smallR\ jet energies are corrected by using energy- and $\eta$-dependent calibration factors derived from simulation and in~situ measurements~\cite{JETM-2018-05}. Only \smallR\ jets with $\pT > 30$~\GeV\ are employed. \LargeR\ jets are calibrated using the local hadronic cell weighting (LCW) scheme~\cite{PERF-2014-07}, which exploits the cluster shape to provide a weighted calibration between the electromagnetic and hadronic scales~\cite{JETM-2018-02}. The effects of \pileup and the underlying event are mitigated by the application of a trimming algorithm~\cite{Krohn2010}: the constituents of the \largeR\ jets are reclustered by the \kt algorithm~\cite{Fastjet} into smaller $R = 0.2$ jets which are removed if they comprise less than 5\% of the \pt of the original \largeR\ jet.
 
The RC \largeR\ jet calibration is propagated from the calibration of the \smallR\ jets that are used as input to the reclustering~\cite{ATLAS-CONF-2017-062}. Most of the \pileup mitigation is thus inherited by the RC \largeR\ jets from the techniques applied to the \smallR\ jets. In addition, a trimming technique is applied to the RC \largeR\ jets that removes all \smallR\ jets comprising less than 5\% of the \pt of the RC \largeR\ jet. RC \largeR\ jets are discarded if they have only one constituent to avoid a dependence on the calibration of the \smallR\ jet mass. Such jets comprise ${\sim}1.5$\%\ of the jets in the sample.
Only RC \largeR\ jets with $\pt > 350$~\GeV\ and $\abseta < 2.0$ are considered when measuring the substructure variables.
 
The use of different types of \largeR\ jets in the two \ttbar\ final states is motivated by the topologies of the \ellplusjets\ and all-hadronic channels and the different background contamination. In the all-hadronic channel the signal is separated from the large multijet background by using a deep-neural-network (DNN) top-quark-tagging algorithm optimized for \largeR\ jets~\cite{JETM-2018-03}. Because this top-quark tagger employs several substructure variables, a tag-and-probe method described in Section~\ref{sec:variables}\ is used to avoid biasing the measured observables in the all-hadronic channel. The DNN top-quark-tagging algorithm has not been trained and tested on RC \largeR\ jets, and consequently can only be applied to the (non-RC) \largeR\ jets.
The \ellplusjets\ topology, on the other hand, is identified by requiring the presence of a semileptonically decaying top quark, which allows the top-quark jet to be identified without any top-quark-tagging requirement.
The use of RC \largeR\ jets in this channel is also motivated by several other factors: the mass resolution of the RC \largeR\ jet, the resolution of the substructure variables, and the stability of the substructure measures in the presence of \pileup. In addition, the usage of RC \largeR\ jets allows a single type of jet calibration (\smallR\ jets) to be applied across the whole event.
While the jet definitions are different in the two channels, the substructure variables
measured in the same jet when it is reconstructed as either a \largeR\ jet or a RC \largeR\ jet are highly correlated, with correlation coefficients ${>}$~90\%.
 
The DL1r multivariate algorithm~\cite{ATLAS:2022qxm} is used to identify \smallR\ jets and \varR\ jets that contain $b$-hadrons. This is a DNN algorithm exploiting information from the large-impact-parameter tracks, the topological decay chain, and the displaced vertices of $b$-hadron decays~\cite{FTAG-2018-01, ATL-PHYS-PUB-2017-013}. The jets are considered $b$-tagged if the value of the discriminant is larger than a threshold that provides 77\%\ efficiency as measured in inclusive \ttbar\ events. The corresponding rejection factors for gluon/light-quark jets and charm-quark jets are approximately 200 and 5, respectively. A lower efficiency of ${\sim}70$\% is observed in the $b$-tagging of \largeR\ top-quark jets in this analysis because of the increased collimation and charged-track density in the boosted top-quark jets.
 
A \largeR\ jet in the all-hadronic channel is labeled \enquote{$b$-matched} if a ghost-matching algorithm~\cite{Cacciari:2008,Cacciari:2008a} matches it with one or more $b$-tagged \varR\ jets.
The ghost-matching procedure treats the objects to be matched as four-vectors of negligible magnitude in the standard jet reconstruction described above.
 
Top-quark tagging of \largeR\ jets in the all-hadronic channel is performed with the DNN
top-quark tagger noted above.
The top-quark tagger's working point is set such that 80\% of simulated top-quark jets are tagged, where the efficiency is determined using jets that fully contain the top-quark decay products.
The tagger's efficiency in the simulated events is corrected to the observed efficiency in data by means of scale factors~\cite{JETM-2018-03,ATL-PHYS-PUB-2020-017},
evaluated in samples enriched in \lplusjets\ \ttbar decays.
 
Tracks are reconstructed in the inner detector using an iterative algorithm~\cite{PERF-2015-08}.
Reconstructed tracks are required to pass quality requirements based on the number of silicon-detector layers crossed and are required to have $\pT > 500$~\MeV\ and $|\eta|<2.5$.
Moreover, to reduce \pileup\ contamination, the tracks must be closely matched to the primary
vertex: the transverse impact parameter calculated relative to
the beam line must satisfy $|d_0| < 2$~mm and the longitudinal impact parameter
$z_0$\ calculated relative to the primary vertex must satisfy $|z_0 \sin(\theta)| < 3$~mm.
The tracks are assigned to the \smallR\ jets and \largeR\ jets using the ghost-association technique and then used to calculate the substructure of the jet to which they are assigned.
The use of tracks, as opposed to the energy flow measured with the calorimeter, provides greater precision
for the substructure variables measured in the \largeR\ jets and RC \largeR\ jets.
 
The reconstruction of electron candidates combines information from the ID with calorimetry information~\cite{PERF-2017-01,EGAM-2018-01}. Tracks from the ID are matched to energy deposits in the electromagnetic calorimeter (ECAL). These tracks must have trajectories consistent with originating from the primary vertex. The ECAL clusters must fall within the central acceptance, $\abseta < 2.47$, and have $\ET>27$~\GeV. Clusters must also be outside the transition region between the barrel and endcap sections of the ECAL, $1.37 < \abseta < 1.52$. The electron candidates are required to have $\pt > 27 $~\GeV\ and are identified using a likelihood-based method requiring that they satisfy
the \enquote{TightLH} working point \cite{PERF-2017-01}.
The charged track matched to the electron candidate must satisfy specific requirements on the
impact parameters: the longitudinal impact parameter $z_0$ calculated relative to the primary vertex must satisty $|z_0 \sin(\theta)| < \SI{0.5}{\mm}$, and the significance of the transverse impact parameter calculated relative to
the beam line must satisfy  $|d_0/\sigma(d_0)| < 5$, where $\sigma(d_0)$ is the expected uncertainty in $d_0$.
The \enquote{FCTight} criteria \cite{PERF-2017-01}\ must also be satisfied to ensure that the electron candidates are well isolated from nearby energy flow.
 
The reconstruction of muon candidates combines charged tracks in the ID with those found in the muon spectrometer~\cite{MUON-2018-03}. The ID tracks must be consistent with coming from the primary vertex. Muon candidates must satisfy the \enquote{Medium} \cite{MUON-2018-03} identification criteria and the \enquote{FCTight} condition is imposed for muon isolation.  In addition, the muons must have impact parameters satisfying $|d_0/\sigma(d_0)| < 3$ and $|z_0 \sin(\theta)| < \SI{0.5}{\mm}$.
Muon candidates are required to have $\pt > 27$~\GeV\ and $\abseta < 2.5$.
 
An overlap removal procedure is applied to reduce any double counting of objects that are independently reconstructed.
An electron candidate is rejected if it shares a track with a muon candidate or if it has a track overlapping with another electron candidate.
For \smallR\ jets that are within $\Delta R(j,e) = 0.2$\ of an electron candidate,
the closest jet fulfilling this criterion is removed. If there are other \smallR\ jets within $\Delta R(j,e) = 0.4$\ of the electron candidate, the electron candidate is removed.
Small-$R$ jets that are within $\Delta R(j,\mu) = 0.2$\ of a muon candidate or share a track with that muon candidate are removed if they comprise fewer than three tracks.
Any muon candidate that is within $\Delta R(j,\mu) = 0.4$\ of a \smallR\ jet is removed.
 
Neutrinos are not measured by the detector so their presence is inferred from the missing transverse momentum, \met.
This is the magnitude of the negative vector sum of the \pt of all selected and calibrated lepton candidates and \smallR\ jets in an event,
as well as a \enquote{soft} term~\cite{PERF-2016-07} calculated from the \pt of ID tracks that are not associated with any of the lepton candidates or \smallR\ jets.
 
\subsection{\ellplusjets\ event selection}
\label{sec:selection_ljets}
 
In the \ellplusjets\ channel, each event is required to have a primary vertex, to have passed a single-lepton trigger and to contain exactly one lepton candidate corresponding to the passed trigger. To reduce contributions from the multijet background and misidentified or nonprompt leptons, requirements are placed on \met and the sum of the \met and transverse mass\footnote{The transverse mass
is defined as $m_{\text{T}}^{W}=\sqrt{2p^{\ell}_{\text{T}}\met[1 - \cos{\Delta\phi}(p^{\ell}_{\text{T}},\met)]}$, where
$p^{\ell}_{\text{T}}$ is the lepton \pT.} \mtw.
Each event is required to have $\met > 20$~\GeV\ and $\met + \mtw > 60$~\GeV. \SmallR\ jets in these events must have $\pt > 30$~\GeV\ and $\abseta < 2.5$. The leading\footnote{If not differently specified, the term leading object always refers to the highest \pT object.} RC \largeR\ jet must contain at least two \smallR\ jets and is required to have $\abseta < 2$ and $\pt > 350$~\GeV.
In the 5$\%$ of selected events where there is more than one RC \largeR\ jet satisfying these criteria, the hadronic top-quark candidate is defined as the one with jet mass closest to the top-quark mass ($\mt = 172.5$~\GeV). This choice maximizes the probability that the top-quark jet is identified correctly.
 
Events are required to contain at least one \smallR\ jet that has been $b$-tagged. Requiring $\Delta R(\ell, b\text{-jet}) \leq 1.5$ ensures that one of these $b$-tagged jets will be close to the lepton candidate. The $b$-tagged \smallR\ jet that is closest to the lepton candidate is chosen as the \enquote{leptonic} $b$-jet. This reflects the requirement that these objects should originate
from the decay of the same top quark.
 
Additional requirements are imposed to further reduce backgrounds.
The hadronic top-quark candidate is required to have a jet mass within $50$ \GeV\ of the top-quark mass to reduce background contributions from boosted $W$ and $Z$ bosons. To further ensure that the correct RC \largeR\ jet is being chosen for measurement, the leptonic $b$-jet cannot be one
of the constituents of the hadronic top-quark candidate.
The lepton candidate is required to satisfy $\Delta R( \ell, \text{hadronic top-quark candidate} )\geq 1.0$ to ensure that it is outside the selected RC \largeR\ jet.
Finally, the invariant mass \mlb\ of the lepton candidate and the leptonic $b$-jet must satisfy $\mlb < 120$~\GeV, reducing contributions from misidentified and nonprompt lepton candidates and from single-top-quark production.

\subsection{All-hadronic event selection}
\label{sec:selection_allhad}
 
The all-hadronic \ttbar\ event selection is divided into two stages, with a preselection consisting of kinematic requirements on the jets, followed by a selection based on \largeR-jet flavor tagging.
 
Events are required to have a primary vertex, satisfy the jet triggers described in Section~\ref{sec:MCttbar}, and have two \largeR\ jets with $\pT > 350$~\GeV\ and $\abseta < 2.0$. The leading \largeR\ jet is required to have $\pT > 500$~\GeV, and both jets are required to have a jet mass within $50$~\GeV\ of the top-quark mass. Events are rejected if they contain an electron or muon candidate with $\pT > 25$~\GeV\ in order to suppress contamination from events with top quarks decaying semileptonically.
This lepton candidate veto also ensures that the two channels are orthogonal.
This defines the preselection for the all-hadronic channel.
 
The preselected events are divided into 16 categories, called \enquote{regions} in the following, depending on whether the leading and second-leading \largeR\ jets are top-quark-tagged and/or $b$-matched. These regions are used in the data-driven estimation of the multijet background described in Section~\ref{sec:ABCD}.
The regions used to measure the substructure are defined by requiring the leading and second-leading \largeR\ jets to be $b$-matched and one of the \largeR\ jets to be top-quark-tagged.
 
Only \largeR\ jets with \pT\ above 500~\GeV\ are used to measure the jet substructure.


\section{Background estimation}
\label{sec:background}

 
The backgrounds in the two channels are estimated using a combination of MC-based
techniques and data-driven measurements.
 
\subsection{MC-estimated backgrounds}
\label{sec:mcbackground}
 
All the backgrounds in the \ellplusjets\ channel, except those from nonprompt and misidentified leptons, are estimated using the MC simulations described in Section~\ref{sec:MCbkg}.
Single-top-quark production represents the largest background, with $tW$ production
being the dominant process.
The $W$+jets process represents the second-largest background, at the same level as the background from events containing nonprompt or misidentified lepton candidates.
Backgrounds from other sources, including the $Z$+jets, $VV(V=Z/W)$, and $\ttV(V=Z/W/H)$ processes, are
negligible.
 
The largest MC-estimated background in the all-hadronic channel
comes from \ttbar\ production where one of the top quarks decays
semileptonically and the charged lepton arising from this decay is not identified.
The next largest background arises from single-top-quark production in the $tW$-channel and $t$-channel.
The $\ttV\ (V=Z/W/H)$ processes produce the smallest backgrounds.
 
These MC-estimated backgrounds comprise ${\sim}6$\%\ and ${\sim}9$\%\ of the \ellplusjets\
and all-hadronic samples, respectively.
 
\subsection{Misidentified- and nonprompt-lepton backgrounds in the \ellplusjets\ channel}
\label{sec:fakesbackground}
 
Some background events in the \ellplusjets\ channel contain an isolated-lepton candidate that does not
originate from a $W$-boson decay.
This type of background includes lepton candidates arising from misidentified jets and leptons coming
from the weak decay of a hadron -- nonprompt (NP) leptons.
These background processes are estimated with a data-driven technique known as the matrix method~\cite{ATLAS-CONF-2014-058}, used in recent differential measurements in the $\ell$+jets channel~\cite{TOPQ-2019-23}.
 
The matrix method is based on \enquote{tight} and \enquote{loose} selections for the
lepton candidates.
The tight selection is the nominal one described in
Section~\ref{sec:objects_and_selection}, while the loose selection has
no isolation requirement and less stringent lepton identification requirements. Loose electron candidates
are reconstructed at the \enquote{MediumLH}~\cite{PERF-2017-01} working point, while the \enquote{Medium}~\cite{MUON-2018-03} working point is used for muon candidates.
The resulting misidentified- and nonprompt-lepton (NP/Misid.) background comprises 2\%\ of the sample.
 
\subsection{Multijet-background estimation in the all-hadronic channel}
\label{sec:multijetbackground}
 
\label{sec:ABCD}
 
The dominant background in the all-hadronic channel arises from multijet production and is estimated using a data-driven technique.
The present measurement uses the \enquote{ABCD} estimate, which was first developed~\cite{TOPQ-2016-09} to measure the \ttbar differential cross-section using an all-hadronic event selection.
This approach is adapted here to estimate the multijet background in bins of \largeR-jet substructure observables.
 
The technique uses the correlations between the mis-tagging rates of
$b$-tagging and top-quark-tagging algorithms for one \largeR\ jet to determine the multijet background.
As described in Section~\ref{sec:selection_allhad}, events passing the preselection are
divided into categories depending on whether the leading and second-leading \largeR\ jets are top-quark-tagged and/or $b$-matched, creating 16 regions with different multijet background contributions, shown in Figure~\ref{fig:FullABCD}. Among these regions, the ones where both leading \largeR\ jets are $b$-matched and at least one is top-tagged represent the signal regions. The two regions defined by requiring both \largeR\ jets to be top-tagged but only one to be $b$-matched are not directly used in the ABCD estimate, but employed as validation regions. The remaining 11 regions are considered as control regions.
 
The estimated contributions from \ttbar signal events and other MC-predicted backgrounds are
subtracted from the total event yields in the control regions, and the remaining event yields
are used as input to
calculate a bin-by-bin estimate of the multijet background in the signal and validation regions,
using the ABCD method, described in detail in Ref.~\cite{TOPQ-2018-11}.
The validation regions, labeled K and M in Figure~\ref{fig:FullABCD}, are employed to validate the modeling of the multijet background with the substructure variables.

\begin{figure}[t]
\begin{center}
\includegraphics[width=0.65\textwidth]{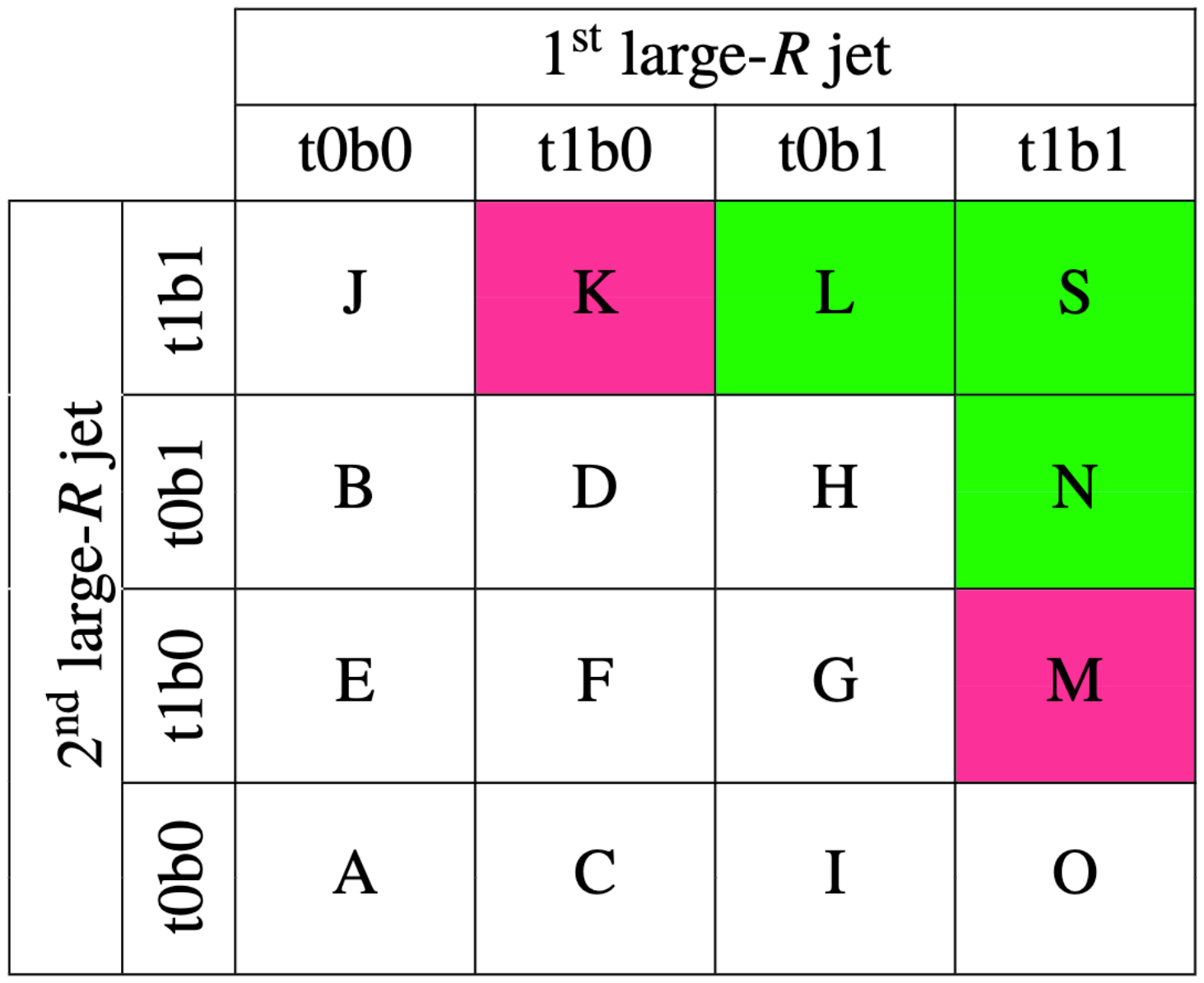}
\end{center}
\caption{Events passing the preselection requirements are divided into 16 regions
according to the $b$-matching and top-quark-tagging status of the leading and
second-leading \largeR\ jets, here labeled simply 1$^\text{st}$ and 2$^\text{nd}$ \largeR\ jet,
respectively.
Labels t1 and t0 indicate whether the \largeR\ jet has passed or failed the top-quark tagger,
respectively.
Labels b1 and b0 indicate whether the \largeR\ jet has or has not been $b$-matched,
respectively.
Regions L, S, and N constitute the regions used to
measure the jet substructure, while K and M are the validation regions.
}
\label{fig:FullABCD}
\end{figure}
\FloatBarrier

\subsection{Comparisons of data with predictions}
\label{sec:control}
 
The event yields for data, predicted \ttbar\ signal, and all backgrounds after applying the selection described in Section~\ref{sec:objects_and_selection} are shown for both channels in Table~\ref{table:Background_yields}.
Since the $Z$+jets and $VV(V=Z/W)$ backgrounds are small in the \ellplusjets\ channel, they are combined into the \enquote{Other} category in the table and following plots.
The event selection is able to separate the signal from the various background sources in both channels, with the signal purity reaching 92$\%$ and 65$\%$ in the \ellplusjets\ and all-hadronic channels, respectively.
 
The predicted yields overestimate the data yields in the \ellplusjets\ and all-hadronic channels by 16\% and 20\%, respectively.
This is consistent with observed yields in the boosted \ttbar\ regime in other measurements~\cite{TOPQ-2018-15,TOPQ-2016-01,TOPQ-2014-15,TOPQ-2016-09,CMS-TOP-18-013},
and is believed to come from missing higher orders in the NLOME+PS differential cross-section calculation, which mostly affect the high-\pt regions.
A correction for this overestimate is made by defining a normalization scaling evaluated as (Data -- Background)/Signal and shown in Table~\ref{table:Background_yields}.
This scaling is used to compare the shapes of the observed distributions with those predicted by the MC calculations and estimated backgrounds.

\begin{table}[t]
\caption{Detector-level event yields for data, simulated \ttbar signal, and background processes. All uncertainties affecting the detector and background modeling are included.
Due to the tag-and-probe method employed by the all-hadronic channel, the numbers here refer to the number of measured \largeR~jets, rather than events present.
The signal purity is defined by the number of data events minus background events divided by the total sample size.
}
 
\label{table:Background_yields}
\center
 
\sisetup{
group-minimum-digits = 4,
table-format = 4.0,
separate-uncertainty = true,
table-align-uncertainty = true,
}
\begin{tabular}[t]{lSS} \toprule{Category} & {Event yields} \hfil & {Number of \largeR\ jets} \hfil \\
& {\ellplusjets\ selection} \hfil & {all-hadronic selection} \hfil\\ \midrule
\textbf{{Data}}  & \textbf{\num{83069}} & \textbf{\num{30524}} \\
\midrule
\textbf{{Predictions}} &  \textbf{\num{97200(3700)}}\phantom{0} & \textbf{\num{36500(1400)}}\phantom{0} \\
{\quad\ttbar (\ellplusjets)} & \num{90600(3400)}\phantom{0} & \num{1610(140)}\phantom{\,0} \\
{\quad\ttbar (all-hadronic)} & {--} & \num{25700(1400)}\phantom{0} \\
{\quad Multijet} & {--} & \num{8100(300)}\phantom{\,0} \\
{\quad Single-top quark} & \num{2200(300)}\phantom{\,0} & \num{710(70)}\phantom{0} \\
{\quad NP/Misid.\ leptons} & \num{1500(600)}\phantom{\,0} \\
{\quad $W$+jets} & \num{1500(700)}\phantom{\,0} & {--} \\
{\quad $t\bar{t}V$} (\ttZ + \ttW + \ttH)  & \num{920(120)} & \num{310(40)}\phantom{0} \\
{\quad Other} & \num{400(200)} & {--} \\
\midrule{Data/Predictions} & \num{0.85(3)} & \num{0.84(3)} \\ \midrule{(Data -- Background)/Signal} & \num{0.84(3)} & \num{0.77(5)} \\ \midrule{Signal purity} & \num{0.92(1)} & \num{0.65(1)} \\ \bottomrule
\end{tabular}
\end{table}
 
Comparisons of the observed kinematic distributions with the ones predicted by the nominal signal MC sample and background estimates in the \ellplusjets\ channel are shown in Figure~\ref{fig:control_plot_ljets_0}.
The figure shows the distributions of the number of RC \largeR\ jets in the event, the \pT\ of the RC \largeR\ jet,
and \mlb.
Comparisons between the observed and predicted kinematic distributions of the leading and second-leading \largeR\ jets in the all-hadronic channel are shown in Figure~\ref{fig:control_plot_allhad_0}.
The shapes of most of the compared distributions are in good agreement,
validating the modeling of the kinematics of the jets in \ttbar\ events and the estimated backgrounds.
The only discrepancy appears in the comparison of the \pT~distributions of the top-quark candidate jets, where the predicted distributions have a flatter \pT~dependence consistent with the increasing overestimate of the \ttbar\ cross-section as a function of top-quark \pT\ as noted above.

\begin{figure}[!ht]
\centering
\subfloat[]{\label{fig:control_plot_ljets_0:1}\includegraphics[width=0.30\textwidth]{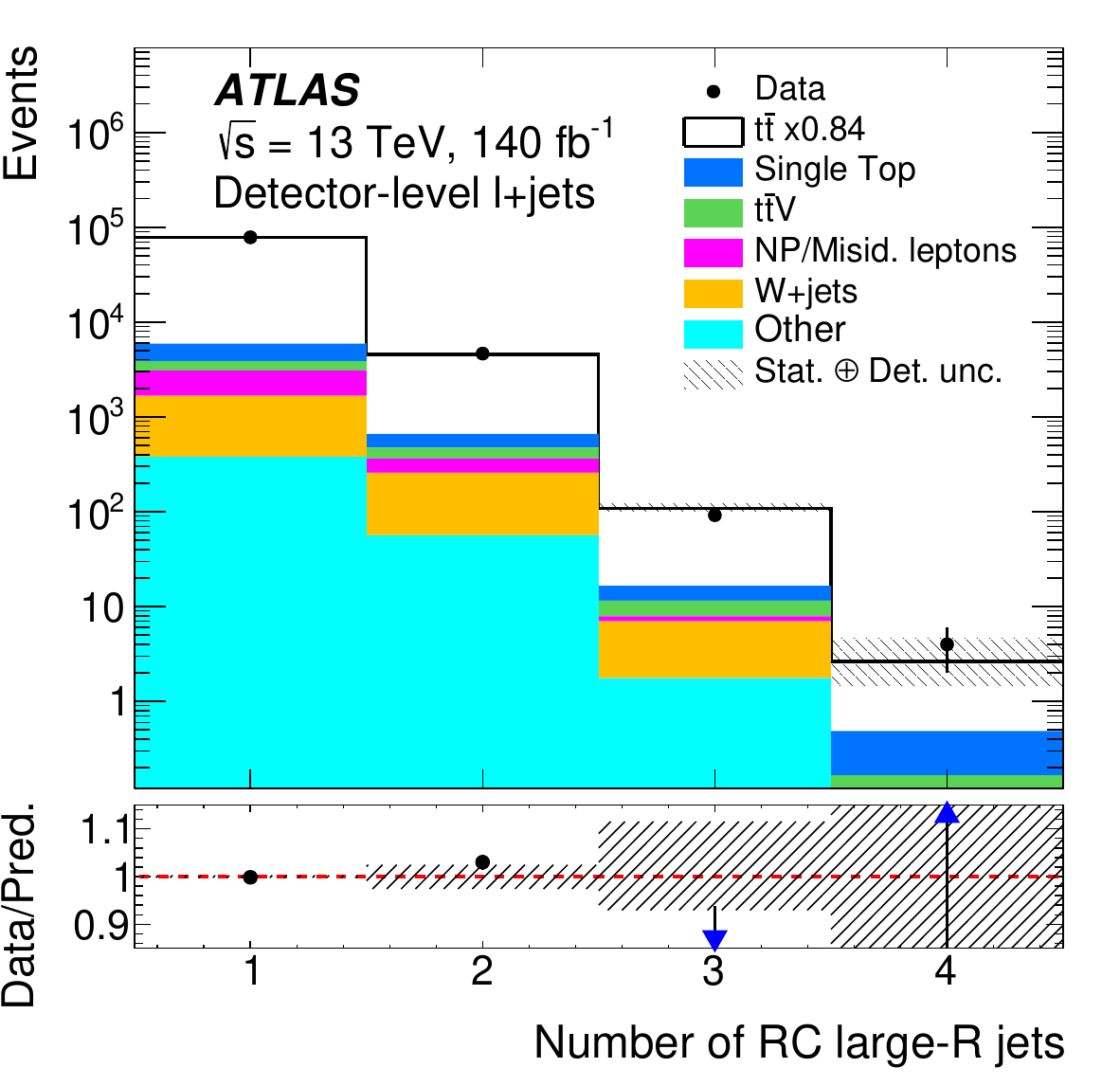}}
\subfloat[]{\label{fig:control_plot_ljets_0:2}\includegraphics[width=0.30\textwidth]{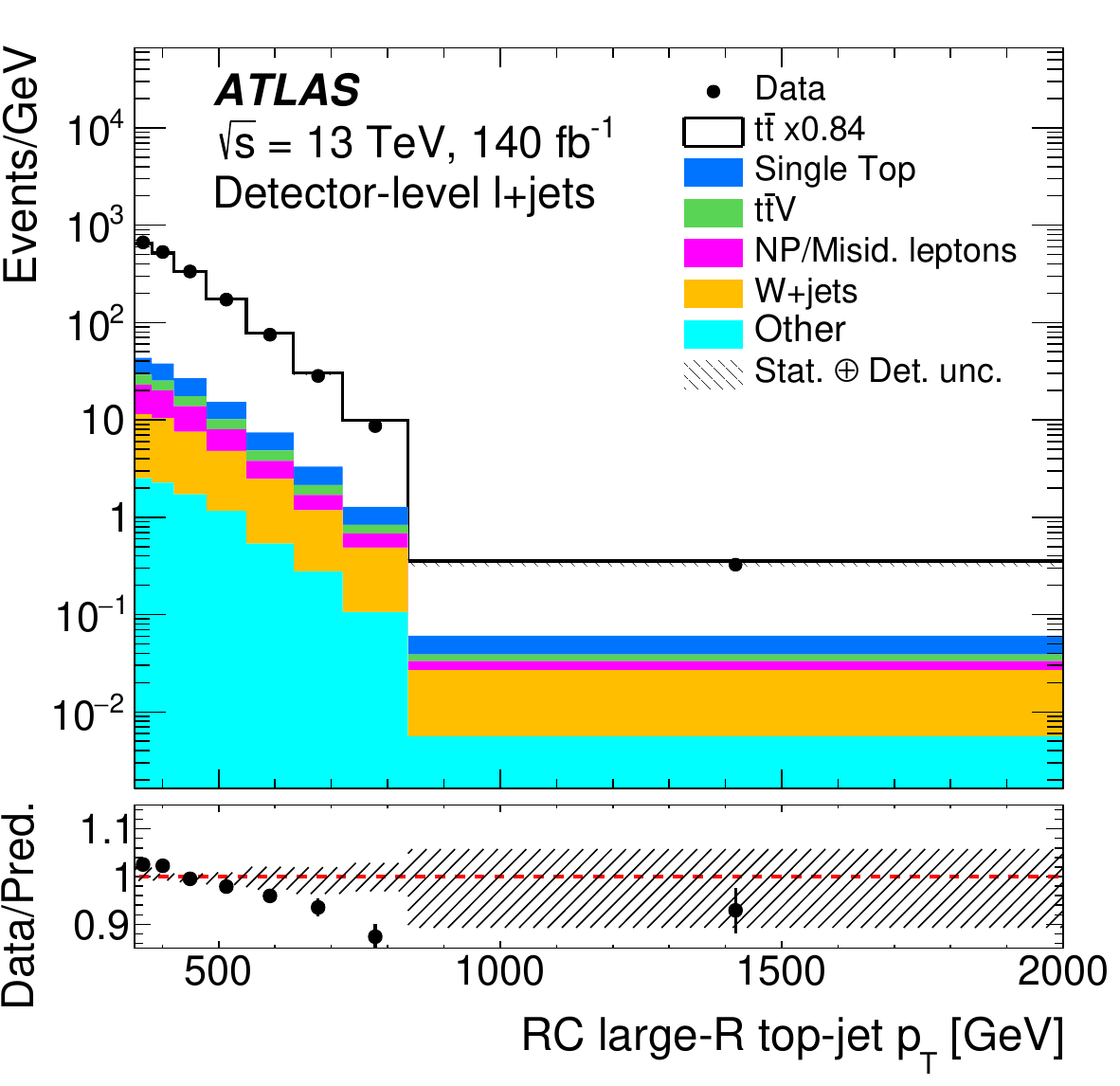}}
\subfloat[]{\label{fig:control_plot_ljets_0:3}\includegraphics[width=0.30\textwidth]{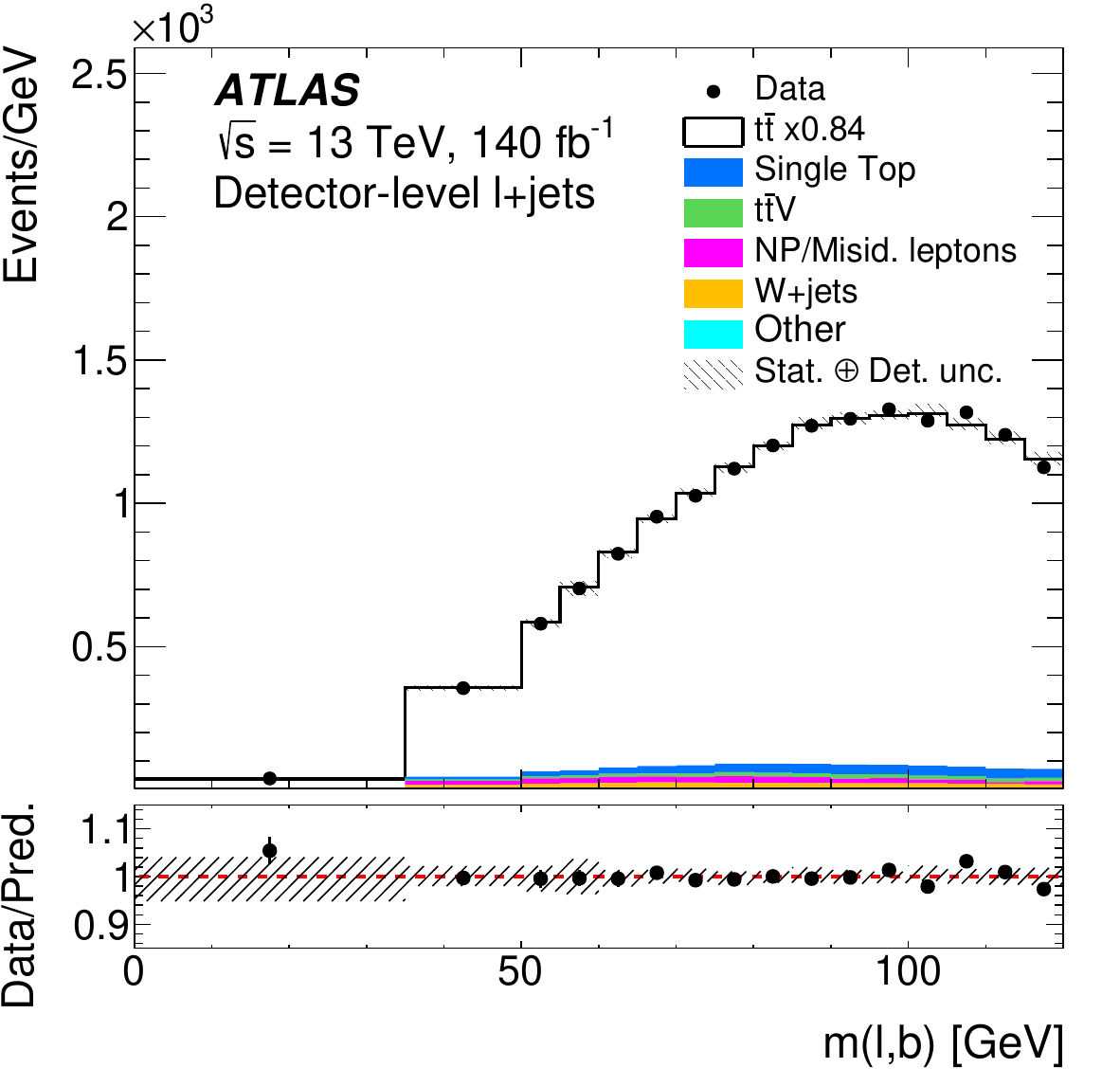}}
\caption{Detector-level distributions in the \ellplusjets\ channel for
\subref{fig:control_plot_ljets_0:1} the number of RC \largeR\ jets,
\subref{fig:control_plot_ljets_0:2} the \pT\ of the  hadronic-top-quark candidate RC \largeR\ jet, and
\subref{fig:control_plot_ljets_0:3} $\mlb$.
The lower graph shows the ratios of the data to the predictions; triangles in \subref{fig:control_plot_ljets_0:1} indicate ratio values outside the shown range.
The hatched band represents the sum of the detector-related uncertainties in the predictions.
The statistical uncertainties are shown for the data points.
Larger values of the observables for the left and middle plots, beyond the histogram range, are included in the last bin.
The predicted $\ttbar$\ yield is adjusted so that the total predicted yield equals the observed number of data events.
The bin contents are divided by the bin widths.
}
\label{fig:control_plot_ljets_0}
\end{figure}

\begin{figure}[!ht]
\centering
\subfloat[]{\label{fig:control_plot_allhad_0:1}\includegraphics[width=0.30\textwidth]{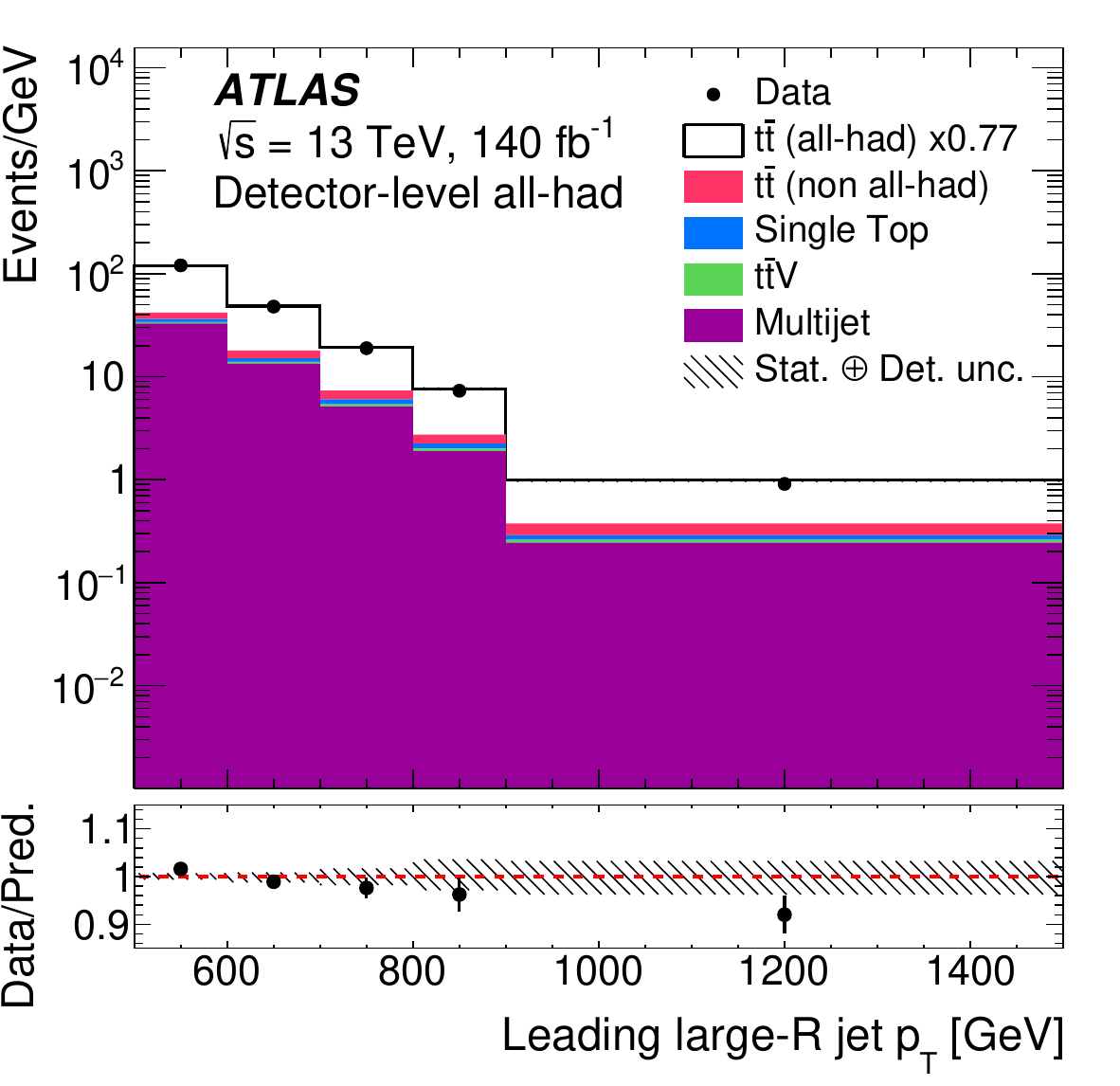}}
\subfloat[]{\label{fig:control_plot_allhad_0:2}\includegraphics[width=0.30\textwidth]{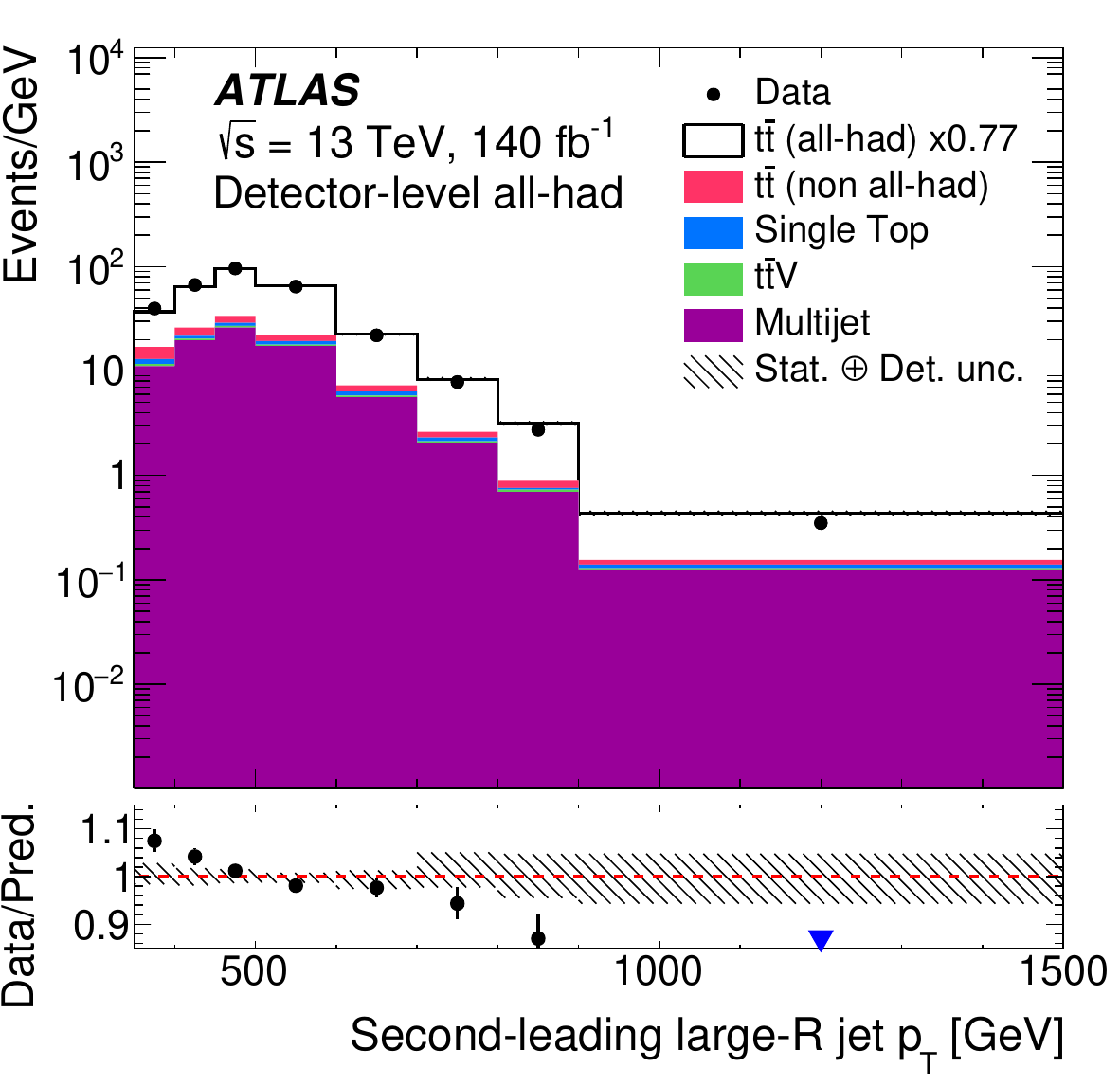}}
\subfloat[]{\label{fig:control_plot_allhad_0:3}\includegraphics[width=0.30\textwidth]{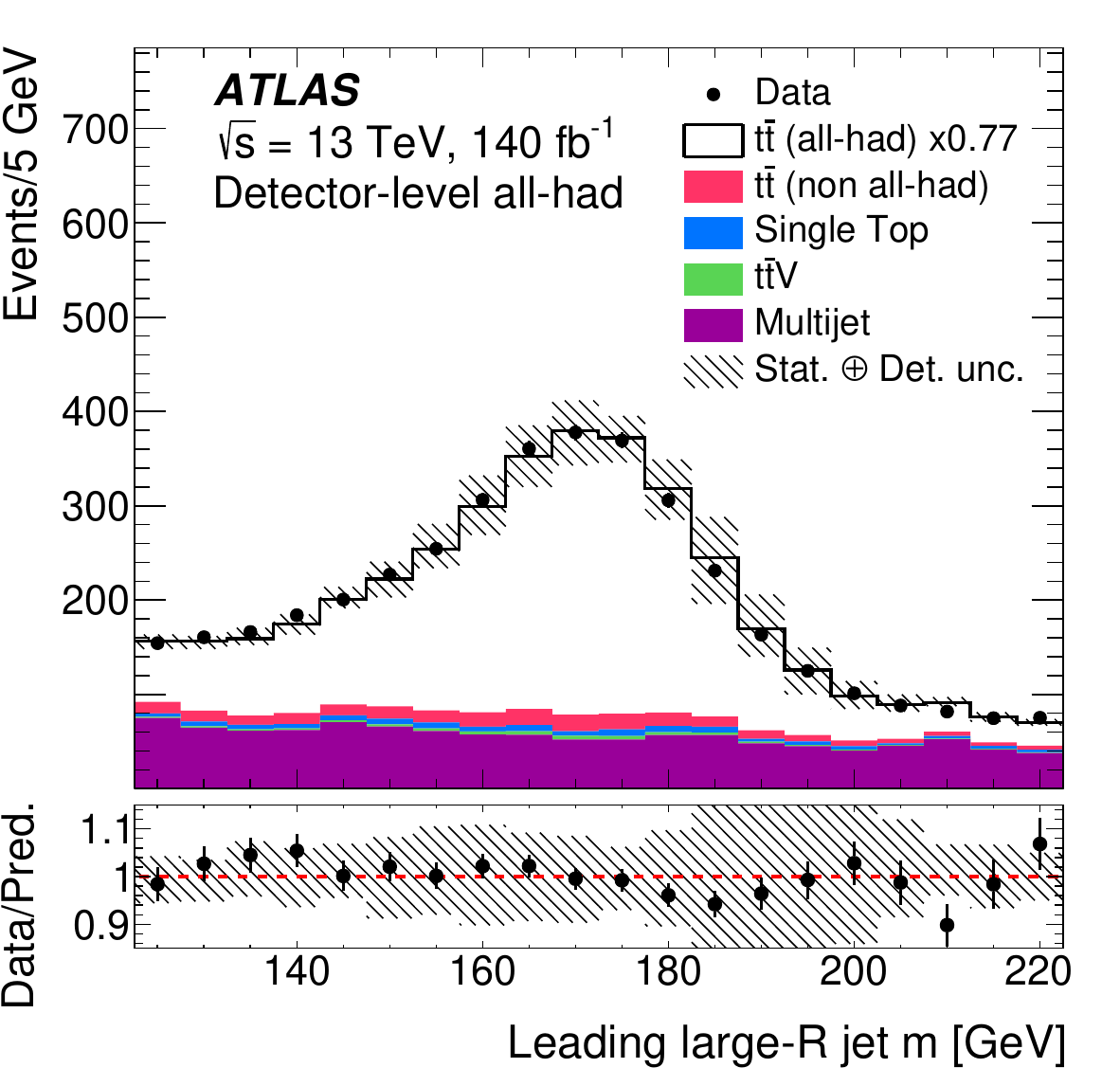}}
\caption{Detector-level distributions in the all-hadronic channel for
\subref{fig:control_plot_allhad_0:1} the \pT\ of the leading \largeR\ jet,
\subref{fig:control_plot_allhad_0:2} the \pT\ of the  second-leading \largeR\ jet, and
\subref{fig:control_plot_allhad_0:3} the mass of the leading \largeR\ jet.
The lower graph shows the ratios of the data to the predictions; the triangle in \subref{fig:control_plot_allhad_0:2} indicates a ratio value below the shown range.
The hatched band represents the sum of the detector-related uncertainties in the predictions.
The statistical uncertainties are shown for the data points.
Larger values of the observables for the left and middle plots, beyond the histogram range, are included in the last bin.
The predicted $\ttbar$\ event yield is adjusted so that the total predicted yield equals the observed number of data events.
The bin contents are divided by the bin widths.
}
\label{fig:control_plot_allhad_0}
\end{figure}


\section{Measured substructure observables}
\label{sec:variables}

The observed substructure variables are calculated using the charged tracks associated with the hadronic top-quark candidate.
In the all-hadronic analysis the charged tracks are ghost-associated with the \largeR\ jet. In this case, the trimming procedure, described in Section~\ref{sec:objects_reco}, is applied concurrently to the ghost-tracks. In the $\ell$+jets channel, tracks are ghost-associated with the \smallR\ jets that constitute the reclustered \largeR\ jet.
 
Several classes of observables are measured:
\begin{itemize}
\item Generalized angularities~\cite{Larkoski_2014_ga}:
\begin{equation*}\label{eq:GALHC}
\lambda_{\beta}^{\kappa}=\sum_{i \in J} z_{i}^{\kappa} \left(\frac{\Delta R(i,\hat{n})}{R}\right)^\beta.
\end{equation*}
Given a jet $J$, $z_{i}$ is the fraction of the jet's \pt carried by the $i^{\text{th}}$ constituent, $\hat{n}$ is the jet axis, and $R$ is the jet radius.
The variables $\kappa$ and $\beta$ are parameters chosen to
weight the \pt and angular terms, respectively.
In this analysis, only $\lambda^2_0$ and $\lambda^1_{0.5}$ are measured and are referred to as the
jet \pt dispersion ($p_{\text{T}}^{\text{d}}$) and the Les Houches Angularity (LHA), respectively.
The jet \pt dispersion is highly correlated with the particle multiplicity ($N$) and thus a scaled \pt dispersion variable is used to mitigate this correlation:
\begin{equation*}\label{eq:scaled_pt_dispersion}
p_{\text{T}}^{\text{d,*}}=\sqrt{ \left(p_{\text{T}}^{\text{d}} - \frac{1}{N}\right)\frac{N}{N-1} }.
\end{equation*}
 
\item Energy correlation functions (ECFs)~\cite{Larkoski_2013, Larkoski_2016}: the general formula for these functions in a jet produced by a hadron collider is
\begin{equation*}\label{eq:ECF}
\operatorname{ECF}(N)=\sum_{i_{1}<i_{2}<\ldots<i_{N} \in J}\left(\prod_{a=1}^{N} p_{\mathrm{T}, i_{a}}\right)\left(\prod_{b=1}^{N-1} \prod_{c=b+1}^{N} \DeltaR(i_{b}, i_{c})\right).
\end{equation*}
Here the sum runs over all constituents of the jet, ordered by index $i$ so that no pairwise calculations are repeated, and $N$ represents the number of prongs.
Ratios of the ECFs can be used to probe $N$-pronged substructure. Here $C_{3}$ and $D_{2}$ are measured:
\begin{equation*}
C_{3}=\frac{\operatorname{ECF}(4) \operatorname{ECF}(2)}{\operatorname{ECF}(3)^{2}},
\end{equation*}
 
\begin{equation*}
D_2=\frac{\operatorname{ECF}(3) \operatorname{ECF}(1)^3}{\operatorname{ECF}(2)^{3}}.
\end{equation*}
 
A jet with a pronounced three-body (two-body) structure will have $C_{3}$ ($D_{2}$) close to 0.
The normalized energy correlation function \ECF\ is also measured. The normalized version is used to reduce the dependence on the jet-constituent multiplicity:
\begin{equation*}
ECF2 = \frac{\text{ECF(2)}}{\text{ECF(1)}^2}.
\end{equation*}
 
\item $N$-subjettiness~\cite{Thaler_2011}: a jet's $N$-subjettiness provides a measure of the degree to which that jet is compatible with comprising $N$ or fewer subjets. It is defined as:
\begin{align*}
\tau_{N}&=\frac{1}{d_{0}} \sum_{k} p_{\mathrm{T}, k} \min \left\{\Delta R_{1, k}, \Delta R_{2, k}, \cdots, \Delta R_{N, k}\right\} ,\\
\text{with }d_{0}&=\sum_{k} p_{\mathrm{T}, k} R_{0}.
\end{align*}
 
The constituents of the jet are reclustered using the exclusive $k_t$ algorithm~\cite{Ellis_1993}, with the winner-takes-all (WTA) recombination scheme. The clustering continues until $N$ subjets are returned. The calculation of $\tau_{N}$ then runs over the constituents of the jet, with $p_{\mathrm{T},k}$ the \pt of the $k^{\text{th}}$ constituent. The variable $R_0$ is the jet radius parameter. The ratios $\tau_{21}\equiv\tau_{2}/\tau_{1}$ and $\tau_{32}\equiv\tau_{3}/\tau_{2}$ are measured, along with $\tau_{3}$.
A jet with a pronounced three-body (two-body) structure will have $\tau_3$ ($\tau_2$) close to 0. The ratios are used to discriminate between jets with different $N$-prong structures, e.g.\ a jet better described by a 3-prong structure than by a 2-prong one will have a $\tau_{32}$ value closer to 0 than to 1.
\end{itemize}
This set of eight substructure variables includes observables sensitive to the modeling of three-prong  ($\tau_{32}$, $C_{3}$) or two-prong ($D_{2}$, $\tau_{21}$) objects, to the distribution of the momentum of the constituents inside the jet (\ptdispersion) and to the broadness of the jet (\LHA). Two of these variables ($D_{2}$, $\tau_{32}$) are the most sensitive in the multivariable discriminator developed by the ATLAS Collaboration~\cite{JETM-2018-03} to identify $W$-boson- and top-quark-initiated jets. Given the importance of $D_{2}$ and $\tau_{32}$ for particle tagging, these observables are also measured as a function of the particle-level top-quark-jet mass (\mtop ) and \pt (\pTtop ).
The \ECF\ and $\tau_{3}$ observables are also measured, since they are relevant in the calculation of $D_{2}$, $C_{3}$ and $\tau_{32}$.
The substructure variable $\tau_{3}$ is selected also because it appears to be poorly described by MC predictions.
These variables are not highly correlated and show sensitivity to showering, hadronization, and final-state radiation.

In the all-hadronic channel, the substructure measurement takes advantage
of the presence of two \largeR\ jets.
It follows a tag-and-probe approach that involves first measuring the substructure distribution of the leading \largeR\ jet by requiring that the second-leading \largeR\ jet is top-quark tagged.
In this configuration, the leading \largeR\ jet is the \enquote{probe} jet and the second-leading jet
is the \enquote{tag} jet.
The substructure distribution of the second-leading jet is then measured in those events
where it satisfies the same kinematic requirements as the leading jet, namely
$\pT > 500$~\GeV, and
where the leading jet is top-quark tagged.
This avoids biasing the substructure of the probe jet through
the substructure-dependent DNN top-quark-tagging algorithm.
This technique requires that there is no event-to-event correlation in the substructure measurement
in the leading and second-leading jets, which is confirmed by MC studies.
 
The distributions of the substructure variables for the leading and second-leading \largeR\ jets are
added together to produce a combined measurement.
The combined distribution is then unfolded using the technique described in
Section~\ref{sec:unfolding}.
This combined measurement increases the number of jets
by ${\sim}50$\%.
 
The observed shapes of the distributions of the substructure observables are compared with the
nominal \ttbar\ model predictions and background estimates for both the \ellplusjets\ and
all-hadronic channels.
While the observed and predicted shapes are generally in good agreement, the
distributions of some observables, such as  \ptdispersion\ or $\tau_{32}$, are poorly described by the predictions.
The latter observable, $\tau_{32}$, is shown in Figure~\ref{fig:control_plot_measure_combined_15}, where only the statistical uncertainty and detector-level systematic uncertainties are included.

\begin{figure}[t]
\centering
\subfloat[]{\label{fig:control_plot_measure_combined_15:1}\includegraphics[width=0.45\textwidth]{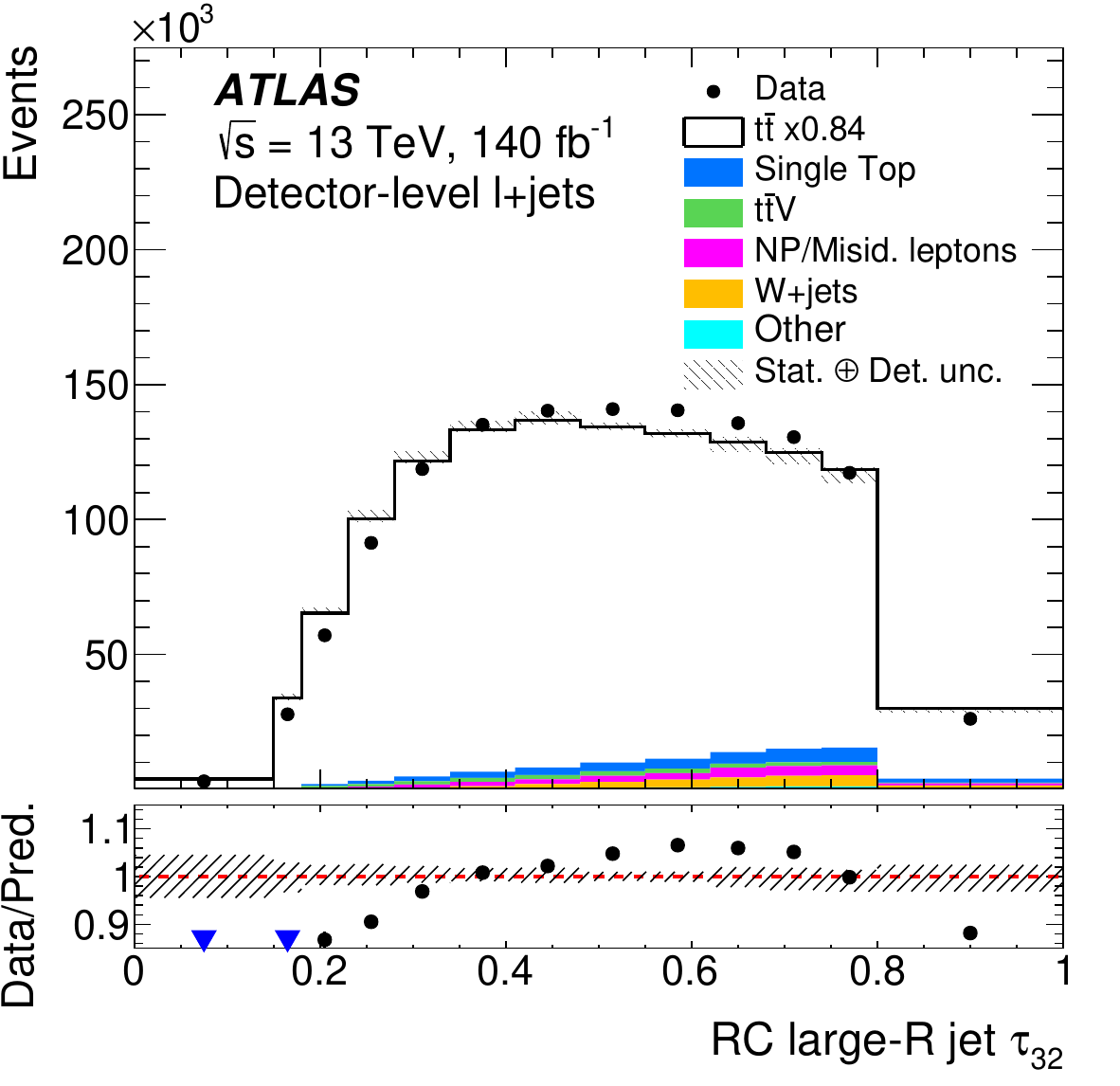}}
\subfloat[]{\label{fig:control_plot_measure_combined_15:2}\includegraphics[width=0.45\textwidth]{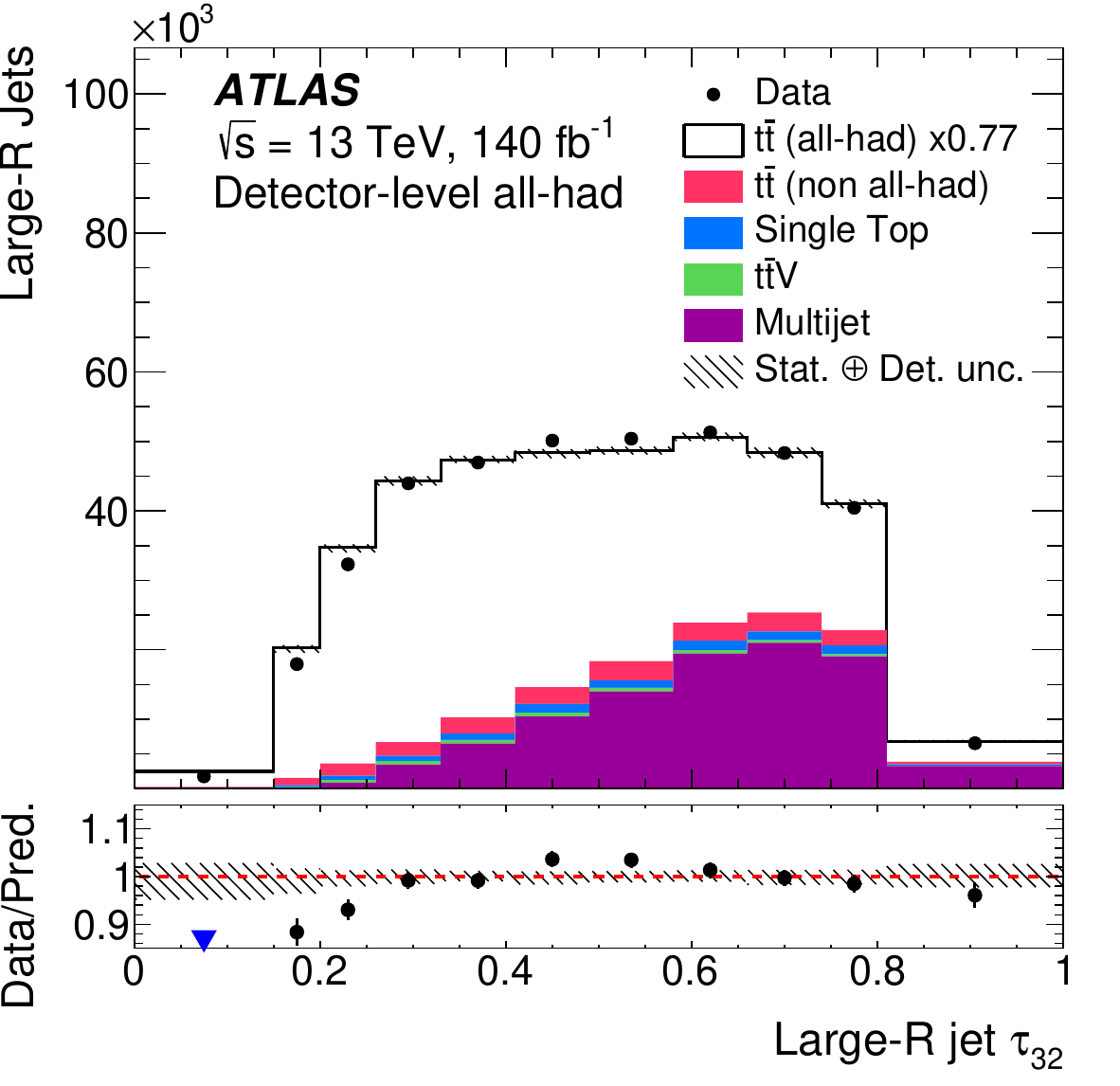}}
\caption{
Detector-level $\tau_{32}$ distributions in the
\subref{fig:control_plot_measure_combined_15:1} $\ell+$jets and
\subref{fig:control_plot_measure_combined_15:2} all-hadronic channel
for data, predicted \ttbar\ signal, and measured backgrounds.
The predicted \ttbar\ signal is normalized so that the total predicted yield agrees with the observed number of events.
The lower pad shows the ratios of the data to the predictions; triangles indicate ratio values below the shown range.
The hatched band represents all the detector-related uncertainties of the predictions.
The modeling uncertainties of the signal are not included in the band.
The data points are presented with their statistical uncertainty.
}
\label{fig:control_plot_measure_combined_15}
\end{figure}


\section{Cross-section measurement}
\label{sec:unfolding}

The observed distributions are affected by the detector and selection acceptance, resolution, and efficiency. An unfolding technique is used to correct for these effects. It unfolds the observed distributions to particle level in a suitably chosen kinematic region called the fiducial phase-space.

\subsection{Particle-level fiducial phase-space}
\label{sec:objects_particle}
 
The particle-level objects are defined using the stable particles (those with $c\tau > 10$~mm where
$\tau$\ is the proper lifetime of the particle) in the MC calculations.
The selection requirements imposed on these objects define the particle-level fiducial phase-space,
and are similar to the detector-level requirements so that  large extrapolations of the measured quantities are avoided.
 
Electrons and muons from the
electroweak decays of top quarks are selected by rejecting those that
come from the decay of hadrons, directly or through $\tau$-lepton decays.
Leptons arising from $\tau$-lepton decays that are not the result of prompt hadron decays
are also accepted.
Leptons are then \enquote{dressed} by summing their four-momenta with any prompt photons within
$\DeltaR = 0.1$.
The dressed leptons are required to have $\pt > 27$~\GeV\ and $\abseta < 2.5$.
Dressed muons and electrons separated by $\Delta R < 0.4$\ from a jet are excluded.
Neutrinos are required to come from a top-quark electroweak decay, and the
\met is calculated from the four-momenta sum of the selected neutrinos.
 
Particle-level \smallR\ and \largeR\ jets are built by clustering all stable particles,
except for the selected dressed leptons, neutrinos not coming from hadron decays, and photons not used in the dressing.
The decay products of hadronically decaying $\tau$-leptons are included.
The RC \largeR\ jets are built by the \antikt jet-clustering algorithm with a radius parameter
of $R=1.0$, using the particle-level \smallR\ jets as input.
The same trimming algorithms as described earlier for \largeR\ jets and RC \largeR\ jets are employed
for the particle-level objects.
 
Particle-level $b$-tagging is performed by ghost-matching $b$-hadrons to jets.
A particle-level \smallR\ or \largeR\ jet with at least one ghost-matched $b$-hadron with $\pt > 5$~\GeV\  is defined as being $b$-tagged.
No specific top-quark tagging is performed on particle-level \largeR\ jets beyond the requirement on the \largeR\ jet mass, and particle-level \largeR\ jets are not required to be matched to parton-level top quarks.
 
The event selection for both channels closely follows the detector-level selection described in Sections~\ref{sec:selection_ljets} and~\ref{sec:selection_allhad}, with the only difference being in the
selection of the jets used to reconstruct the variables in the all-hadronic channel, where the tag-and-probe method is not applied at particle level. Top-quark candidates in the event are required to be $b$-tagged, and the ones with $\pT > 500$~\GeV\ are employed to reconstruct the jet substructure.
The reconstruction of the jet substructure at particle level closely follows the definition employed at detector level.
In the all-hadronic case, only the charged components of the \largeR\ jets are considered as input when reconstructing the observables.
In the \ellplusjets\ case, the substructure observables are built by using the charged constituents of the \smallR\ jets as components of the RC \largeR\ jets. In both channels, only charged particles with  \pt above 500~\MeV\ and $|\eta|<2.5$\ are considered.

\subsection{Iterative Bayesian-unfolding method}
\label{sec:unfolding_iterbayes}
 
The iterative Bayesian-unfolding algorithm~\cite{DAgostini:1995}, as implemented in RooUnfold~\cite{Adye:2011gm}\ is employed to unfold the observed data distributions to the particle-level
distributions.
The differential cross-section (both single and double) in bin $j$ of a particle-level distribution is
represented heuristically as
\begin{eqnarray}
\label{eqn:unf}
\frac{\dif\sigma_{j}}{\dif X_{j}} = \frac{1}{\int\!{\cal{L}} \dif t \cdot \Delta X_j} \frac{1}{f^{\mathrm{eff}}_{j}} \sum_{i} M_{ij}^{-1} f_{\mathrm{acc}}^{i} (D^{i}-B^{i}),
\end{eqnarray}
where $i$ runs over the detector-level bins, $X$ represents one or a pair of
variables, $\int\!{\cal{L}} \dif t$\ is the integrated luminosity of the sample,  and
$\Delta X_j$\ represents the width of bin $j$.
The variable $D_i$\ represents the data histogram obtained after the event selection,
and $B_i$\ represents the distributions of the estimated backgrounds.
By concatenating the bins of a given substructure observable for each region of \pTtop\ or \mtop, a one-dimensional distribution can be constructed in order to apply this formula to the double-differential distributions~\cite{TOPQ-2018-15}.
The functions $f^{\mathrm{eff}}$\ and $f_{\mathrm{acc}}$ are unfolding corrections, derived using the
nominal \POWPY[8]\ prediction for $t\bar{t}$ production and decay.
The efficiency correction $f^{\mathrm{eff}}$\ accounts for events passing the
particle-level phase-space requirements but not the detector-level requirements.
The acceptance correction $f_{\mathrm{acc}}$\ accounts for events that pass the detector-level
selection but
fail the particle-level fiducial phase-space requirements.

Examples of the $f_{\mathrm{acc}}$ and $f^{\mathrm{eff}}$ corrections in the \ellplusjets\
and all-hadronic channels are illustrated in
Figure~\ref{fig:corr_ljets_and_allhad}. The efficiency is driven by the use of algorithms which are not fully efficient, such as $b$- and top- tagging. The acceptance is driven by the resolution on the quantities used for the selection requirements. As a consequence, the unfolding corrections differ between the two channels due to the differences in the selection: the cuts on \met and \mlb\ decrease the acceptance in the \ellplusjets\ channel, while the requirements on the number of $b$-tagged jets and the top-tagging lowers the efficiency in the all-hadronic channel.
 
\begin{figure}[t]
\centering
\subfloat[]{\label{fig:corr_ljets_and_allhad:1}\includegraphics[width=0.40\textwidth]{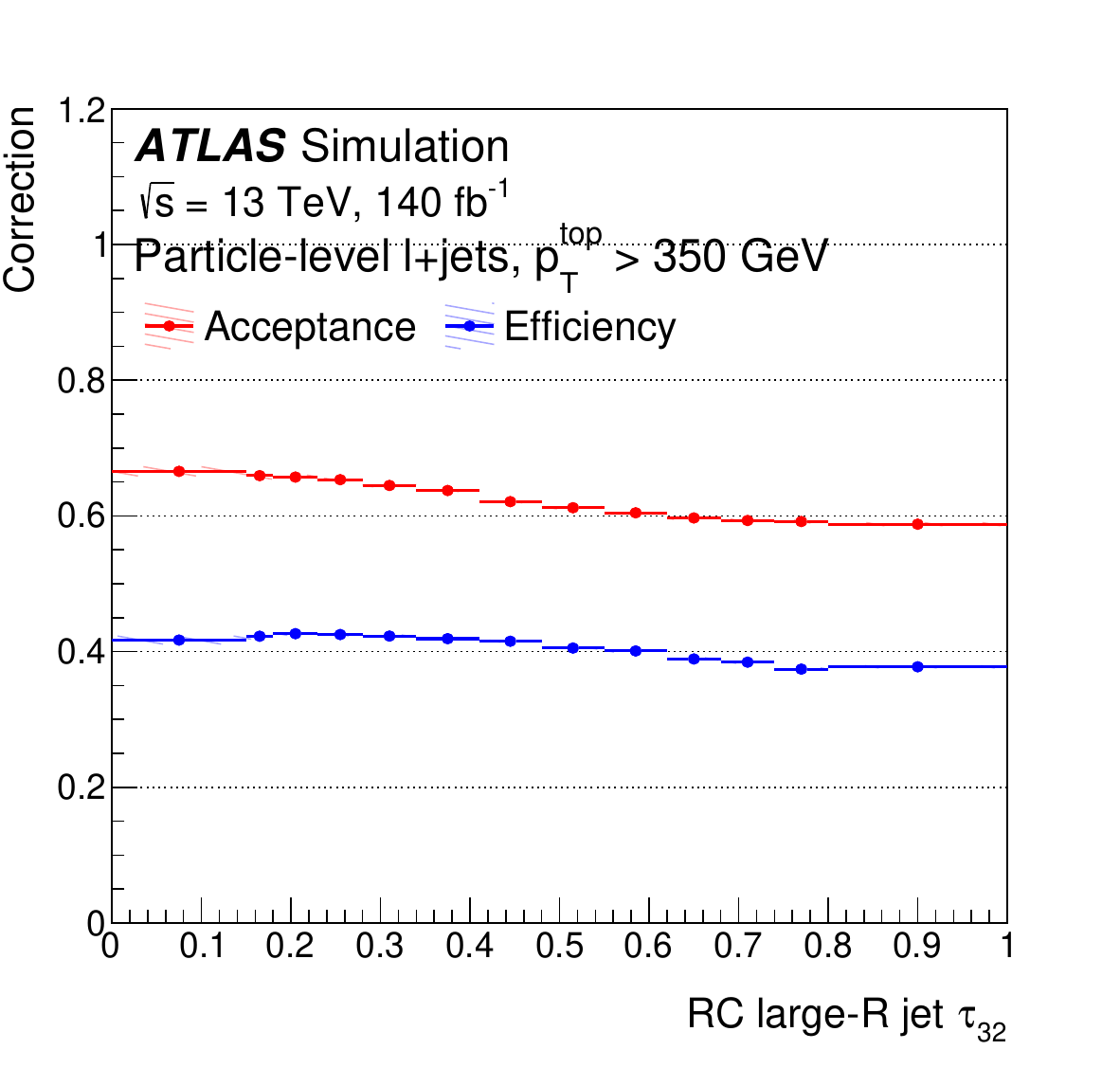}}
\subfloat[]{\label{fig:corr_ljets_and_allhad:2}\includegraphics[width=0.40\textwidth]{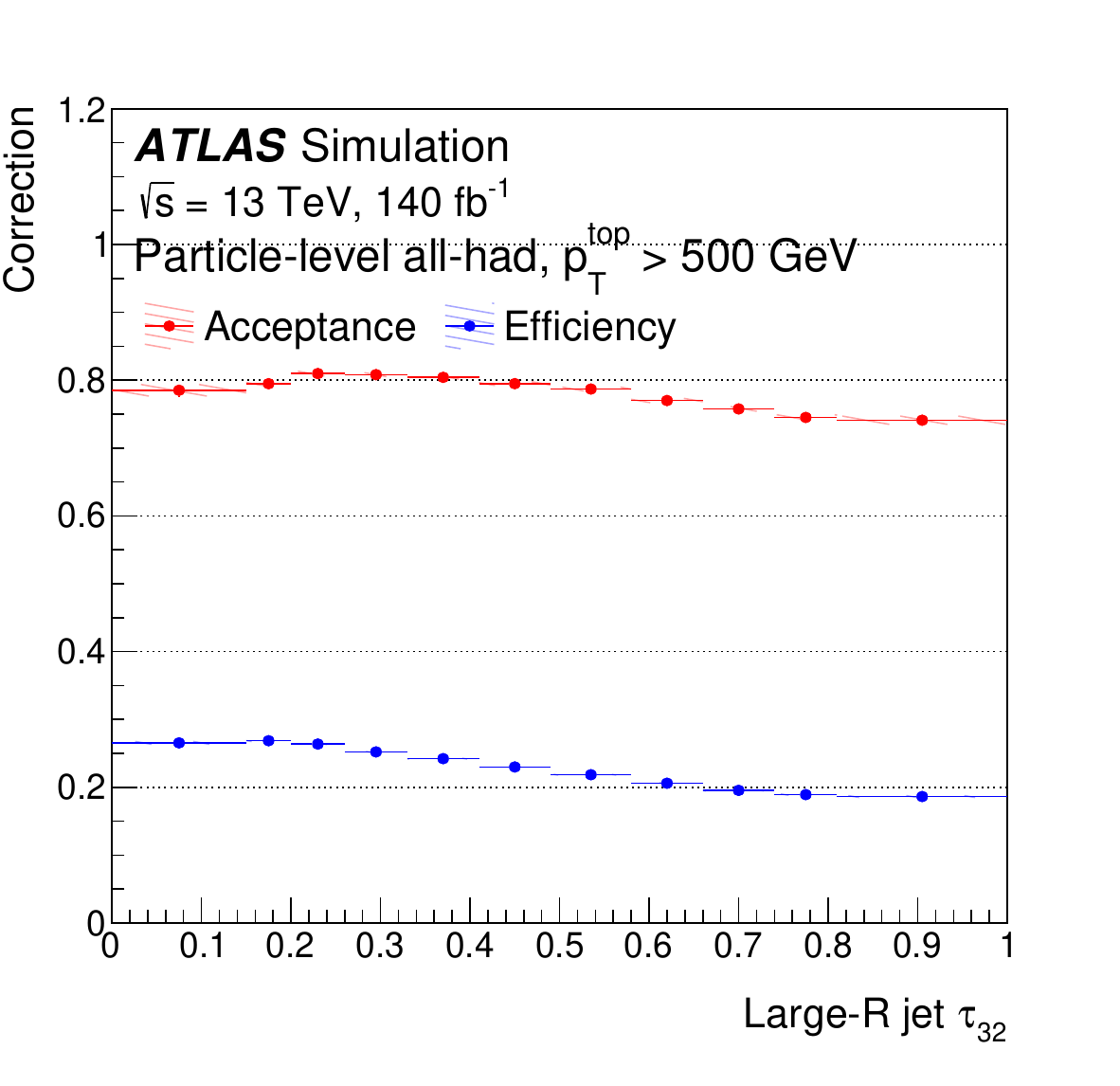}}
\caption{
Acceptance and efficiency corrections for the $\tau_{32}$ variable in the
\subref{fig:corr_ljets_and_allhad:1} \ellplusjets\ and
\subref{fig:corr_ljets_and_allhad:2} all-hadronic channels.}
\label{fig:corr_ljets_and_allhad}
\end{figure}
 
The migration matrix $M$\ quantifies the smearing that arises from detector-resolution effects.
It is evaluated by using MC events that pass both the detector-level and particle-level fiducial phase-space selections and binning them into the matrix, where the rows and columns represent the particle-level and detector-level bins, respectively.
Each row is normalized to sum to unity.
In Eq.~(\ref{eqn:unf}), $M^{-1}$ represents the effects of the iterative Bayesian-unfolding procedure that is employed to unfold the detector-level distributions.
The amount of regularization introduced by this unfolding method is determined by the number of iterations. A large number of iterations corresponds to fully unregularized matrix-inversion. The number of iterations employed in this analysis for all distributions is six, chosen as a compromise between the bias and the statistical fluctuations. The absence of a non-negligible bias is verified with the methods described in Section~\ref{sec:unf_val}.
 
The bin widths chosen for the substructure distributions optimize the robustness of the unfolding procedure while retaining as many bins as possible to resolve detailed features in the unfolded distributions.
The first criterion is addressed by employing bin widths that result in unfolding matrices that are largely diagonal by requiring that at least 40\%\ of the jets fall in the diagonal bins.
 
The migration matrices for the observable ${\tau_{32}}$\ in
the \ellplusjets\ and all-hadronic channels are shown in Figure~\ref{fig:migration}, and
are similar to the other migration matrices.
 
\begin{figure}[t]
\centering
\subfloat[]{\label{fig:migration:1}\includegraphics[width=0.40\textwidth]{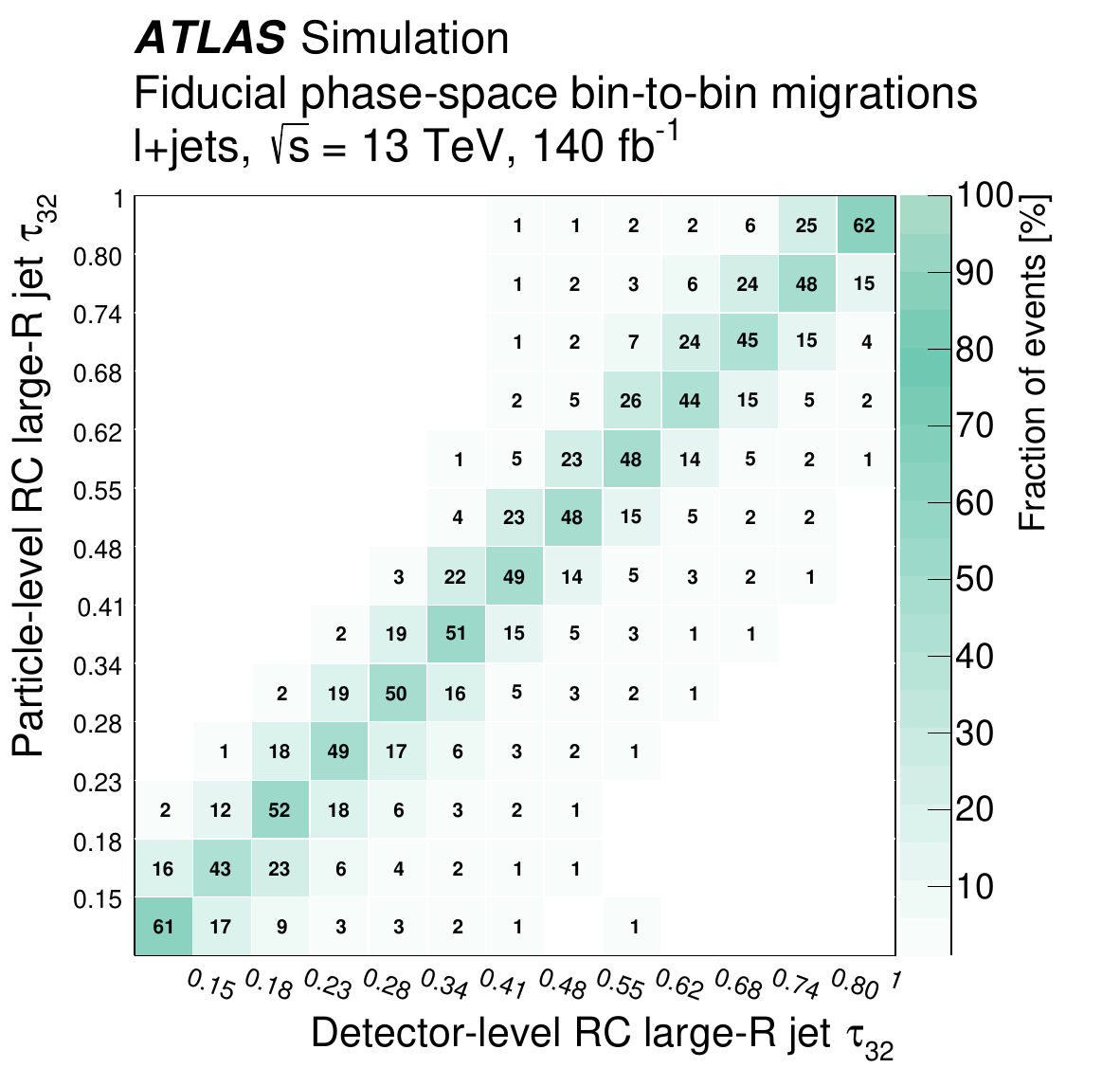}}
\subfloat[]{\label{fig:migration:2}\includegraphics[width=0.40\textwidth]{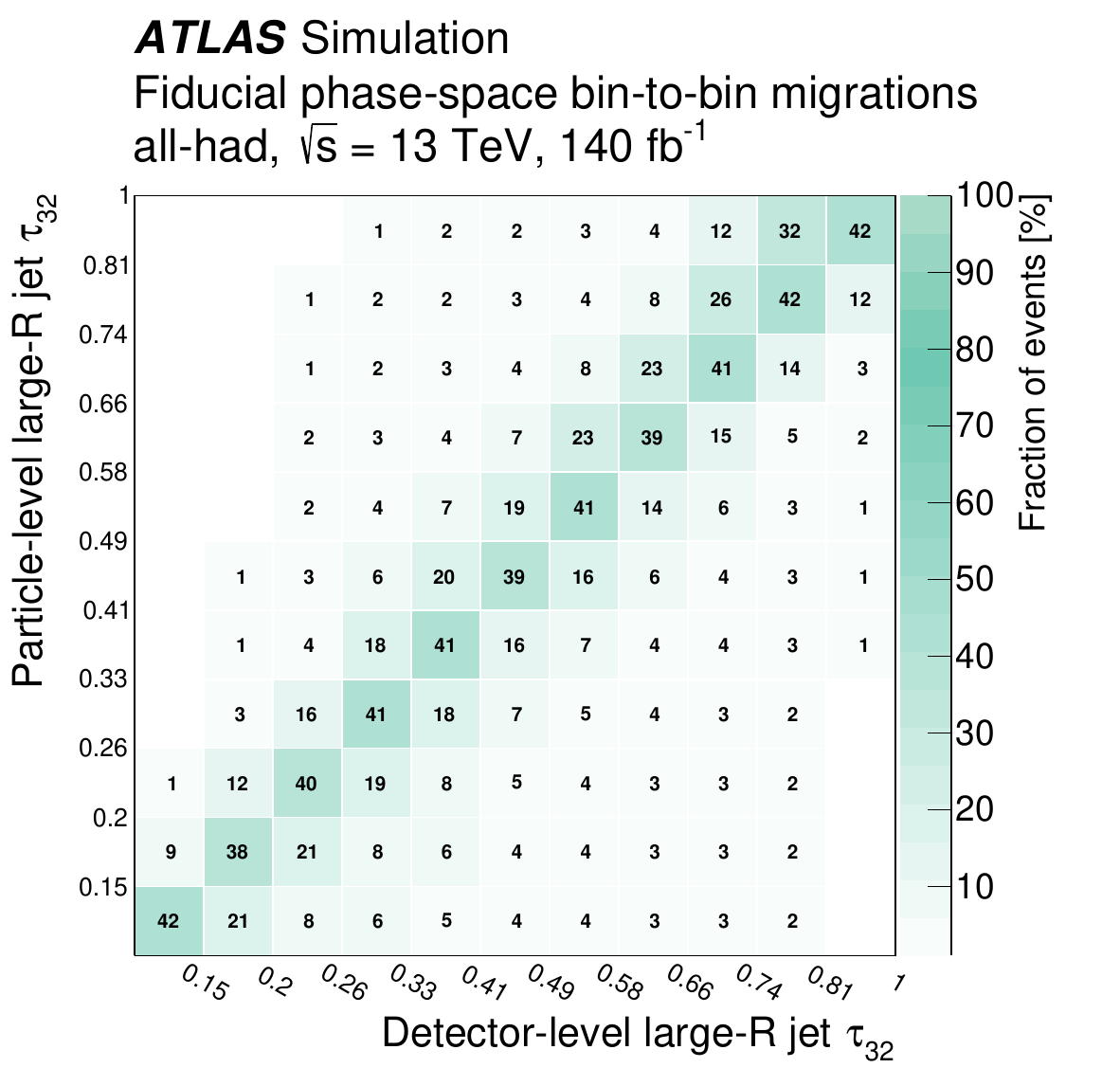}}
\caption{
Migration matrices for the variable $\tau_{32}$ in the
\subref{fig:migration:1} \ellplusjets\ channel and
\subref{fig:migration:2} all-hadronic channel.}
\label{fig:migration}
\end{figure}

\subsection{Unfolding validation}
\label{sec:unf_val}
Tests are performed to ensure the unfolded distributions have the correct
statistical properties, reproduce
input distributions that differ from the nominal ones used to produce the
corrections and unfolding matrices, and are stable within
the expected statistical uncertainty.
In order to check for stability amid statistical fluctuations, each bin of the nominal MC distributions is smeared according to a Poisson distribution, with mean equal to the bin content, to create pseudo-experiments.
The distributions of the unfolded variables in the pseudo-experiments are then compared with
the particle-level input distributions to verify that the unfolding procedure is unbiased
and the statistical uncertainty of the unfolded distributions matches the
precision as observed in the ensemble of pseudo-experiments.
 
Additional tests are performed to verify the absence of bias originating from the choice of MC model used to derive the corrections in Eq.~(\ref{eqn:unf}).
Nine pseudo-experiments are created for each observable by reweighting the nominal MC distributions, at detector level and particle level, with nine different weighting functions. Each function is based on a polynomial fit to the ratio of the data to the nominal MC prediction of the measured observables. An additional pseudo-experiment is based on the disagreement between the data and the nominal prediction on the top-quark jet \pT, in this case a linear fit is employed and the slope of the fit is doubled to derive the weight function used for the test. The fit is performed separately in the two channels and in the all-hadronic channel separately for the leading and sub-leading jet. The pseudo-experiments are then unfolded using the nominal set of corrections.
The reweighted unfolded distributions are compared with the weighted distributions at particle level.
The procedure correctly reproduces most of the weighted distributions, verifying the robustness of the unfolding procedure.
The only exception is the reweighting as a function of top-quark \pt, where a relatively small degree of nonclosure is observed. This nonclosure, due to the top-quark \pt shape, is included as an additional source of uncertainty and described in Section~\ref{sec:systematics_contributions_signal}.


\section{Systematic uncertainties}
\label{sec:systematics}

The systematic uncertainties in the unfolded distributions are separated into
uncertainties arising from the
reconstruction and calibration of the jet and lepton candidates,
the background estimates, and the modeling of the \ttbar\ final state.
 
Uncertainties related to modeling of the detector and background are propagated to the unfolded distributions
by creating pseudo-experiments from simulated signal and background events in which a single source of uncertainty is varied by
one standard deviation.
The resulting distributions are then
treated as if they were data and unfolded using the nominal unfolding corrections, as described in Section~\ref{sec:unfolding_iterbayes}.
 
The signal-modeling uncertainties affect the measured cross-section through the unfolding corrections. To estimate these uncertainties, the alternative \ttbar\ simulations at detector level are used to create pseudo-data, which are then unfolded using the nominal unfolding correction (efficiency, acceptance, migration matrix, and background subtraction). The nonclosure between the unfolded pseudo-data and the alternative particle-level prediction is taken as an uncertainty in the measured cross-section.
 
\subsection{Detector uncertainties}
\label{sec:systematics_contributions_objects}
 
Lepton reconstruction and identification uncertainties are estimated by comparing the identification performance in data and MC samples and adjusting the performance in the MC samples to match that in data.
The uncertainties from these comparisons are used as the systematic uncertainties associated with the lepton reconstruction in both channels.
Uncertainties in the lepton-momentum scale and resolution are determined by comparing the performance measured in data and in MC samples~\cite{EGAM-2018-01, MUON-2018-03}.
 
The jet energy scale (JES) uncertainties for \smallR\ and \largeR\ jets are determined from
test beam data, simulation, and
in situ measurements of the \pT\ balance between jets and recoiling objects in $Z$+jet,
$\gamma$+jet, and multijet events~\cite{JETM-2018-02,JETM-2018-05}.
The jet energy resolution (JER) uncertainty is measured in dijet events as the width of the best-fit \pt-asymmetry distribution.
The JES and JER uncertainties associated with the RC \largeR\ jets are derived from the uncertainties in the constituent \smallR\ jets.
 
Uncertainties in the \met modify both the magnitude and direction of its vector.
They are determined from the uncertainties in the momentum calibration and resolution of the leptons and jets used to calculate this observable,
as well as an uncertainty in the soft term~\cite{PERF-2016-07}.
 
The jet mass scale (JMS) uncertainty for \largeR\ jets  is estimated by fitting
the top-quark-jet and \Wboson-jet mass peaks in \ellplusjets\ \ttbar events and by
measuring the relative response for calorimeter-based jets and track-jets in data
and MC samples of dijet events~\cite{JETM-2018-02,ATLAS-CONF-2020-022}.
The jet mass resolution (JMR) uncertainty for \largeR\ jets is derived from the
same fits of the top-quark-jet and \Wboson-jet mass peaks~\cite{ATLAS-CONF-2020-022}.
The JMR is conservatively set to $20\%$ outside the phase space covered by these measurements.
 
The uncertainty in the efficiency of the JVT algorithm is determined from studies
of its performance in data samples and MC events~\cite{PERF-2014-03}.
 
The $b$-tagging performance is measured in data events and MC samples, and
any difference is used to correct the MC $b$-tagging efficiency.
The uncertainty in this calibration is used as the systematic uncertainty
in the $b$-tagging efficiency~\cite{FTAG-2018-01,ATL-PHYS-PUB-2017-013}.
An efficiency uncertainty for jets with $\pT > 300$~\GeV\ is
estimated by extrapolating the efficiency measured at lower \pT\ to
these higher-\pT\ jets~\cite{ATL-PHYS-PUB-2021-003}.
 
The efficiency and mis-tagging rate of the DNN top-quark tagger, and their uncertainties, are derived
using hadronically decaying top quarks in boosted \ttbar events, $\gamma$+jet events, and
dijet events~\cite{JETM-2018-03}.
The uncertainty includes effects arising from JES and modeling uncertainties
that affect the substructure of the top-quark jets.
 
Several sources of systematic uncertainty are considered for the tracks used in the substructure measurements~\cite{ATL-PHYS-PUB-2015-051}.
The uncertainty in the fake rate is derived from template fits to data and simulated events performed in a control region enriched in fake tracks.
An additional uncertainty is used to account for differences in the impact parameter resolution between simulated and real data events.
This uncertainty is estimated in bins of jet $\eta$ and \pT\ and applied to simulated events in the form of a random smearing of the transverse and longitudinal track impact parameters.
Uncertainties in efficiencies and alignment of the tracking detectors are also considered.
In addition, uncertainties in the reconstruction of tracks associated with high-\pT\ jets are also included~\cite{PERF-2015-08}.
The track uncertainties are propagated to the track requirements employed to select the tracks used to identify the primary vertex and measure the jet substructure.

\subsection{\ellplusjets-channel background modeling}
\label{sec:systematics_contributions_ljetsbkg}

Several sources of uncertainty affect the single-top-quark background, the largest background process in the \ellplusjets\ channel.
The uncertainty associated with the initial-state-radiation (ISR) model is determined by modifying the nominal \POWPY[8]\ calculation by varying \muR\ and \muF\  independently by factors of 0.5 and 2.0 in the
matrix-element calculation,
and by varying the scales in the showering model according to the Var3c eigentune of the A14 tune.
The final-state-radiation (FSR) uncertainty is determined using a separate \POWPY[8]\ sample where the renormalization scale is varied from its nominal value by factors of 0.5 and 2.
The uncertainties due to the \PYTHIA scale variations are correlated with the corresponding \ttbar\ variations.
The uncertainty arising from the scheme used to remove interference between the \ttbar\ and $tW$\ processes is determined by taking the difference between the DR and DS approaches.
The uncertainty due to the choice of showering and hadronization scheme is determined by comparing the nominal prediction with an alternative model using the \HERWIG\ generator.
The uncertainty in the NNLO cross-section for the single-top-quark processes arising from the choice of PDF set and $\alphas$\ is 4.2$\%$ for the $t$-channel process,
5.4$\%$ for the $tW$ process, and 3.9$\%$ for the $s$-channel process~\cite{PhysRevD.81.054028,PhysRevD.73.034001}.
 
The uncertainties in the $W$+jets background estimates are evaluated using two alternative MC samples where \muR\ and \muF\ in the matrix-element and showering models are varied simultaneously by factors of 0.5 and 2.0~\cite{ATL-PHYS-PUB-2017-006}.
The uncertainty in the cross-section calculation for the associated-production processes \ttZ, \ttW, and \ttH\ are 13.3$\%$, 12$\%$, and 9.9$\%$, respectively~\cite{deFlorian:2016spz}.
The uncertainty in the sum of these backgrounds is set to 13.3$\%$,
which is the largest of the three individual uncertainties.
 
The uncertainty in the cross-sections for the $Z$+jets and diboson processes is conservatively estimated to be 50$\%$~\cite{ATL-PHYS-PUB-2017-006}.
However, these backgrounds contribute ${<}1$\%\ and thus are negligible.
 
The uncertainty in the misidentified-lepton background is assessed by comparing the estimates from the matrix-element method, described in Section~\ref{sec:fakesbackground},
with an alternative method.
In this method, the shape of the distributions of the misidentified-lepton events is derived from a MC estimate,
obtained using all-hadronic \ttbar\ and $V$+jets samples, while the normalization is obtained through a fit of the observed $\met$ and $\met + \mtw$~distributions.
The uncertainty in the misidentified-lepton estimate is below $1$\%,
except in some poorly populated regions, where it can reach 5\%.
 
\subsection{All-hadronic-channel background modeling}
\label{sec:systematics_contributions_allhadbkg}
 
The main uncertainty in the multijet background estimate is due to the data statistical uncertainty in the control regions introduced in Section~\ref{sec:ABCD} and used in the ABCD method.
All the detector- and signal-modeling uncertainties are propagated through the multijet estimation,
in the subtraction of the estimated signal and MC backgrounds from the control regions.
The largest uncertainty in the multijet background arises from the uncertainty in the
top-quark mis-tagging rate in the MC-subtracted backgrounds.

The uncertainties associated with the non-all-hadronic background, the
single-top-quark background, and the background due to the
associated-production processes \ttW, \ttZ, and \ttH\ arise from
their cross-section uncertainties and are incorporated into the overall
uncertainty in these background estimates.
 
\subsection{Signal modeling}
\label{sec:systematics_contributions_signal}
 
The uncertainties arising from the modeling of the \ttbar production and decay processes are evaluated by using alternative MC samples to create pseudo-experiments and unfolding them with the nominal unfolding matrices.
The differences between the unfolded results and the predictions of the alternative MC samples are taken as the modeling uncertainty.
 
The uncertainties in the modeling of ISR in the \ttbar process are estimated using the samples where the scales \mur\ and \muf\  in the matrix element are varied independently by factors of 0.5 and 2.0 in the \POWHEG\ calculations, and two samples employing the Var3cUp/Down eigentune of the A14 tune of \PYTHIA[8].
The FSR uncertainty is estimated using the samples where the renormalization scale regulating FSR is changed by factors of 0.5 and 2.0,
corresponding to \alphaFSR($m_Z$) = 0.142 (FSR Down) and \alphaFSR($m_Z$) = 0.114 (FSR Up), respectively.
The uncertainty associated with the choice of \hdamp\ parameter value is estimated using a sample with \hdamp\ a factor of 2 larger than the nominal value.
The resulting changes in the unfolded results are symmetrized as an up-and-down systematic
uncertainty. All uncertainties pertaining to radiative corrections are combined into \emph{IFSR}.
 
The \emph{Parton Shower} uncertainty arises from the choice of parton-shower, hadronization, underlying-event,
and color-reconnection models. It is estimated by employing the \HERWIG[7]\
parton-showering and hadronization model instead of the nominal \PYTHIA[8]\ model, with the resulting uncertainty being symmetrized.
The \emph{Hard Scattering} uncertainty arises from the matching method, and is estimated by comparing the results of a \POWPY[8]\ MC sample with the one created using the \AMCatNLOPY\ MC generator.
This uncertainty takes into account a different choice of parton-matching scheme and
renormalization scheme~\cite{ATL-PHYS-PUB-2020-023}.
To exclude the differences arising from the matrix-element corrections applied to all emissions in the nominal \POWPY[8]\ sample,
a dedicated \POWPY[8]\ simulation with these corrections turned off is used for this comparison.
The uncertainty arising from the choice of PDF is estimated using the 30 eigenvectors decomposed from the Hessian representation of the \PDFforLHC[30]\ PDF uncertainties~\cite{Butterworth:2015oua}.

An additional source of uncertainty is the effect of the top-quark mass uncertainty on the measured distributions.
The uncertainty considered for the top-quark mass is 0.5~\GeV, reflecting the precision of the top-quark mass measurements reached by the ATLAS~\cite{TOPQ-2017-03} and CMS~\cite{CMS-TOP-14-022}\ collaborations.
The resulting uncertainty is evaluated by creating pseudo-experiments with an alternative top-quark mass and propagating it through the analysis.
This uncertainty is negligible in all the measured differential cross-sections.

Finally, a modeling uncertainty arises from the observed difference between the observed and predicted shapes of the top-quark $\pT$ distribution.
This difference, between the data and the nominal \POWPY[8]\ prediction, is used to define a reweighting function,
which is then applied to the nominal prediction to obtain an alternative sample.
Pseudo-experiments created from this alternative sample are then propagated through the unfolding procedure,
and the difference between these results and those from the nominal procedure is taken as the uncertainty. This uncertainty was already mentioned as unfolding nonclosure in section~\ref{sec:unf_val}.
 
\subsection{Summary of systematic uncertainties}
\label{sec:systematics_display}
 
The resulting systematic uncertainties affect each unfolded variable differently and are determined bin-by-bin, although their scales are comparable.
As an example, the sizes of the systematic uncertainties in the normalized differential cross-section as a function of $\tau_{32}$ are shown in Figure~\ref{fig:systs}.
The uncertainties are generally below 5$\%$ and around 5$\%$ in the \ellplusjets\ channel and all-hadronic channel, respectively,
and reach ${\sim}40$\%\ at low and high $\tau_{32}$, where the differential cross-section
is smaller, especially at low values of $\tau_{32}$.
The dominant uncertainty varies bin-by-bin.
The FSR and the parton-shower and hadronization models are the sources of the largest signal-modeling
uncertainties,
while the JES/JER and charged-track uncertainties yield the largest detector-level uncertainties.
 
\begin{figure}[t]
\centering
\subfloat[]{\label{fig:systs:1}\includegraphics[width=0.45\textwidth]{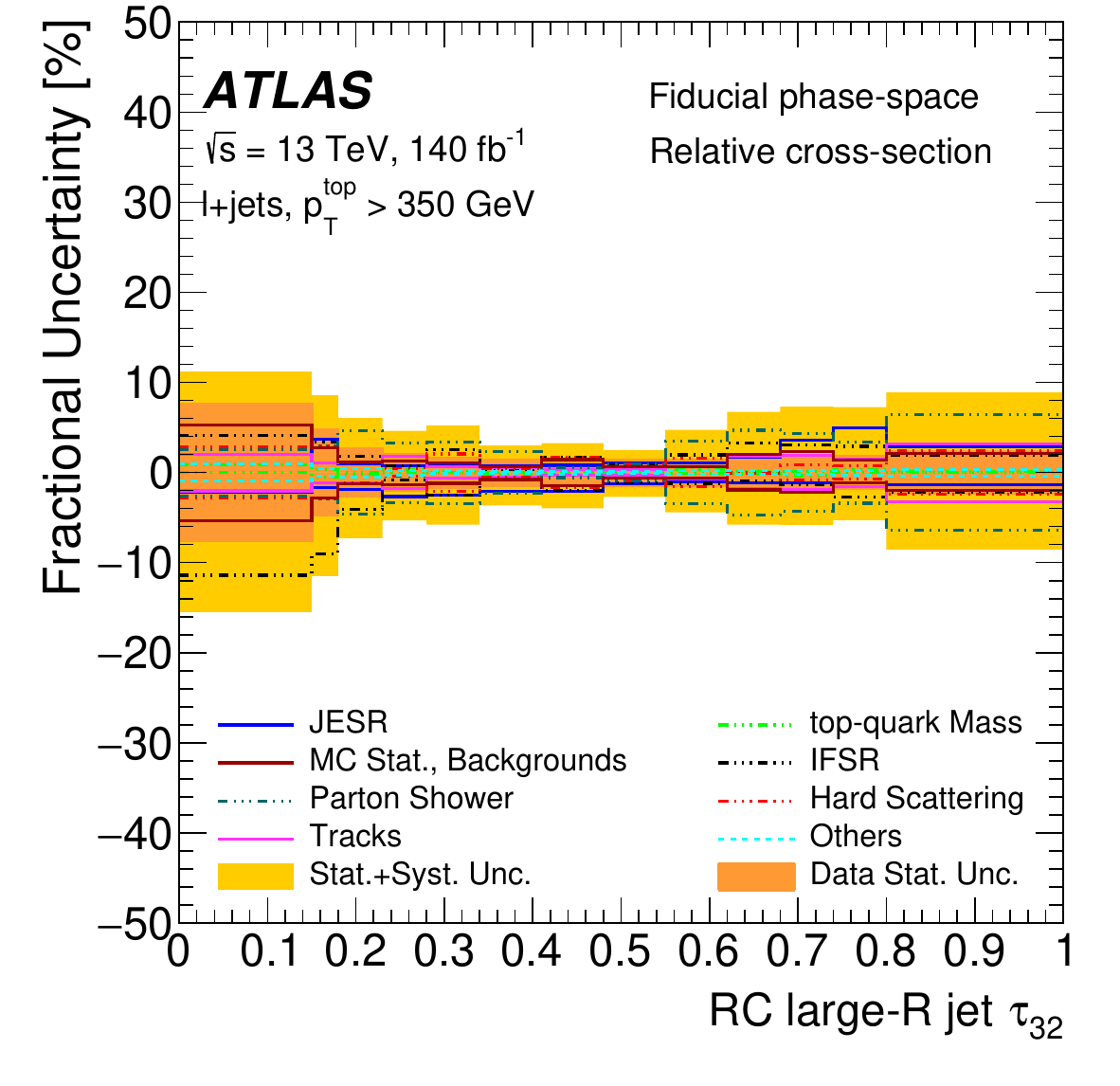}}
\subfloat[]{\label{fig:systs:2}\includegraphics[width=0.45\textwidth]{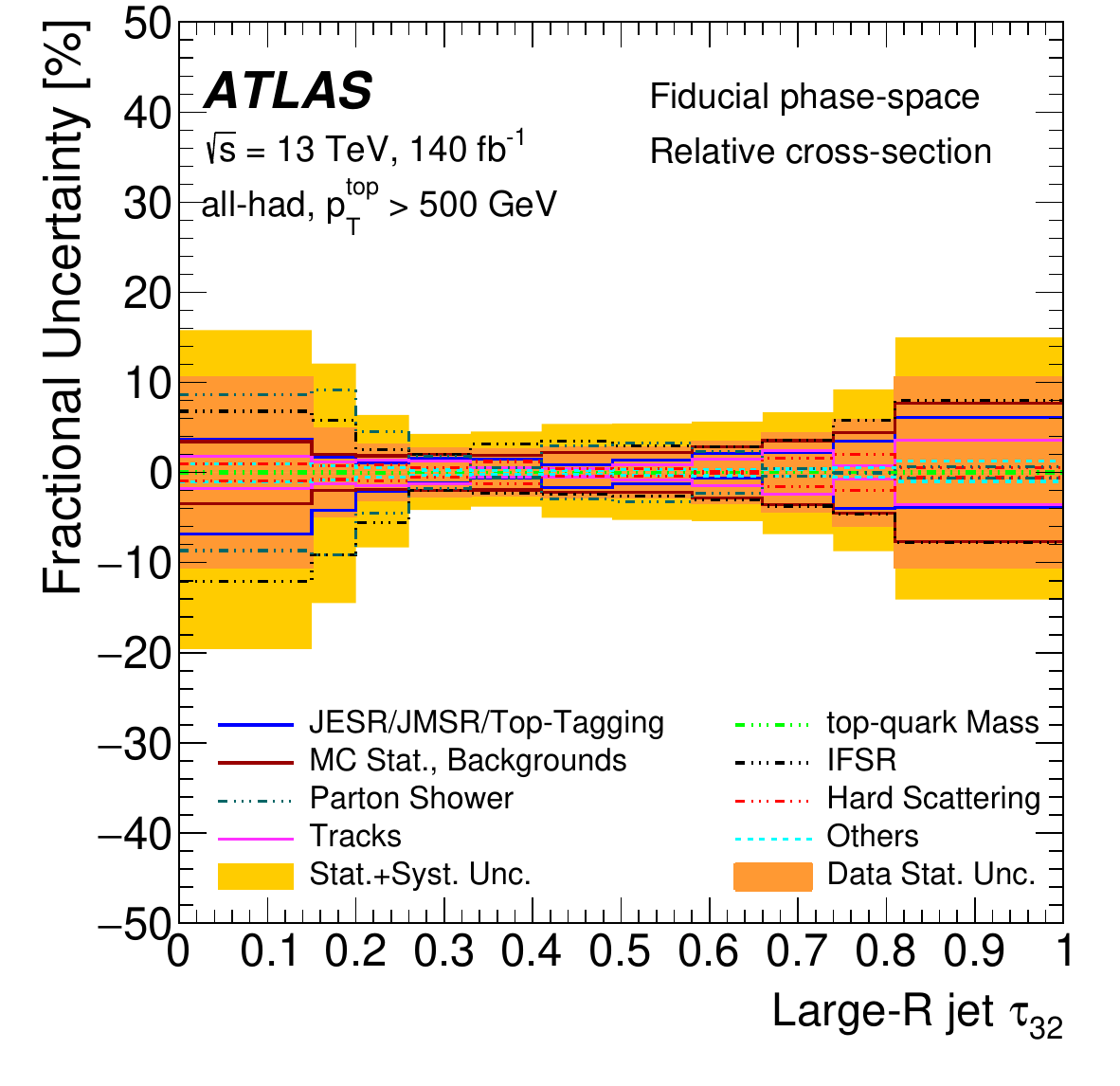}}
\caption{
Fractional uncertainties in the  particle-level differential $t\bar{t}$ production cross-section as a function of $\tau_{32}$.
The uncertainties are evaluated in the
\subref{fig:systs:1} \ellplusjets\ channel and
\subref{fig:systs:2} all-hadronic channel.
The orange band represents the data statistical uncertainty, while the yellow band represents the total uncertainty.
The uncertainties in the top-quark $\pT$ mismodeling, leptons, $\met$, flavor tagging,
\pileup and PDF have a very limited impact and are grouped in the plot under the label \enquote{Others}.
}
\label{fig:systs}
\end{figure}


\section{Results}
\label{sec:results}

The normalized particle-level fiducial phase-space differential cross-sections as a function of the
substructure observables are shown for both
the $\ell$+jets and all-hadronic channels.
A set of double-differential cross-sections is also presented.
The results are compared with several NLOME+PS predictions of the Standard Model, including several
tuning variations of the nominal \POWPY[8]\ prediction.
 
A quantitative comparison with the predictions is shown in Section~\ref{sec:results_covariance},
where $\chi^2$\ and $p$-value calculations are presented.
 
\subsection{Single-differential cross-sections}
\label{sec:results_ljets}
 
The normalized particle-level single-differential cross-sections for the
\ellplusjets\ and all-hadronic channels are shown in
Figures~\ref{fig:particle_level_results_t32}--\ref{fig:particle_level_results_LHA}.
 
The unfolded distributions are generally similar in shape between the two channels, although
the distributions from the all-hadronic channel tend to be narrower,
consistent with the SM predictions and arising from the higher \pT-interval
used in the all-hadronic channel.
The precision of the unfolded distributions allows qualitative and quantitative
comparisons with SM predictions.
The effect of the mismodeling of the top-quark \pT, introduced in Section~\ref{sec:control}, on the measured distributions is covered by the dedicated uncertainty introduced in Section~\ref{sec:systematics_contributions_signal}. This uncertainty is non-negligible only in the observables correlated with the top-quark \pT\ (\ECF, \LHA, \ptdispersion ) and relevant only in the tails of the distributions, where it reaches a 10$\%$ effect in the all-hadronic channel.
 
The $\tau_{32}$\ distributions show a rapid increase around 0.2 followed by a plateau out to about 0.8 for
the \ellplusjets\ channel and a decrease out to 0.8 for the all-hadronic channel.
The general behavior reflects the three-body nature of the top-quark decay.
The SM nominal \POWPY[8]\ and FSR Up predictions tend to overestimate the $\tau_{32}$\ distributions at low values and
correspondingly underestimate them at higher values.
The prediction made by the \POWHER[7]\ MC generator is in best agreement with the data.
The same behavior is also observed for the $\tau_{3}$\ prediction.
 
The $\tau_{21}$\ distributions show a rising trend from 0.0 out to 0.5, where they
then fall off quickly.
This reflects the general absence of two-body structure in these jets.
The MC predictions are in good agreement with the data, with a slight overestimation of the
distributions for $\tau_{21}$\ below ${\sim}0.3$.
 
The \ellplusjets\ and all-hadronic channels show a similar trend in the
$C_3$\ predictions from several generators, with the SM predictions that tend to overestimate the data at low values of $C_3$ underestimating it in the right tail of the distribution.
The $D_2$\ SM predictions are in good agreement with the \ellplusjets\ channel's distribution while the
predicted distribution in the all-hadronic channel is somewhat broader than what is observed.
 
While the measured \LHA\ and \ECF\ distributions agree within the uncertainty
band with all the predictions in most of the bins, a large discrepancy is observed in the \ptdispersion\ distribution
for the \POWPY[8]\ prediction in both channels, while \POWHER[7]\ seems to provide a better description of this observable.

 
\begin{figure}[t]
\centering
\subfloat[]{\label{fig:particle_level_results_t32:1}\includegraphics[width=0.45\textwidth]{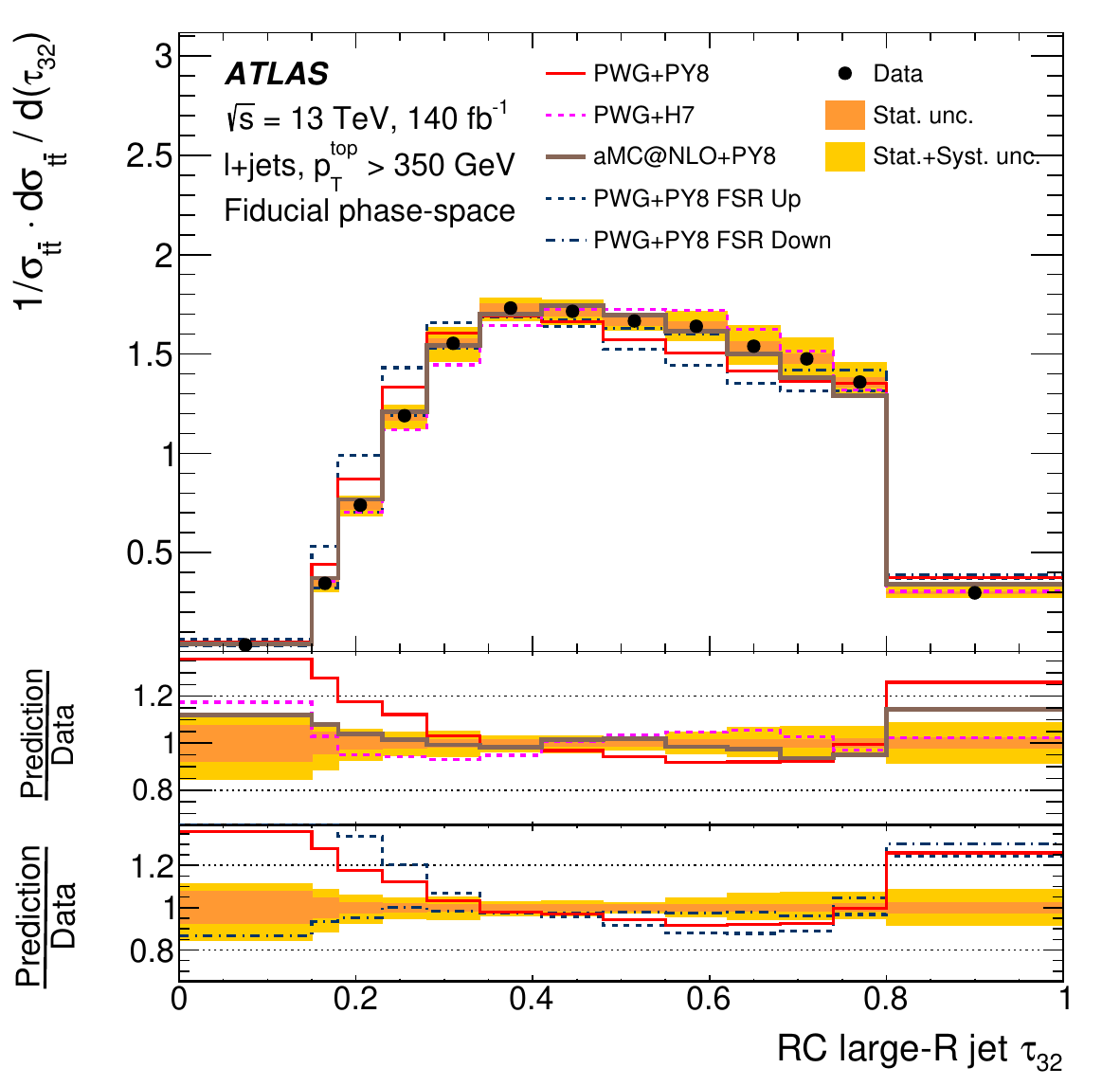}}
\subfloat[]{\label{fig:particle_level_results_t32:2}\includegraphics[width=0.45\textwidth]{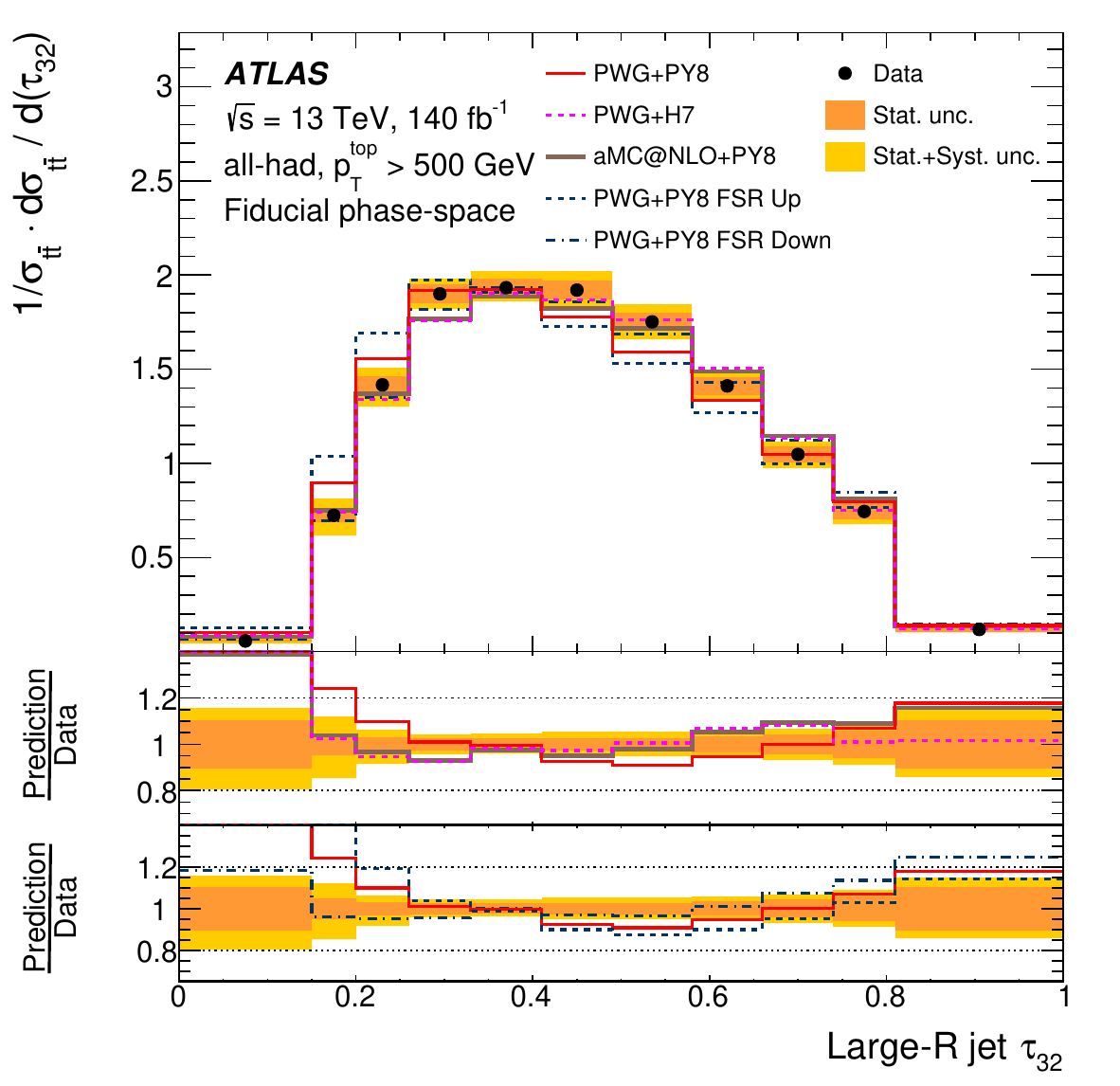}}
\caption{
Particle-level normalized differential cross-sections as a function of $\tau_{32}$\ for the data and several NLOME+PS MC predictions.
The unfolded results shown here are in the
\subref{fig:particle_level_results_t32:1} $\ell+$jets and
\subref{fig:particle_level_results_t32:2} all-hadronic channels.
The lower pads show the ratios of the predictions to the data.
The yellow band represents the total uncertainty of the measured differential cross-section, while the orange band shows the statistical component.
}
\label{fig:particle_level_results_t32}
\end{figure}
 
\begin{figure}[t]
\centering
\subfloat[]{\label{fig:particle_level_results_t21:1}\includegraphics[width=0.45\textwidth]{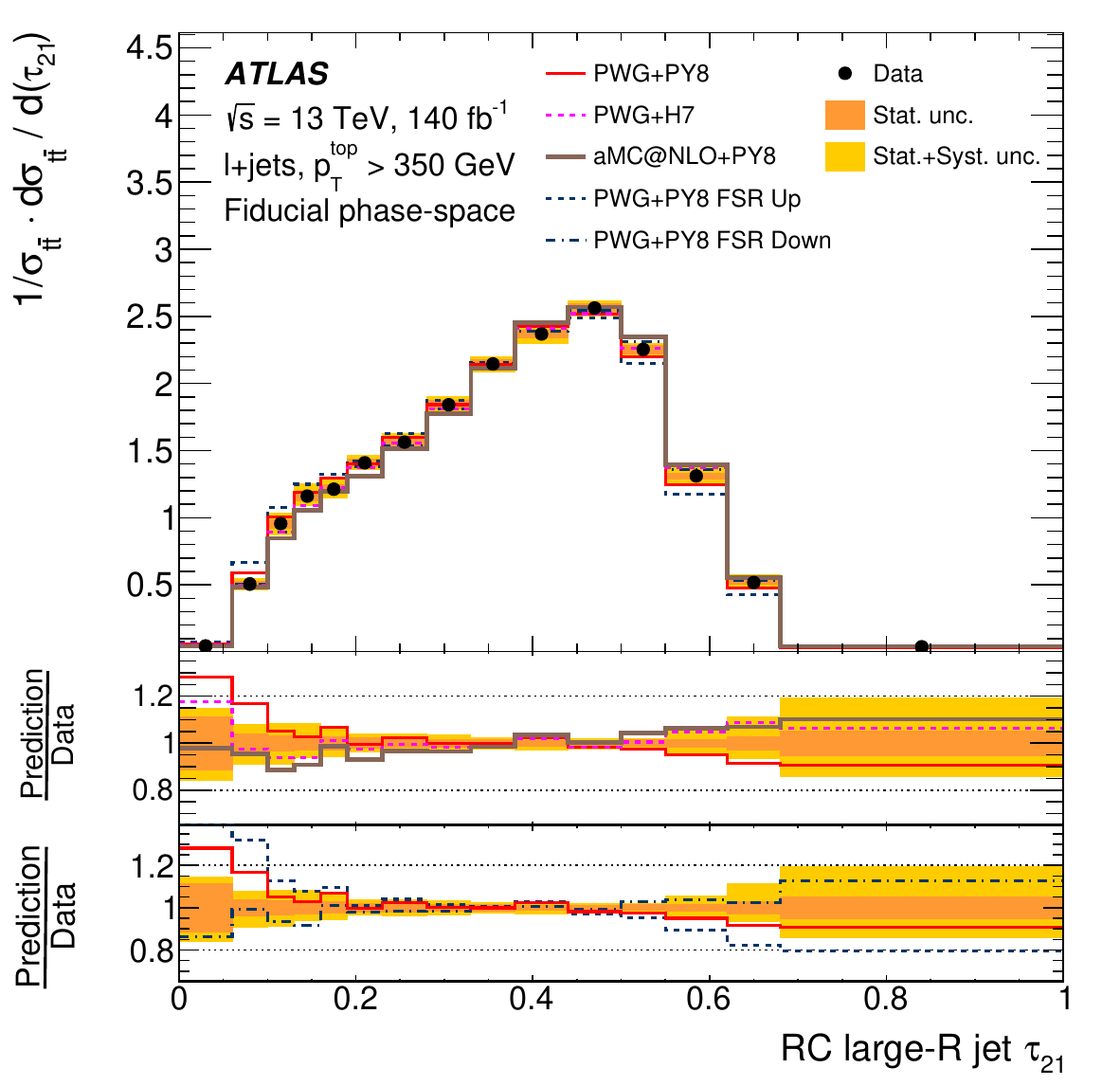}}
\subfloat[]{\label{fig:particle_level_results_t21:2}\includegraphics[width=0.45\textwidth]{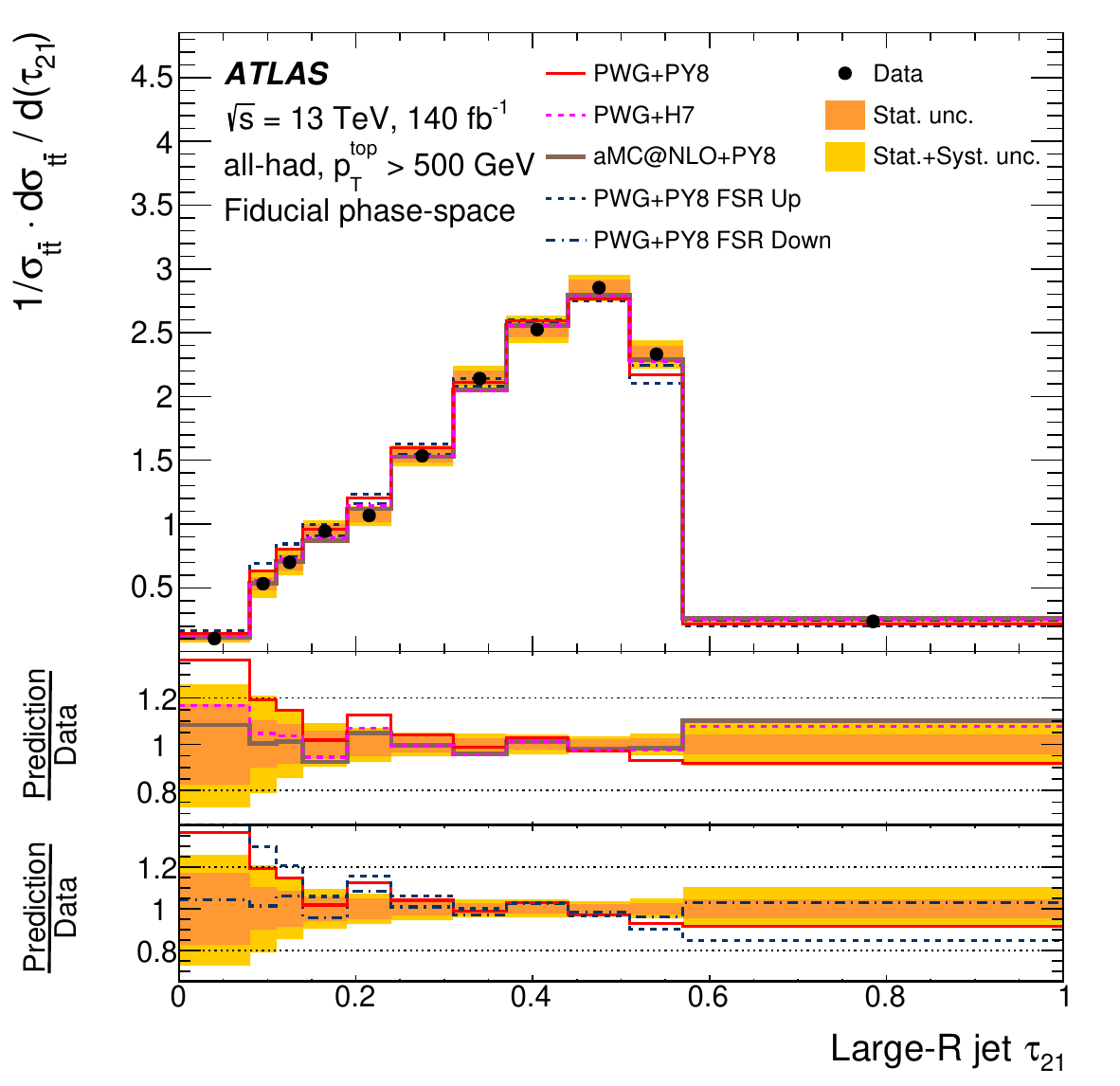}}
\caption{
Particle-level normalized differential cross-sections as a function of $\tau_{21}$\ for the data and several NLOME+PS MC predictions.
The unfolded results shown here are in the
\subref{fig:particle_level_results_t21:1} $\ell+$jets and
\subref{fig:particle_level_results_t21:2} all-hadronic channels.
The lower pads show the ratios of the predictions to the data.
The yellow band represents the total uncertainty of the measured differential cross-section, while the orange band shows the statistical component.
}
\label{fig:particle_level_results_t21}
\end{figure}
 
\begin{figure}[t]
\centering
\subfloat[]{\label{fig:particle_level_results_t3:1}\includegraphics[width=0.45\textwidth]{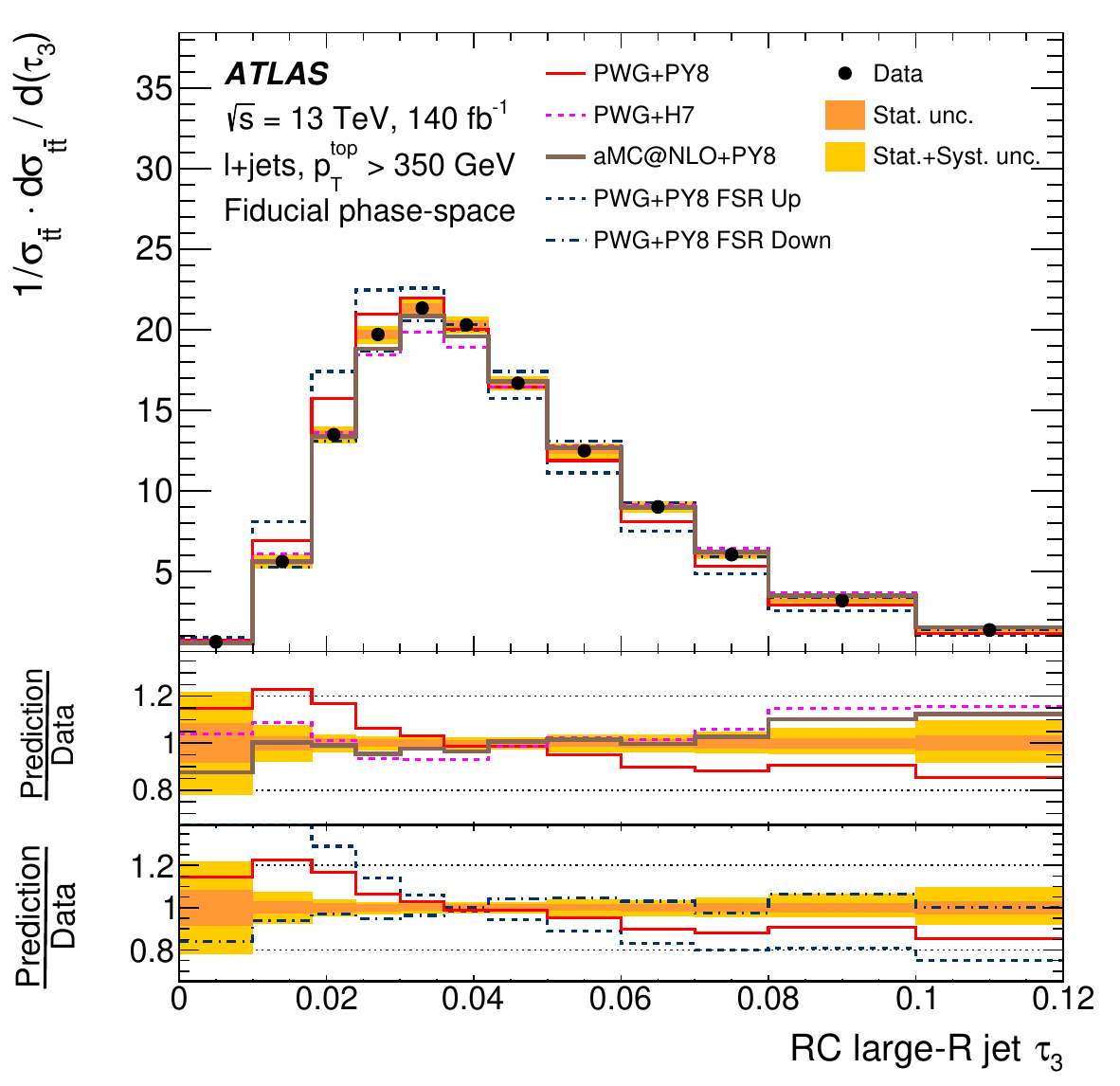}}
\subfloat[]{\label{fig:particle_level_results_t3:2}\includegraphics[width=0.45\textwidth]{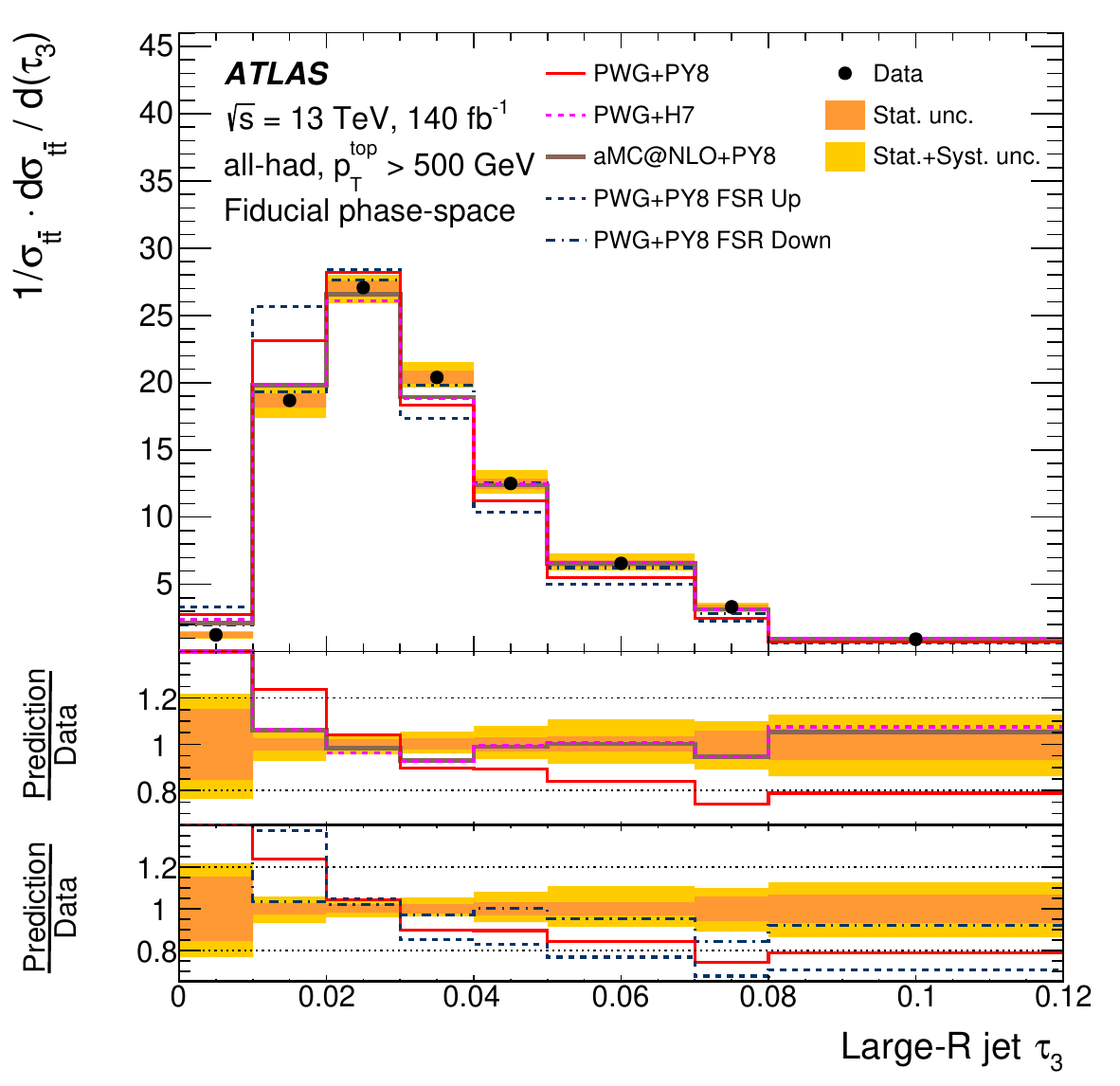}}
\caption{
Particle-level normalized differential cross-sections as a function of $\tau_3$\ for the data and several NLOME+PS MC predictions.
The unfolded results shown here are in the
\subref{fig:particle_level_results_t3:1} $\ell+$jets and
\subref{fig:particle_level_results_t3:2} all-hadronic channels.
The lower pads show the ratios of the predictions to the data.
The yellow band represents the total uncertainty of the measured differential cross-section, while the orange band shows the statistical component.
}
\label{fig:particle_level_results_t3}
\end{figure}
 
\begin{figure}[t]
\centering
\subfloat[]{\label{fig:particle_level_results_ECF2:1}\includegraphics[width=0.45\textwidth]{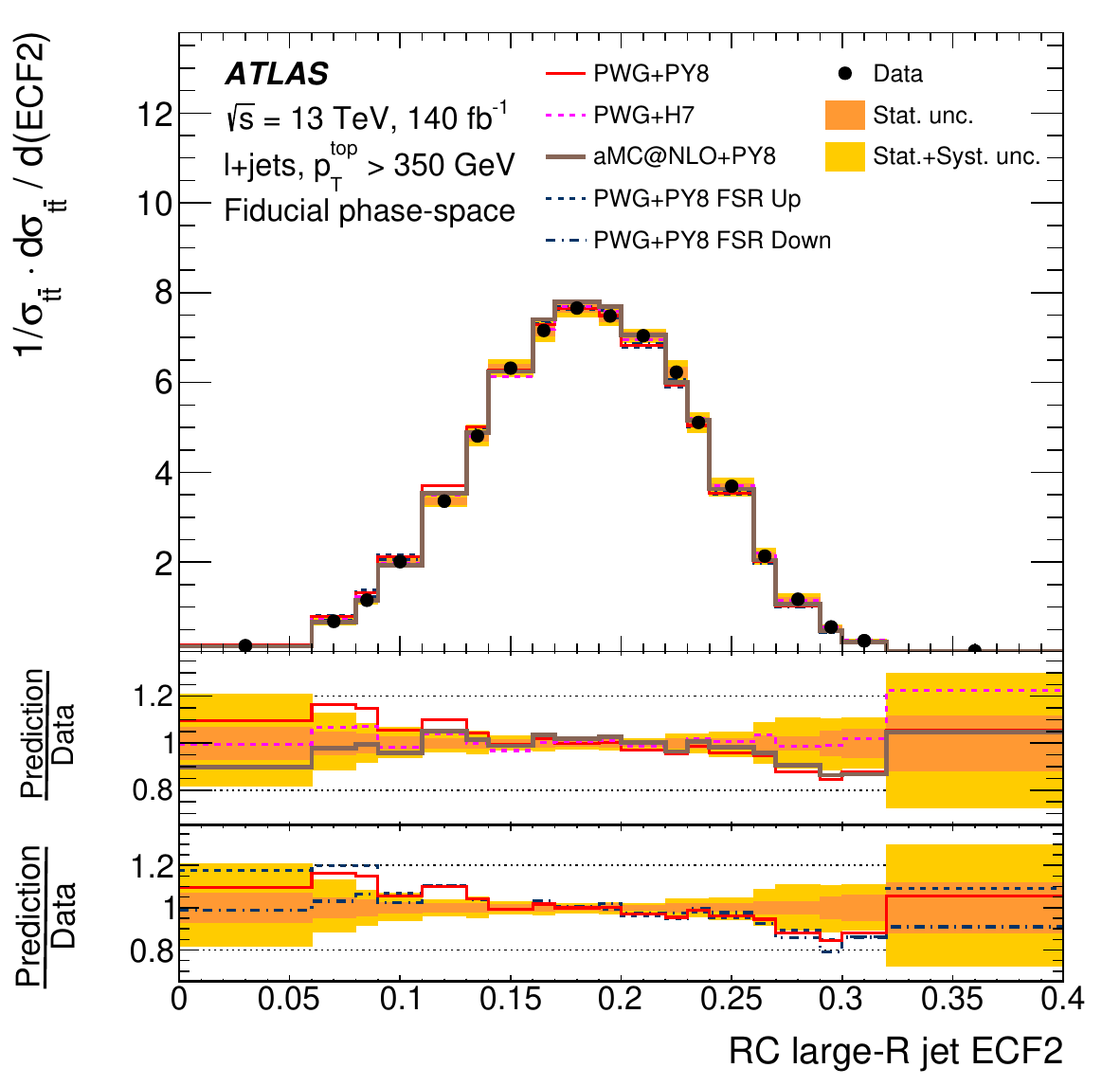}}
\subfloat[]{\label{fig:particle_level_results_ECF2:2}\includegraphics[width=0.45\textwidth]{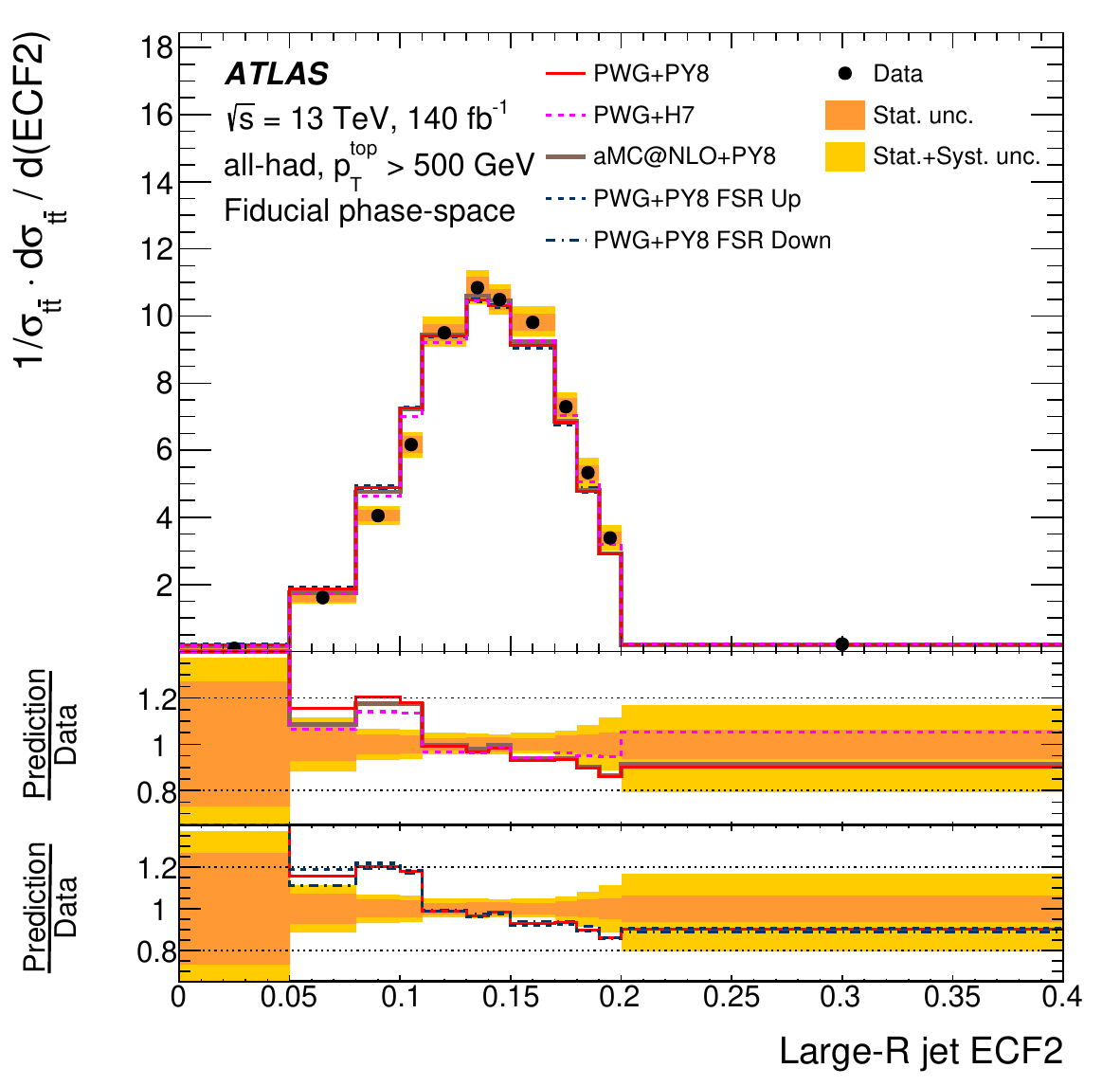}}
\caption{
Particle-level normalized differential cross-sections as a function of \ECF\ for the data and several NLOME+PS MC predictions.
The unfolded results shown here are in the
\subref{fig:particle_level_results_ECF2:1} $\ell+$jets and
\subref{fig:particle_level_results_ECF2:2} all-hadronic channels.
The lower pads show the ratios of the predictions to the data.
The yellow band represents the total uncertainty of the measured differential cross-section, while the orange band shows the statistical component.
}
\label{fig:particle_level_results_ECF2}
\end{figure}
 
\begin{figure}[t]
\centering
\subfloat[]{\label{fig:particle_level_results_D2:1}\includegraphics[width=0.45\textwidth]{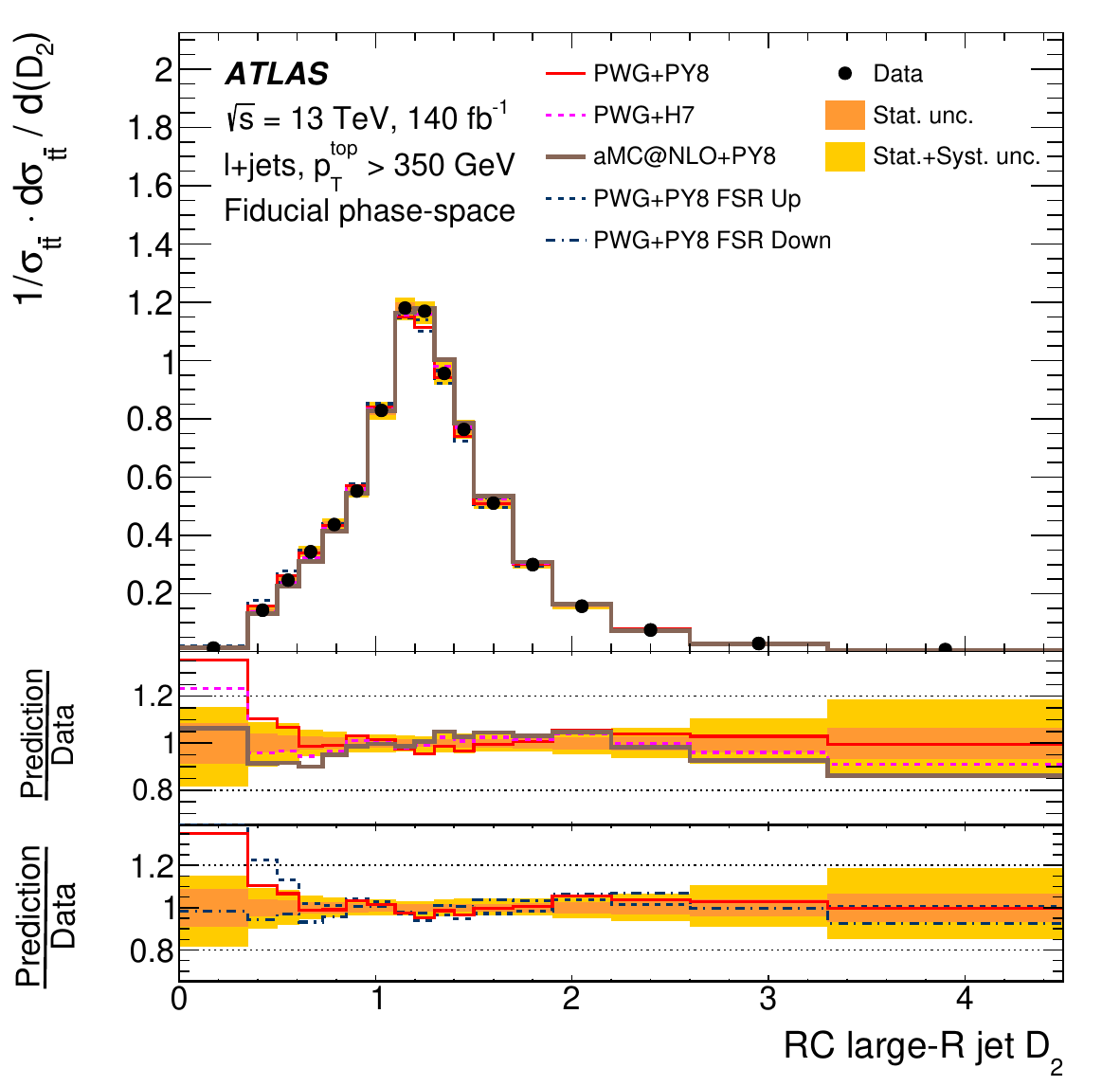}}
\subfloat[]{\label{fig:particle_level_results_D2:2}\includegraphics[width=0.45\textwidth]{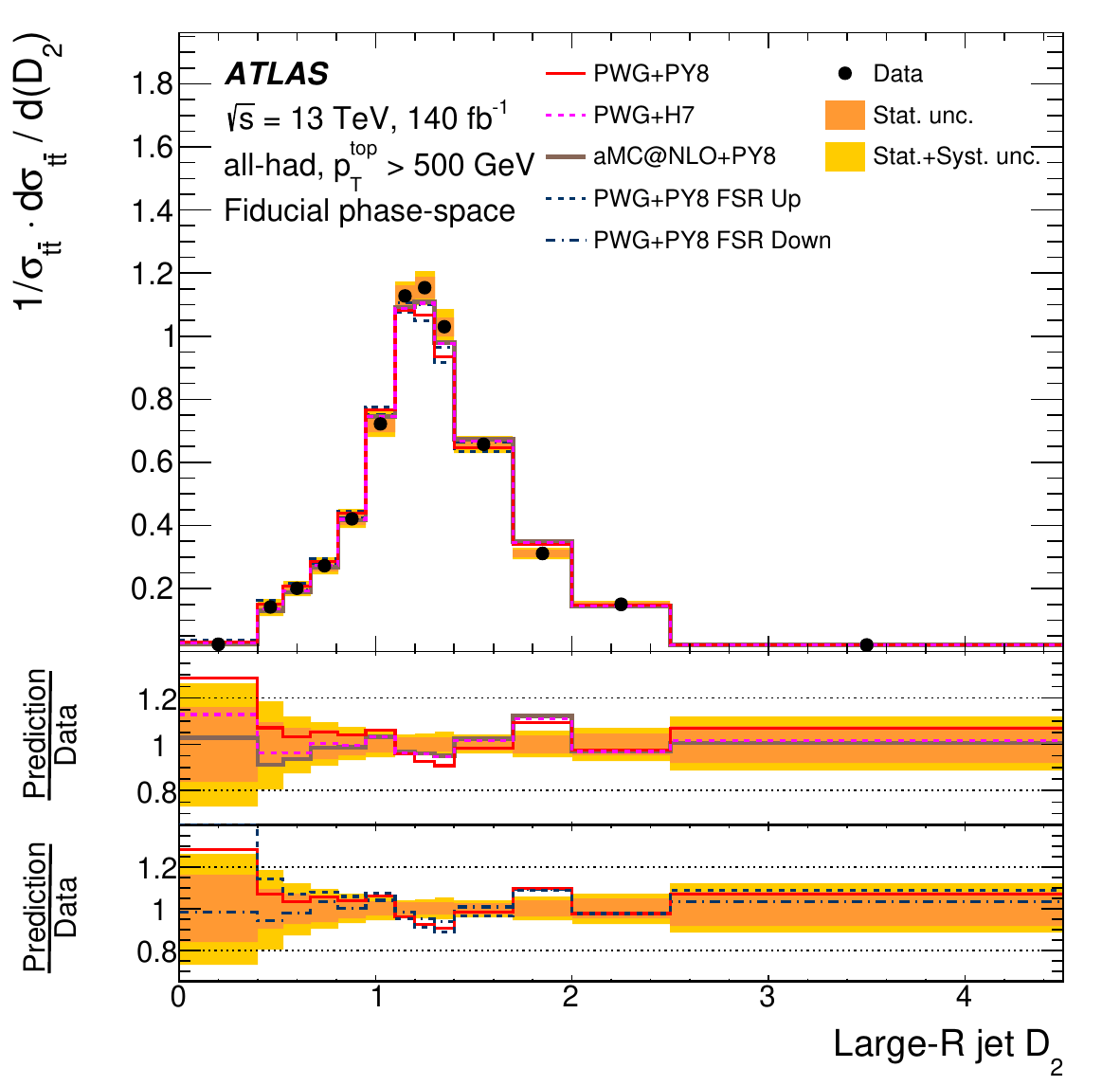}}
\caption{
Particle-level normalized differential cross-sections as a function of $D_2$\ for the data and several NLOME+PS MC predictions.
The unfolded results shown here are in the
\subref{fig:particle_level_results_D2:1} $\ell+$jets and
\subref{fig:particle_level_results_D2:2} all-hadronic channels.
The lower pads show the ratios of the predictions to the data.
The yellow band represents the total uncertainty of the measured differential cross-section, while the orange band shows the statistical component.
}
\label{fig:particle_level_results_D2}
\end{figure}
 
\begin{figure}[t]
\centering
\subfloat[]{\label{fig:particle_level_results_C3:1}\includegraphics[width=0.45\textwidth]{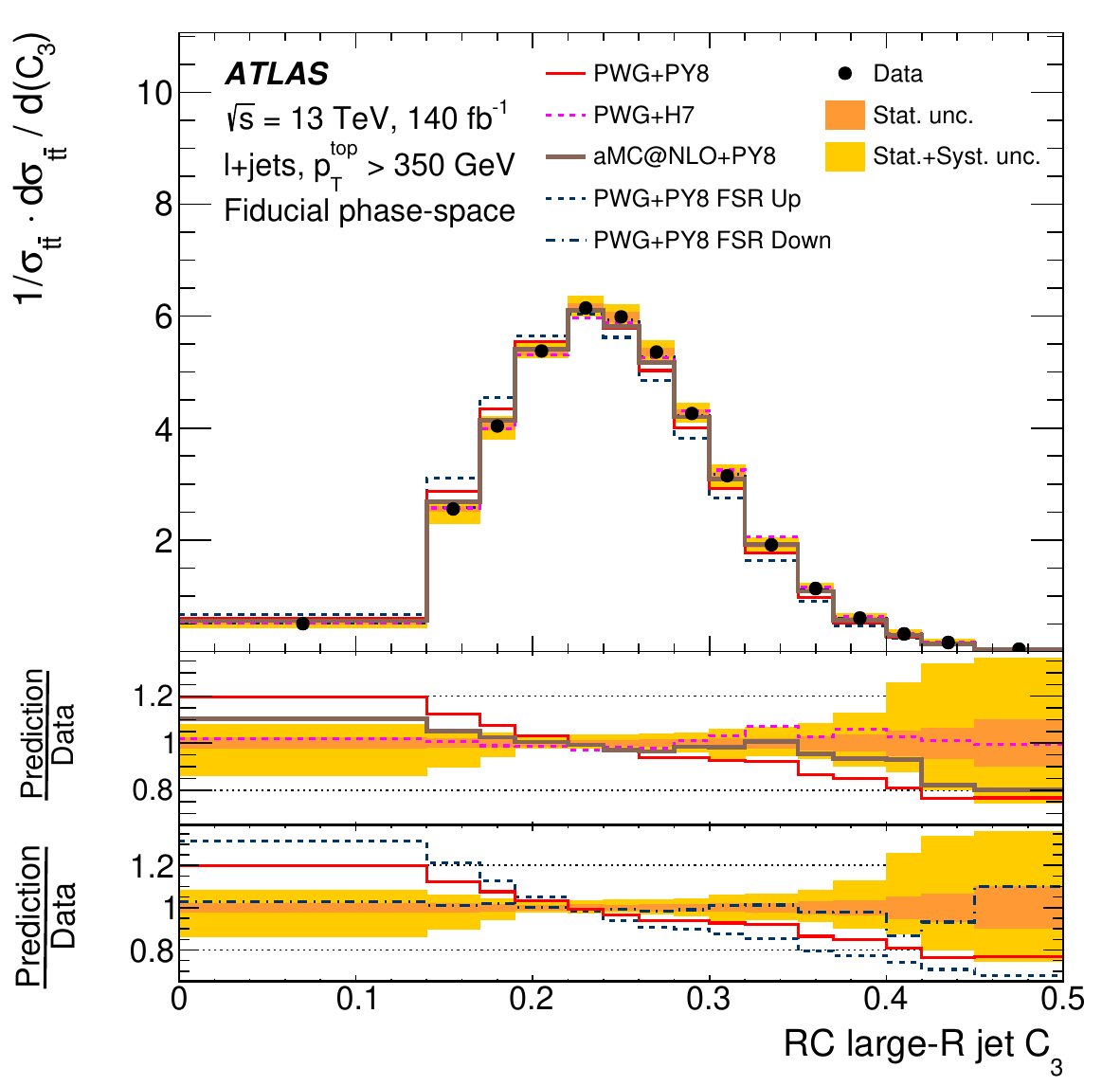}}
\subfloat[]{\label{fig:particle_level_results_C3:2}\includegraphics[width=0.45\textwidth]{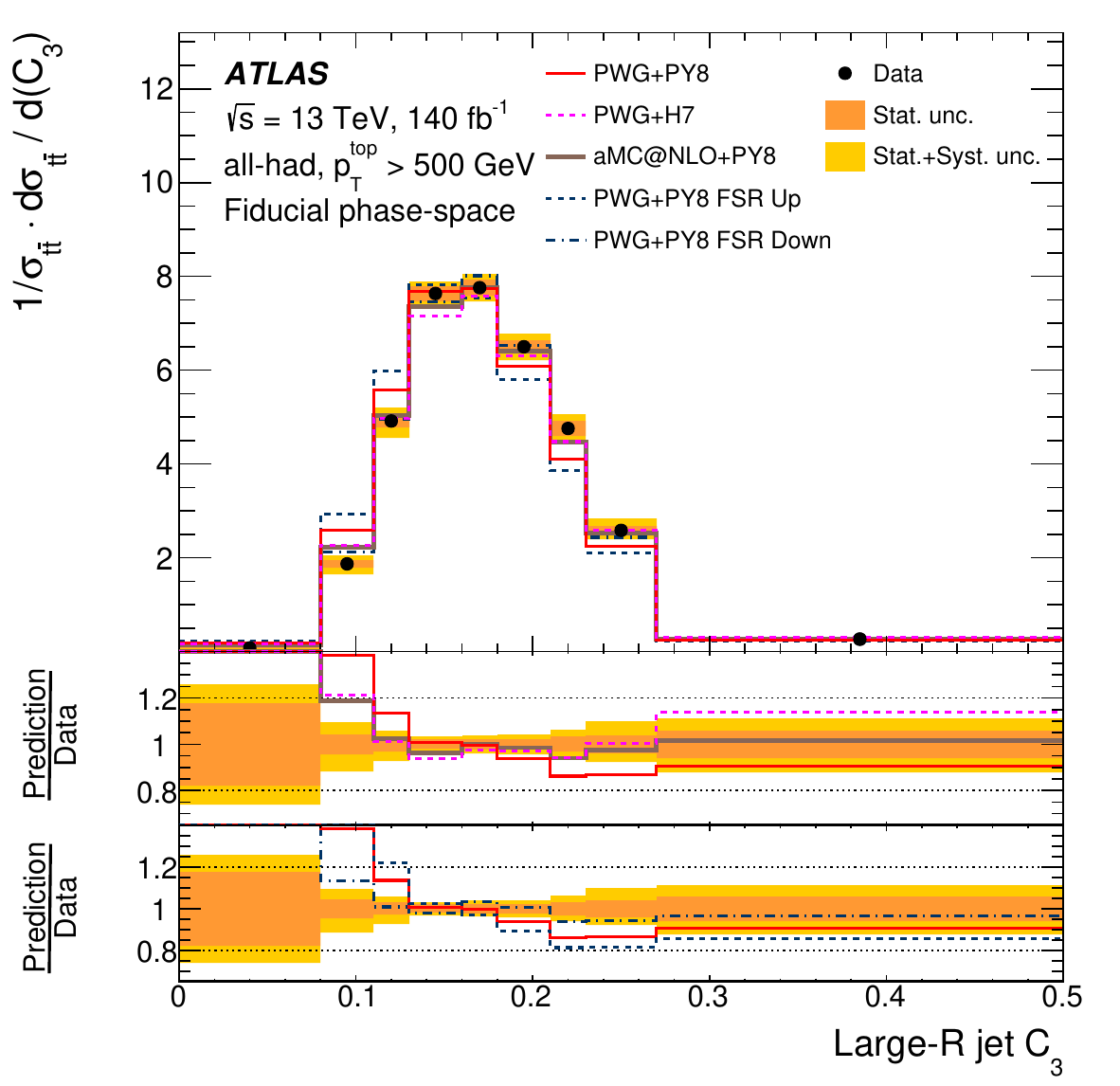}}
\caption{
Particle-level normalized differential cross-sections as a function of $C_3$\ for the data and several NLOME+PS MC predictions.
The unfolded results shown here are in the
\subref{fig:particle_level_results_C3:1} $\ell+$jets and
\subref{fig:particle_level_results_C3:2} all-hadronic channels.
The lower pads show the ratios of the predictions to the data.
The yellow band represents the total uncertainty of the measured differential cross-section, while the orange band shows the statistical component.
}
\label{fig:particle_level_results_C3}
\end{figure}
 
\begin{figure}[t]
\centering
\subfloat[]{\label{fig:particle_level_results_ptds:1}\includegraphics[width=0.45\textwidth]{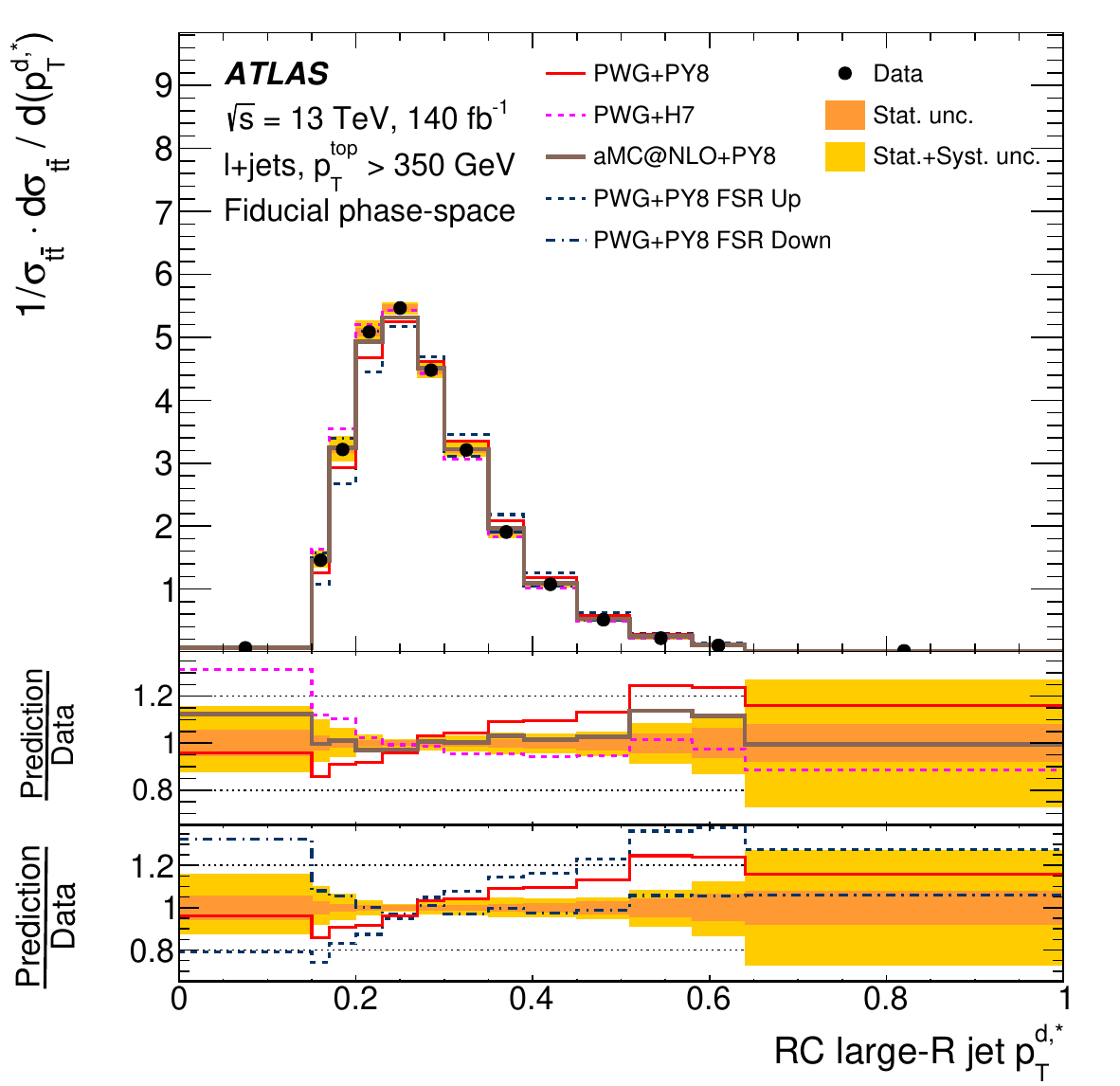}}
\subfloat[]{\label{fig:particle_level_results_ptds:2}\includegraphics[width=0.45\textwidth]{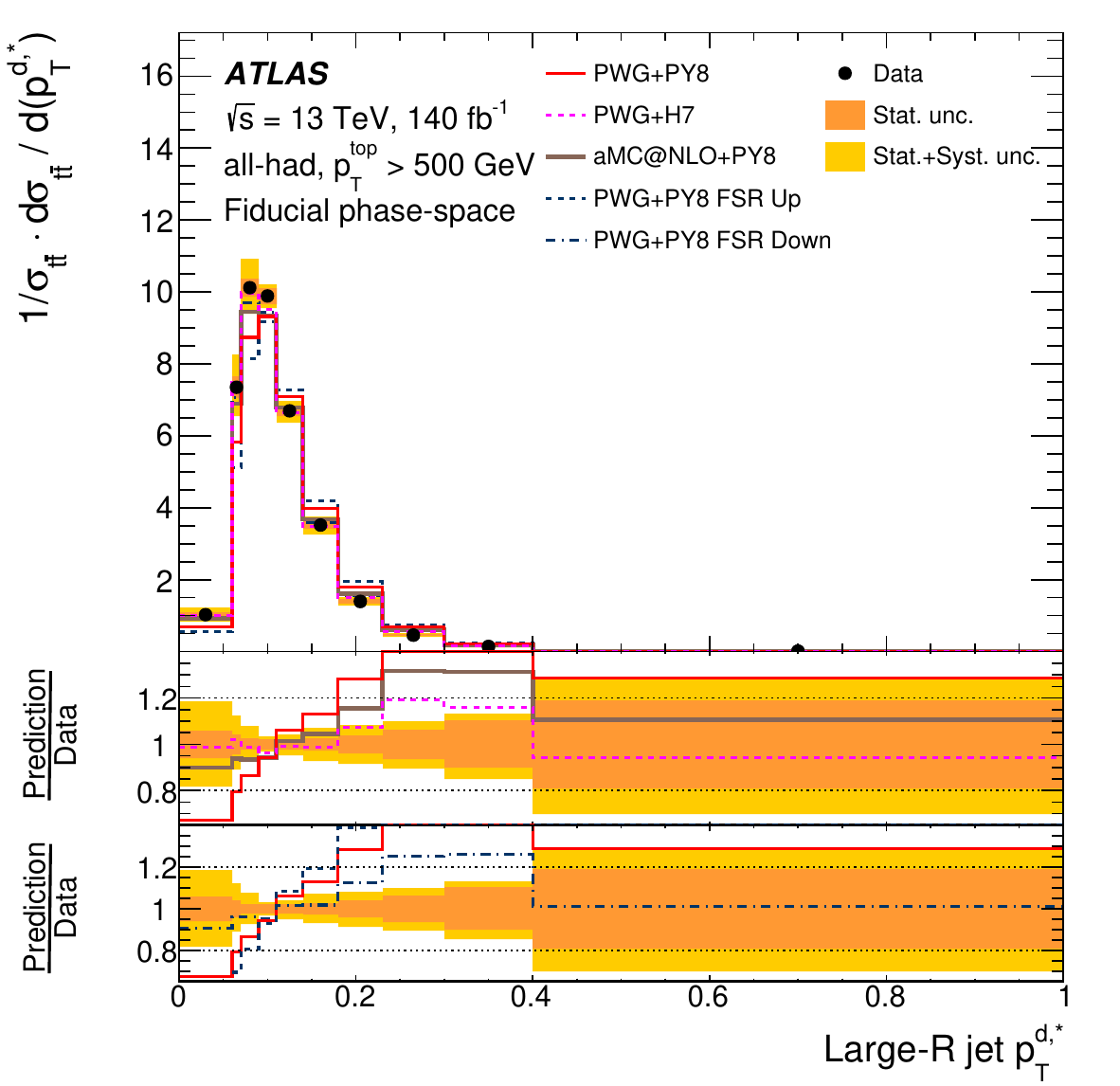}}
\caption{
Particle-level normalized differential cross-sections  as a function of jet \pT\ dispersion for the data and several NLOME+PS MC predictions.
The unfolded results shown here are in the
\subref{fig:particle_level_results_ptds:1} $\ell+$jets and
\subref{fig:particle_level_results_ptds:2} all-hadronic channels.
The lower pads show the ratios of the predictions to the data.
The yellow band represents the total uncertainty of the measured differential cross-section, while the orange band shows the statistical component.
}
\label{fig:particle_level_results_ptdisp}
\end{figure}
 
\begin{figure}[t]
\centering
\subfloat[]{\label{fig:particle_level_results_LHA:1}\includegraphics[width=0.45\textwidth]{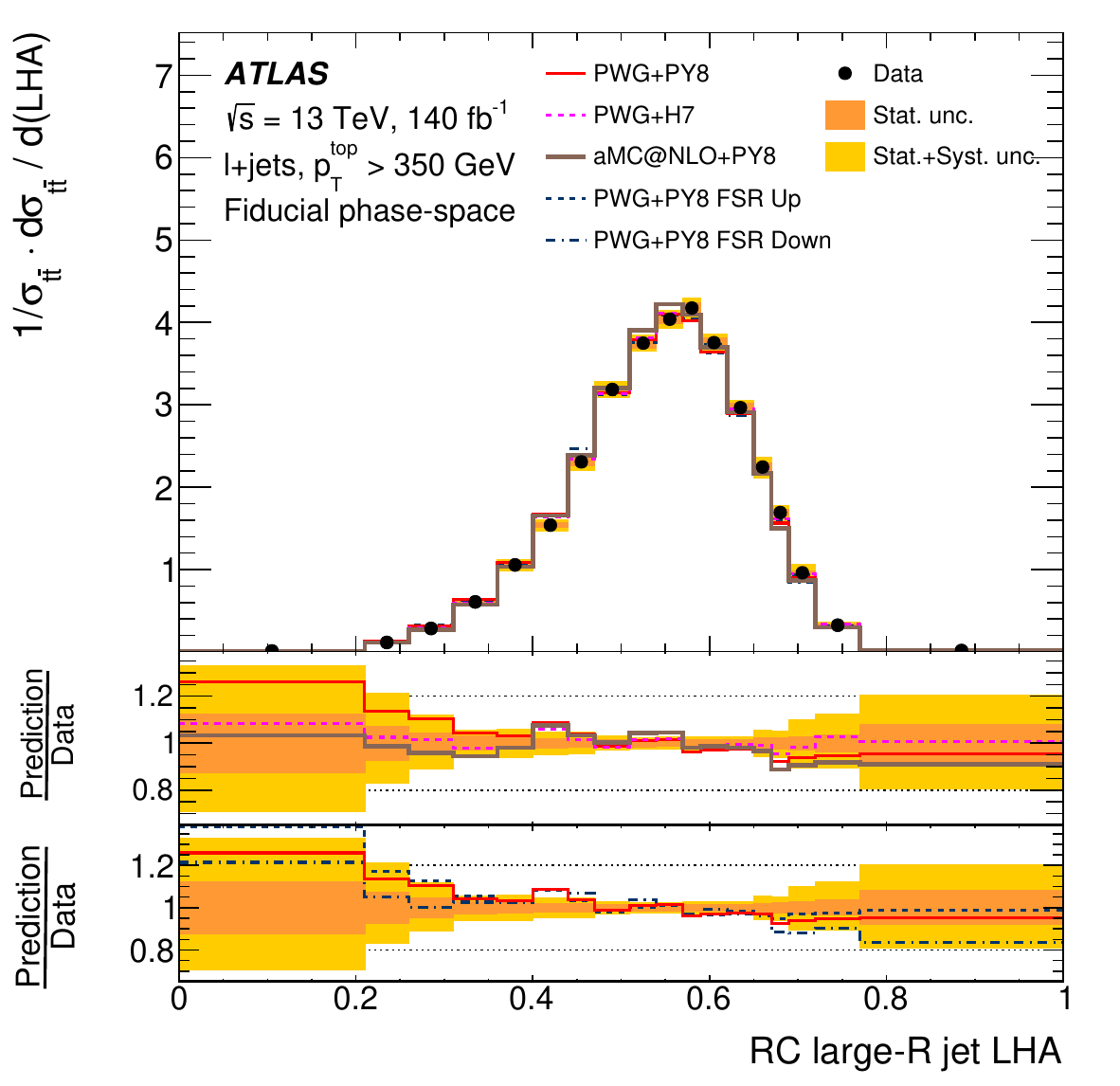}}
\subfloat[]{\label{fig:particle_level_results_LHA:2}\includegraphics[width=0.45\textwidth]{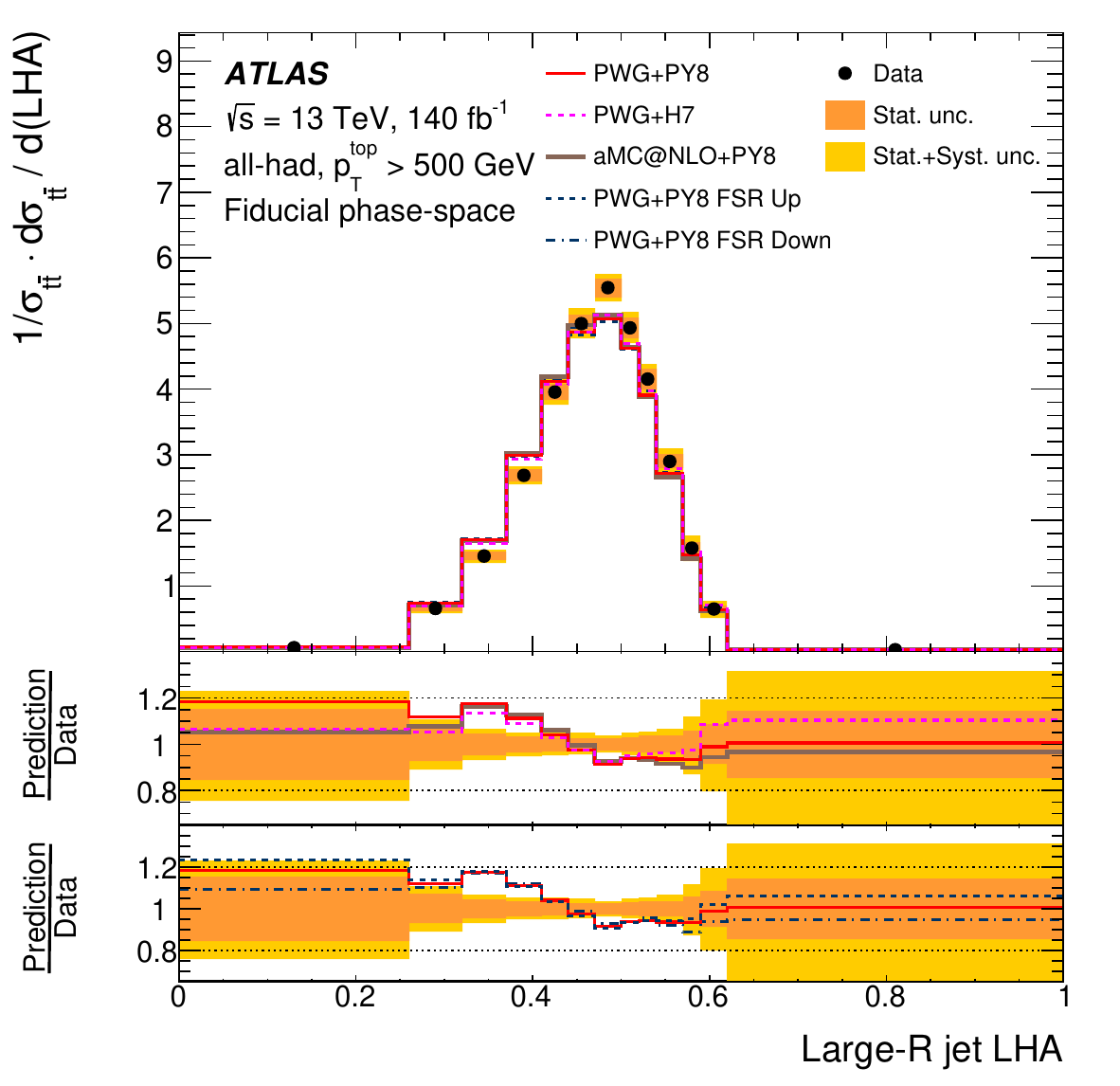}}
\caption{
Particle-level normalized differential cross-sections as a function of \LHA\ for the data and several NLOME+PS MC predictions.
The unfolded results shown here are in
\subref{fig:particle_level_results_LHA:1} the $\ell+$jets and
\subref{fig:particle_level_results_LHA:2} the all-hadronic channels.
The lower pads show the ratios of the predictions to the data.
The yellow band represents the total uncertainty of the measured differential cross-section, while the orange band shows the statistical component.
}
\label{fig:particle_level_results_LHA}
\end{figure}


\FloatBarrier
\subsection{Double-differential cross-sections}
\label{sec:results_allhad}
 
The normalized particle-level double-differential cross-sections for the \ellplusjets\ channel are
shown in Figures~\ref{fig:particle_level_results_ljets_dd_0}--\ref{fig:particle_level_results_ljets_dd_3},
while the corresponding results for the all-hadronic channel are displayed in Figures~\ref{fig:particle_level_results_allhad_dd_0}--\ref{fig:particle_level_results_allhad_dd_3}. The normalization considers the whole spectrum, so each bin is divided by the integral of the double-differential distribution.
 
The four figures for each channel present the correlations between
$\tau_{32}$\ and \mtop ,
$\tau_{32}$\ and \pTtop ,
$D_{2}$\ and \mtop , and
$D_{2}$\ and \pTtop .
The $\tau_{32}$\ and $D_{2}$\ substructure variables have the power to separate jets with three-body or two-body substructure from jets with simpler topologies and have been employed in various taggers.
Therefore, their correlations with \mtop\ and \pTtop\ are of particular interest.
 
The distributions of $\tau_{32}$\ versus \mtop\ predicted by the various MC generators
generally show poor agreement in the central \mtop\ interval and better agreement at both low and high \mtop, where there are large systematic uncertainties, due to the jet energy resolution.
Agreement between data and predictions is somewhat better in the all-hadronic channel
than in the \ellplusjets\ channel, although similar trends are observed in both.
 
The distribution of $\tau_{32}$\ versus \pTtop\ and the predictions show similar trends and are in
reasonable agreement at values of $\tau_{32}$\ between 0.4 and 0.7.
The MC predictions are generally larger than the data at lower and higher $\tau_{32}$.
 
The predictions of $D_{2}$\ correlations with \mtop\ and \pTtop\ from the various
MC generators are generally in better agreement with the unfolded data, although
there are considerable variations between the predictions at both low and high $D_{2}$.
However, these variations are smaller than those seen in the comparisons of the $\tau_{32}$\
predictions.
 
Overall, these correlations show that the MC predictions tend to have more
three-body-like substructure than observed in the data.
This disagreement does not have a strong \pTtop\ dependence, but appears to be larger in the central \mtop\ region.
The results for the all-hadronic and \ellplusjets\ channels are similar, although
agreement is generally better in the all-hadronic channel.


\begin{figure}[t]
\centering
\subfloat[]{\label{fig:particle_level_results_ljets_dd_0:1}\includegraphics[width=0.6\textwidth]{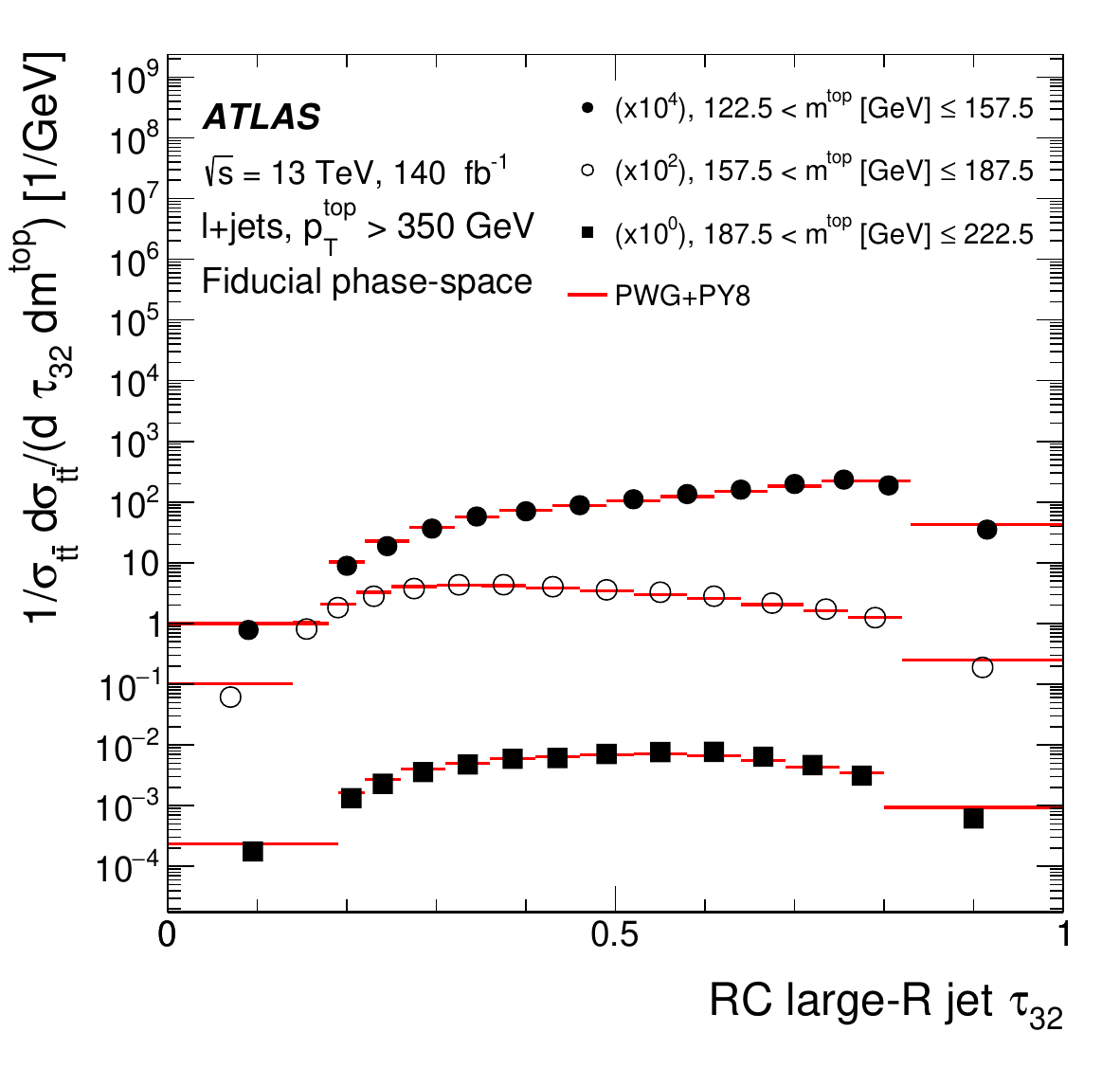}}\\
\subfloat[]{\label{fig:particle_level_results_ljets_dd_0:2}\includegraphics[width=\textwidth]{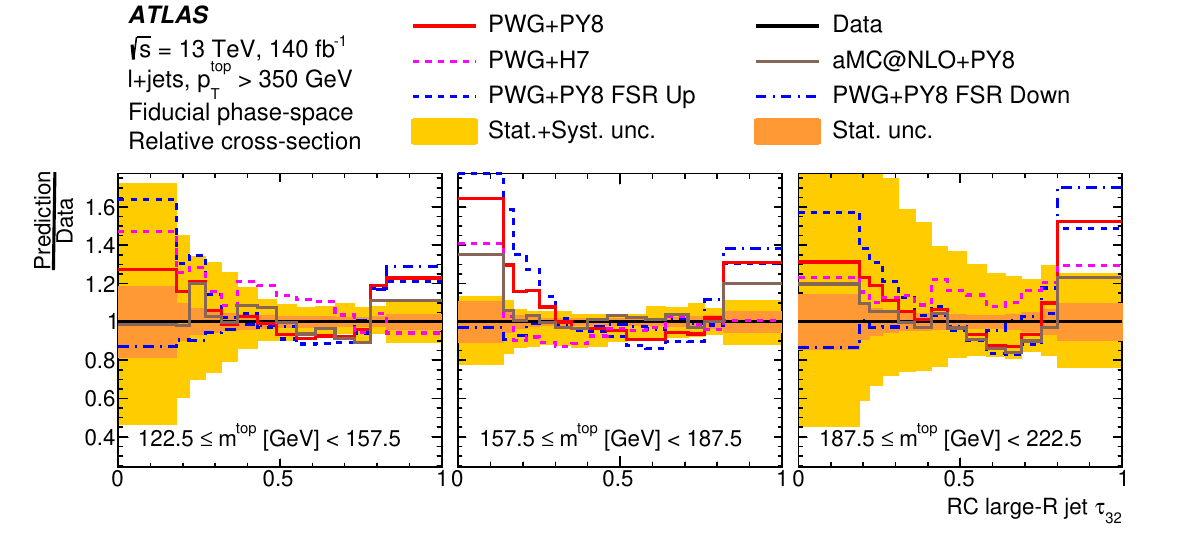}}
\caption{
Particle-level differential $t\bar{t}$ production cross-section as a function of $\tau_{32}$\ and jet mass,
for several NLOME+PS predictions of $t\bar{t}$ signal and the data.
The events shown here are from the \ellplusjets\ channel:
\subref{fig:particle_level_results_ljets_dd_0:1} shows the data and the nominal prediction, while
\subref{fig:particle_level_results_ljets_dd_0:2} shows the ratios of the predictions to the data.
The lighter band represents the total uncertainty of the measured differential cross-section, while the darker band shows the statistical component.
}
\label{fig:particle_level_results_ljets_dd_0}
\end{figure}
 
\begin{figure}[t]
\centering
\subfloat[]{\label{fig:particle_level_results_ljets_dd_1:1}\includegraphics[width=0.6\textwidth]{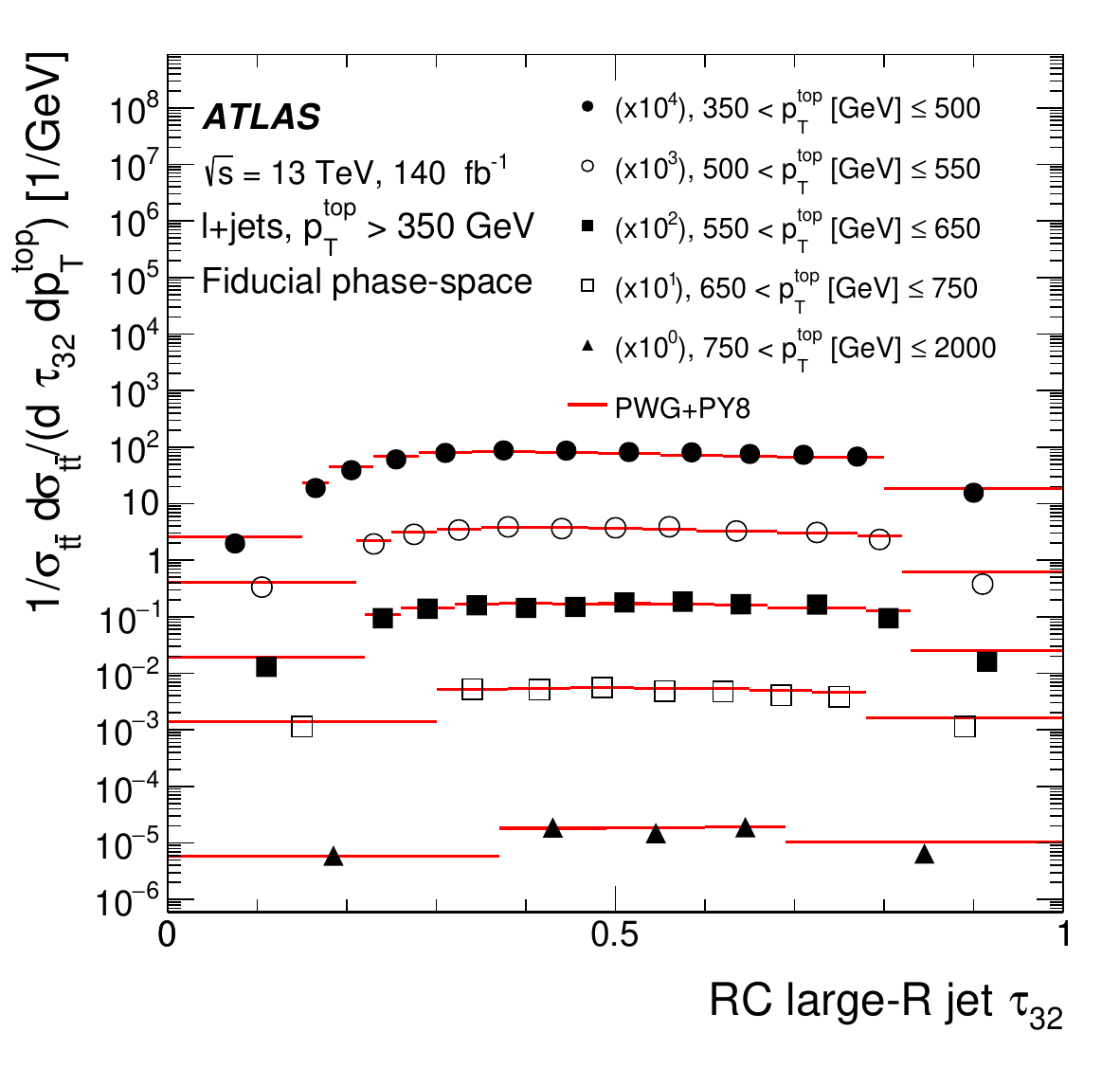}}\\
\subfloat[]{\label{fig:particle_level_results_ljets_dd_1:2}\includegraphics[width=\textwidth]{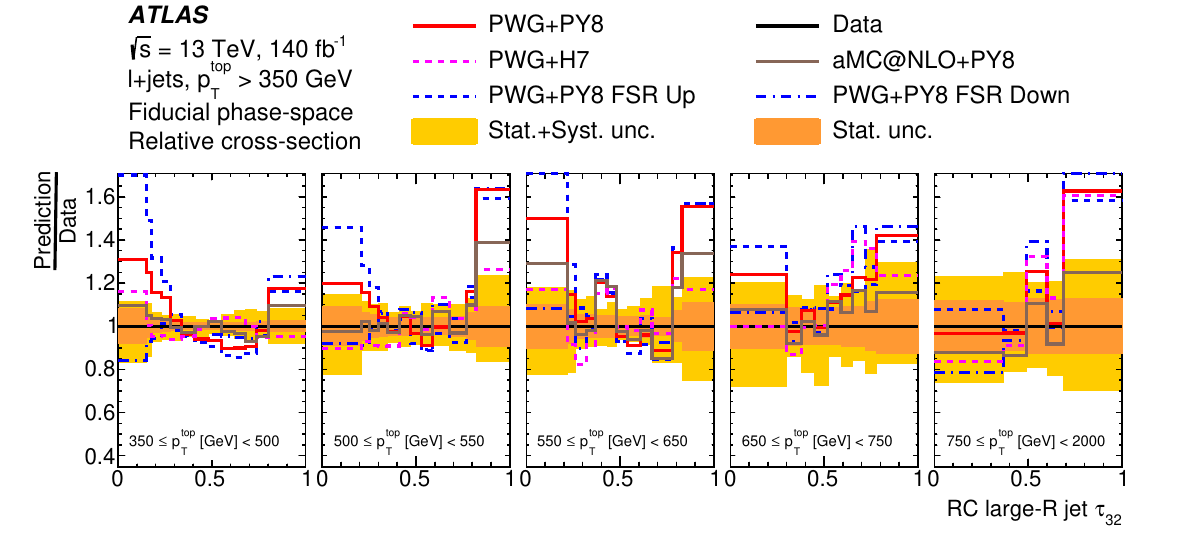}}
\caption{
Particle-level differential $t\bar{t}$ production cross-section as a function of $\tau_{32}$\ and jet \pT,
for several NLOME+PS predictions of $t\bar{t}$ signal and the data.
The events shown here are from the \ellplusjets\ channel:
\subref{fig:particle_level_results_ljets_dd_1:1} shows the data and the nominal prediction, while
\subref{fig:particle_level_results_ljets_dd_1:2} shows the ratios of the predictions to the data.
The lighter band represents the total uncertainty of the measured differential cross-section, while the darker band shows the statistical component.
}
\label{fig:particle_level_results_ljets_dd_1}
\end{figure}
 
\begin{figure}[t]
\centering
\subfloat[]{\label{fig:particle_level_results_ljets_dd_2:1}\includegraphics[width=0.6\textwidth]{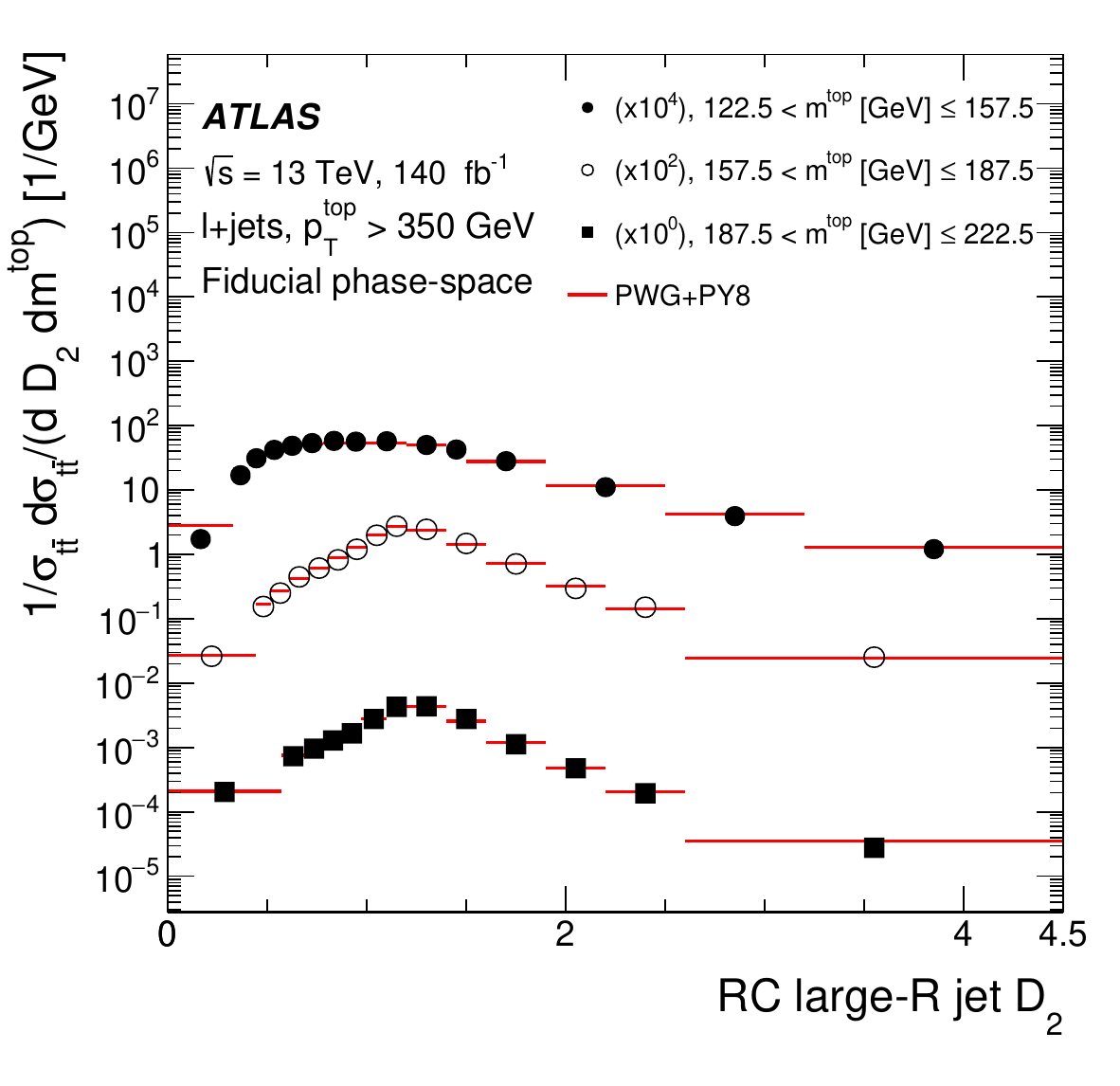}}\\
\subfloat[]{\label{fig:particle_level_results_ljets_dd_2:2}\includegraphics[width=\textwidth]{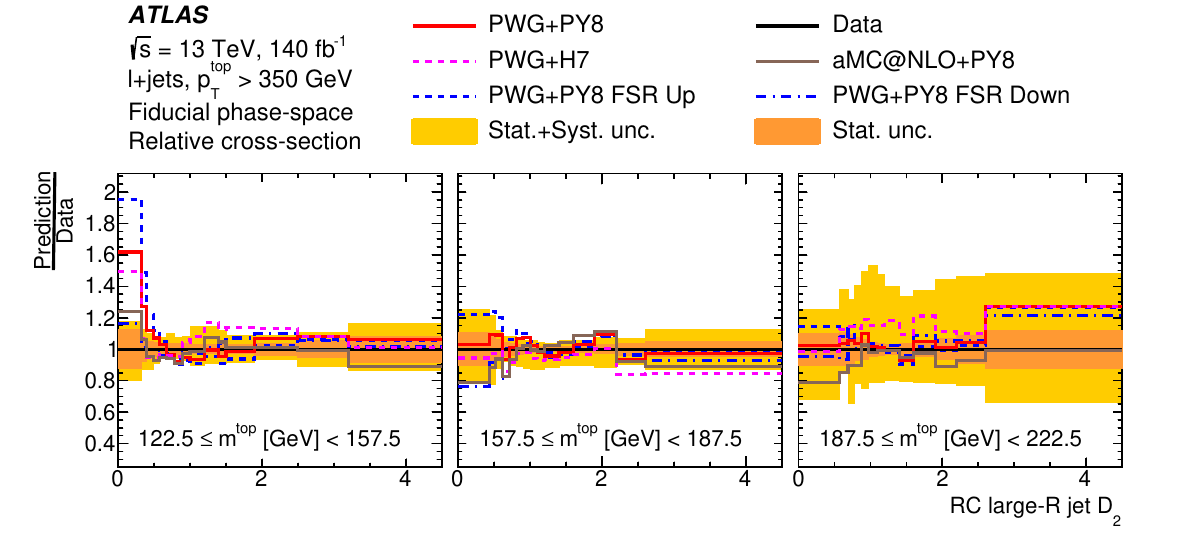}}
\caption{
Particle-level differential $t\bar{t}$ production cross-section as a function of $D_{2}$\ and jet mass,
for several NLOME+PS predictions of $t\bar{t}$ signal and the data.
The events shown here are from the \ellplusjets\ channel:
\subref{fig:particle_level_results_ljets_dd_2:1} shows the data and the nominal prediction, while
\subref{fig:particle_level_results_ljets_dd_2:2} shows the ratios of the predictions to the data.
The lighter band represents the total uncertainty of the measured differential cross-section, while the darker band shows the statistical component.
}
\label{fig:particle_level_results_ljets_dd_2}
\end{figure}
 
\begin{figure}[t]
\centering
\subfloat[]{\label{fig:particle_level_results_ljets_dd_3:1}\includegraphics[width=0.6\textwidth]{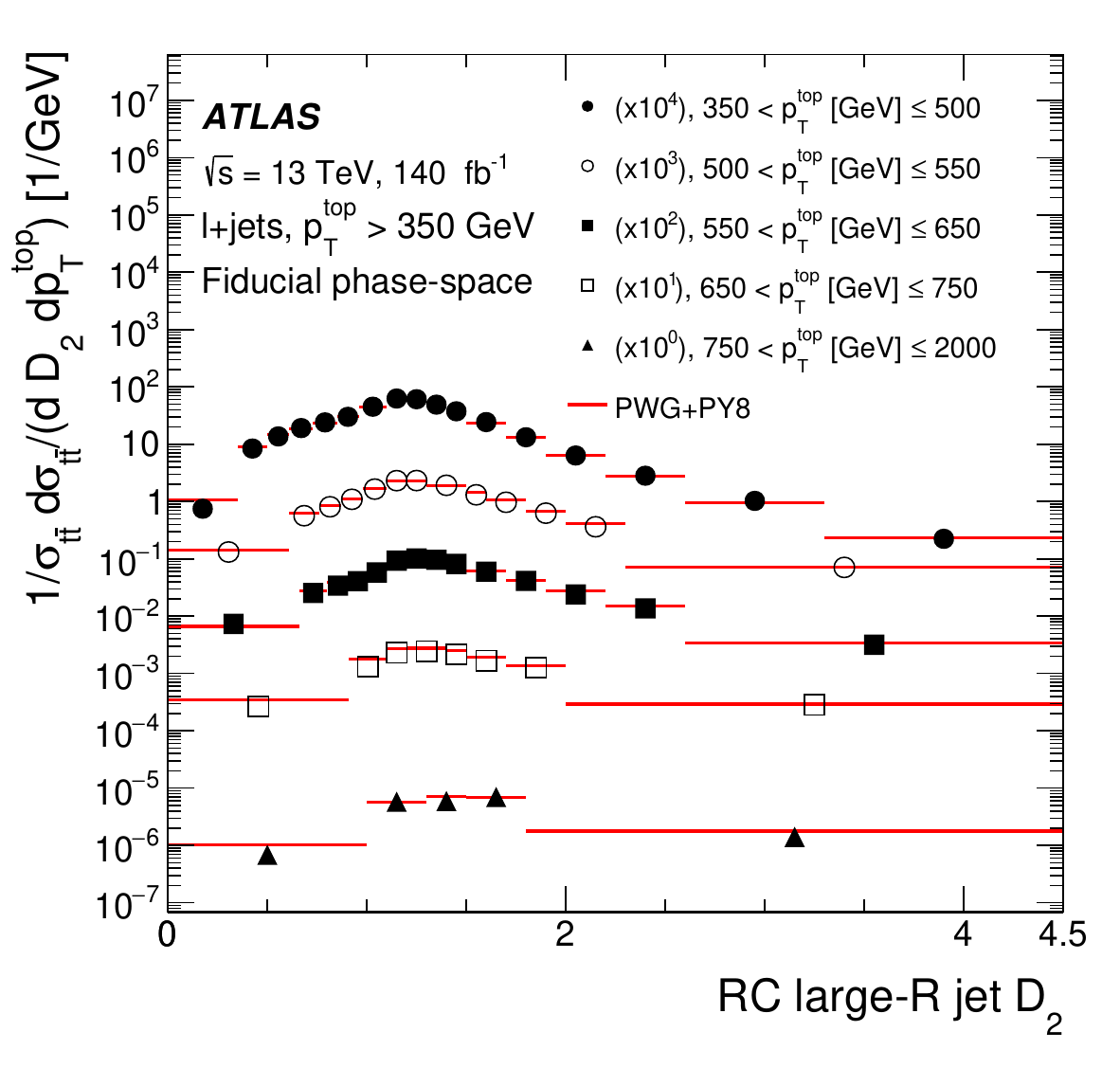}}\\
\subfloat[]{\label{fig:particle_level_results_ljets_dd_3:2}\includegraphics[width=\textwidth]{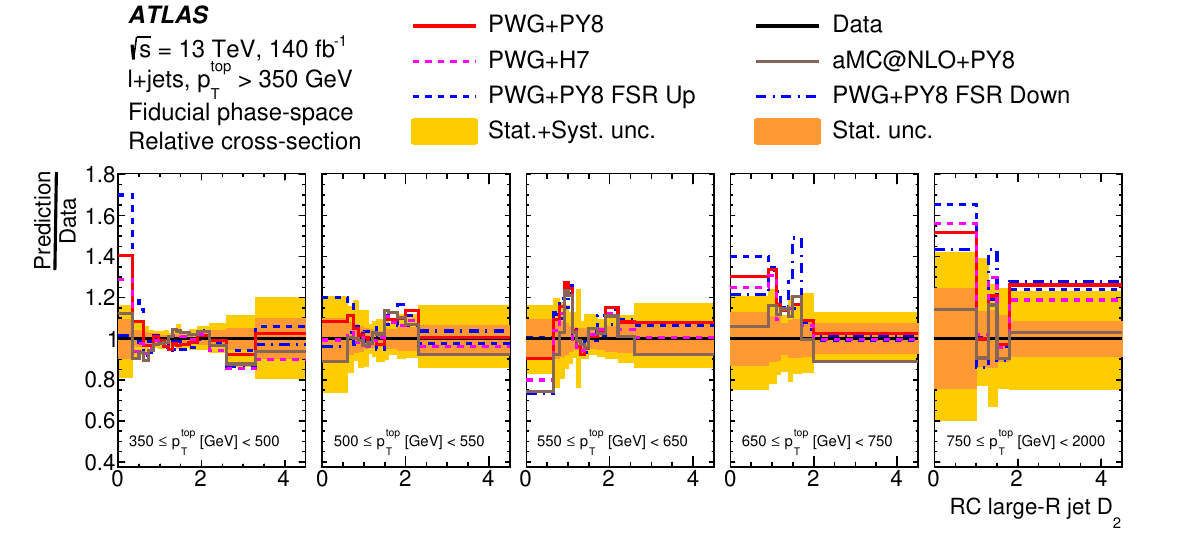}}
\caption{
Particle-level differential $t\bar{t}$ production cross-section as a function of $D_{2}$\ and jet \pT,
for several NLOME+PS predictions of $t\bar{t}$ signal and the data.
The events shown here are from the \ellplusjets\ channel:
\subref{fig:particle_level_results_ljets_dd_3:1} shows the data and the nominal prediction, while
\subref{fig:particle_level_results_ljets_dd_3:2} shows the ratios of the predictions to the data.
The lighter band represents the total uncertainty of the measured differential cross-section, while the darker band shows the statistical component.
}
\label{fig:particle_level_results_ljets_dd_3}
\end{figure}
 
\begin{figure}[t]
\centering
\subfloat[]{\label{fig:particle_level_results_allhad_dd_0:1}\includegraphics[width=0.6\textwidth]{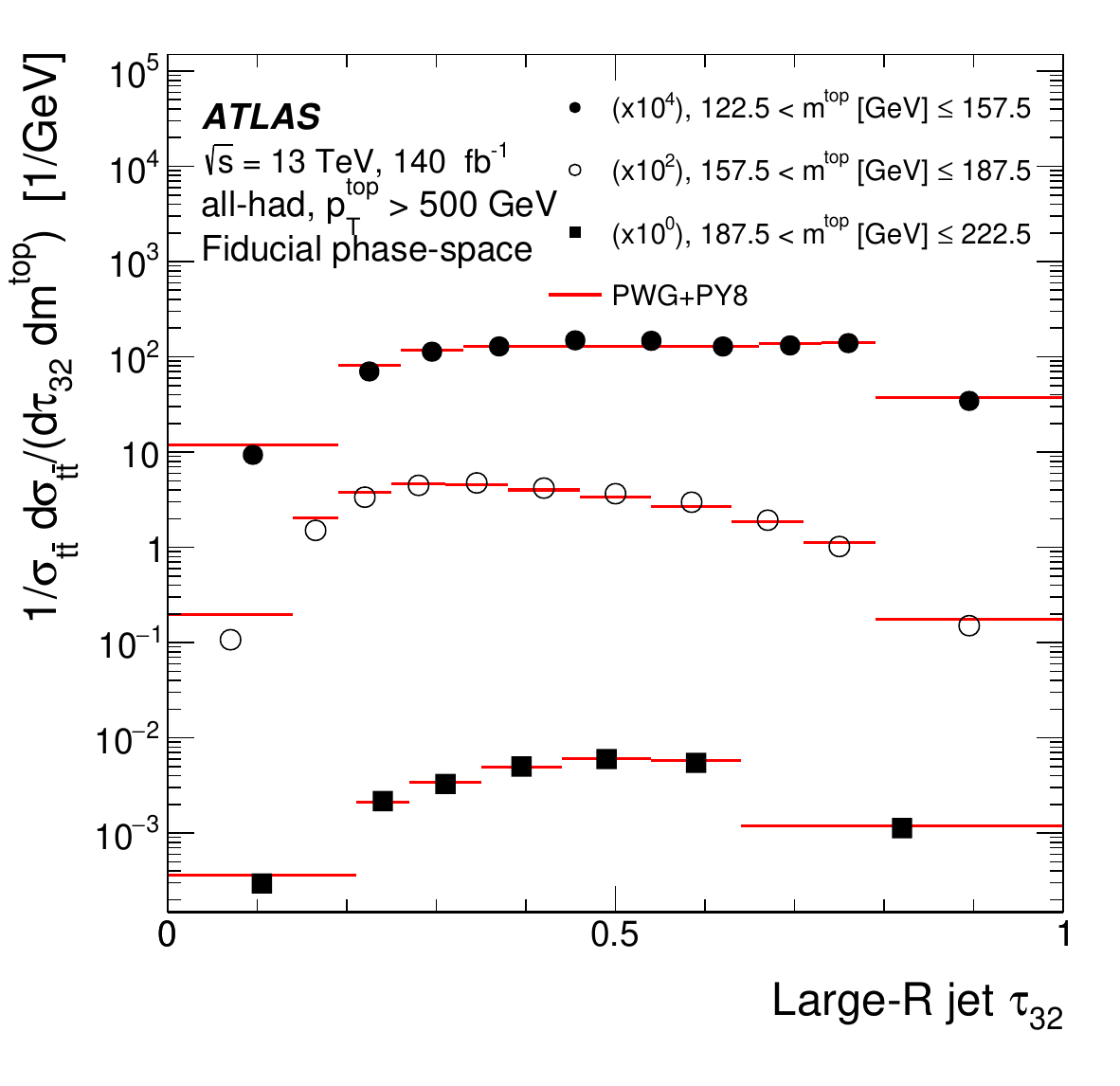}}\\
\subfloat[]{\label{fig:particle_level_results_allhad_dd_0:2}\includegraphics[width=\textwidth]{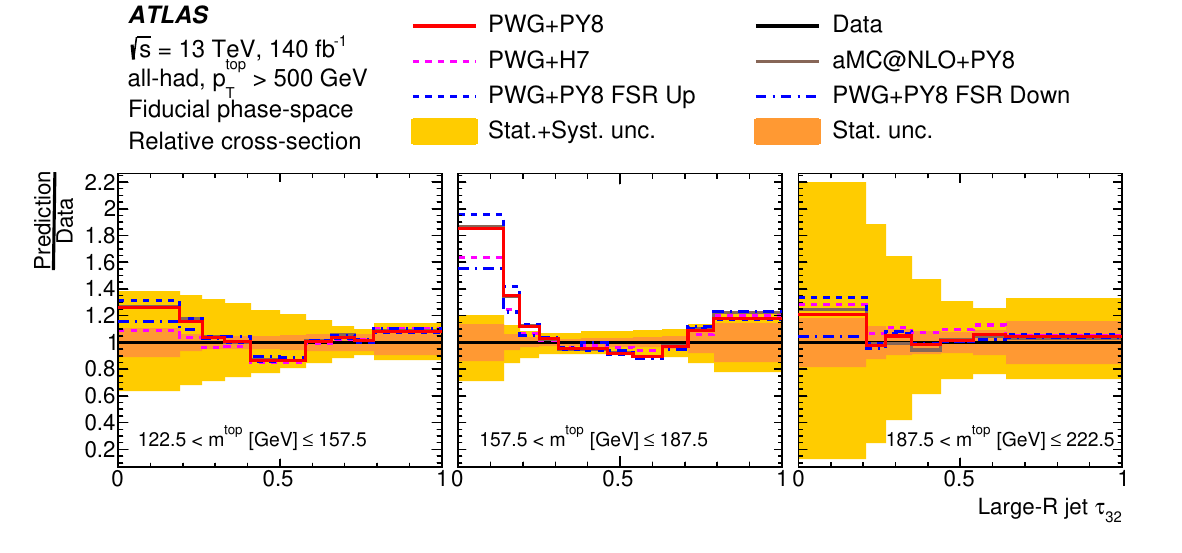}}
\caption{
Particle-level differential $t\bar{t}$ production cross-section as a function of $\tau_{32}$\ and jet mass,
for several NLOME+PS predictions of $t\bar{t}$ signal and the data.
The events shown are from the all-hadronic channel:
\subref{fig:particle_level_results_allhad_dd_0:1} shows the data and the nominal prediction, while
\subref{fig:particle_level_results_allhad_dd_0:2} shows the ratios of the predictions to the data.
The lighter band represents the total uncertainty of the measured differential cross-section, while the darker band shows the statistical component.
}
\label{fig:particle_level_results_allhad_dd_0}
\end{figure}
 
\begin{figure}[t]
\centering
\subfloat[]{\label{fig:particle_level_results_allhad_dd_1:1}\includegraphics[width=0.6\textwidth]{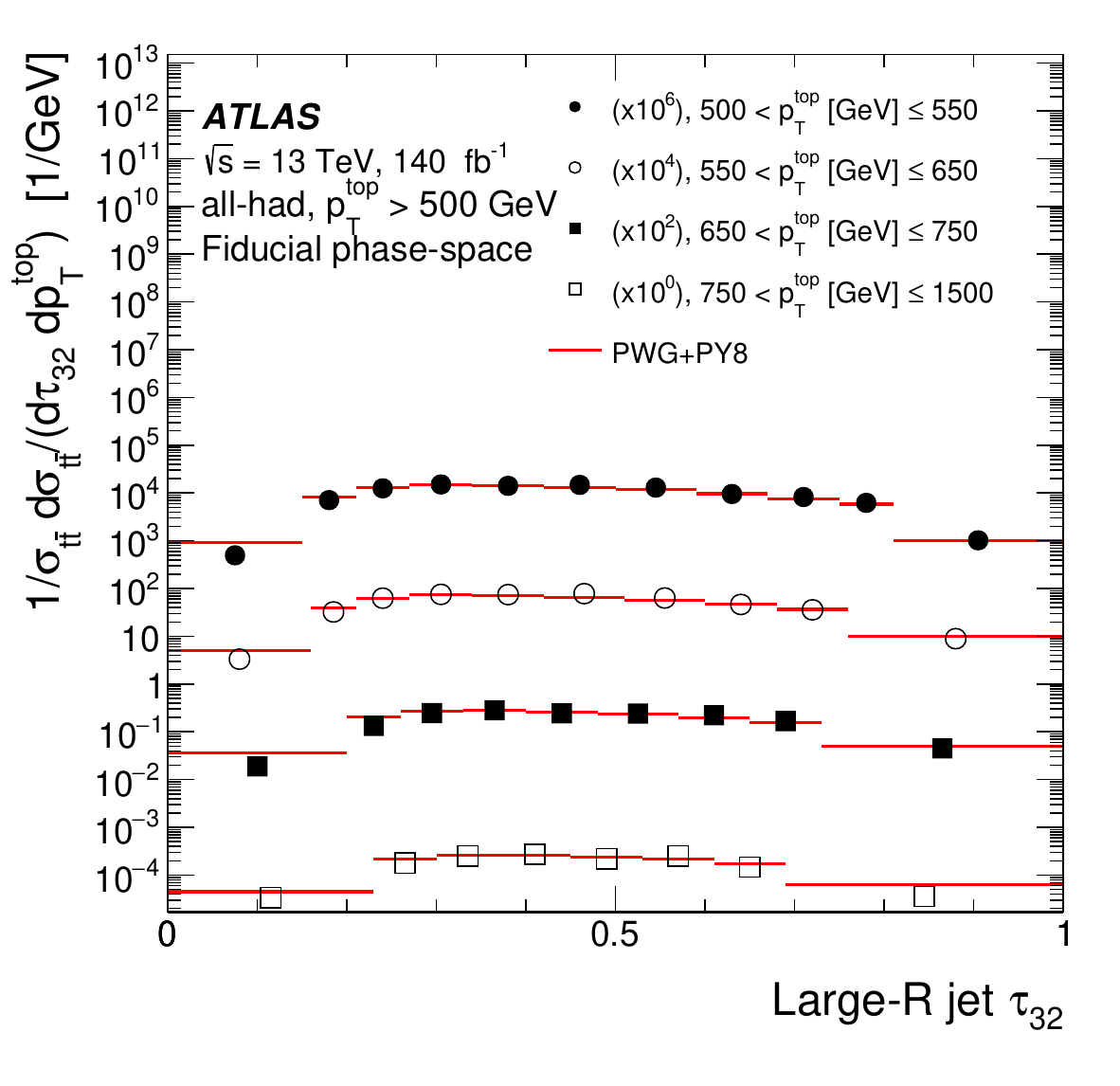}}\\
\subfloat[]{\label{fig:particle_level_results_allhad_dd_1:2}\includegraphics[width=\textwidth]{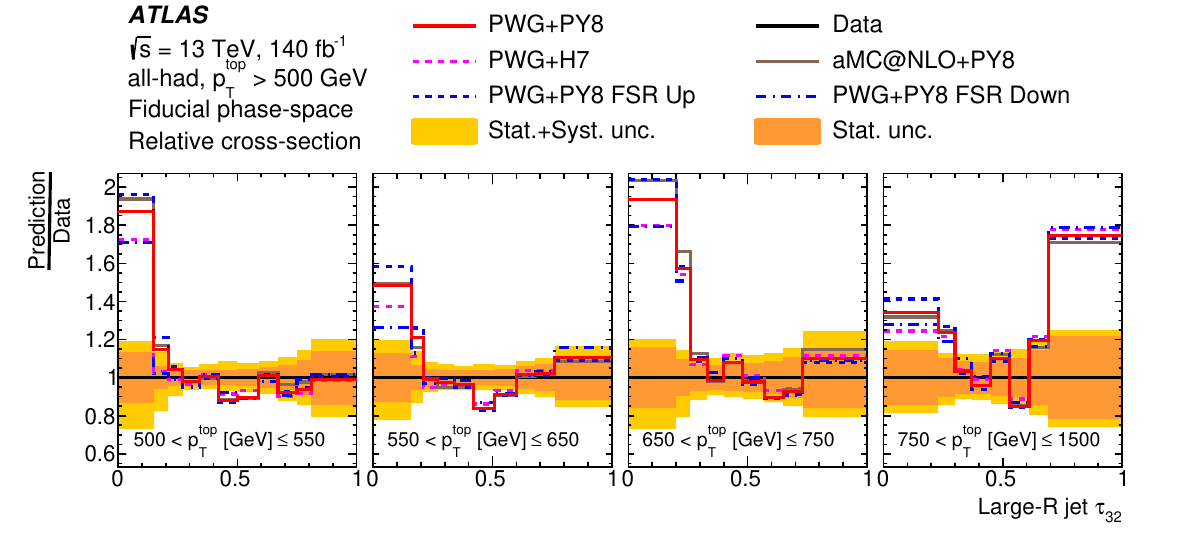}}
\caption{
Particle-level differential $t\bar{t}$ production cross-section as a function of $\tau_{32}$\ and jet \pT,
for several NLOME+PS predictions of $t\bar{t}$ signal and the data.
The events shown here are from the all-hadronic channel:
\subref{fig:particle_level_results_allhad_dd_1:1} shows the data and the nominal prediction, while
\subref{fig:particle_level_results_allhad_dd_1:2} shows the ratios of the predictions to the data.
The lighter band represents the total uncertainty of the measured differential cross-section,
while the darker band shows the statistical component.
}
\label{fig:particle_level_results_allhad_dd_1}
\end{figure}
 
\begin{figure}[t]
\centering
\subfloat[]{\label{fig:particle_level_results_allhad_dd_2:1}\includegraphics[width=0.6\textwidth]{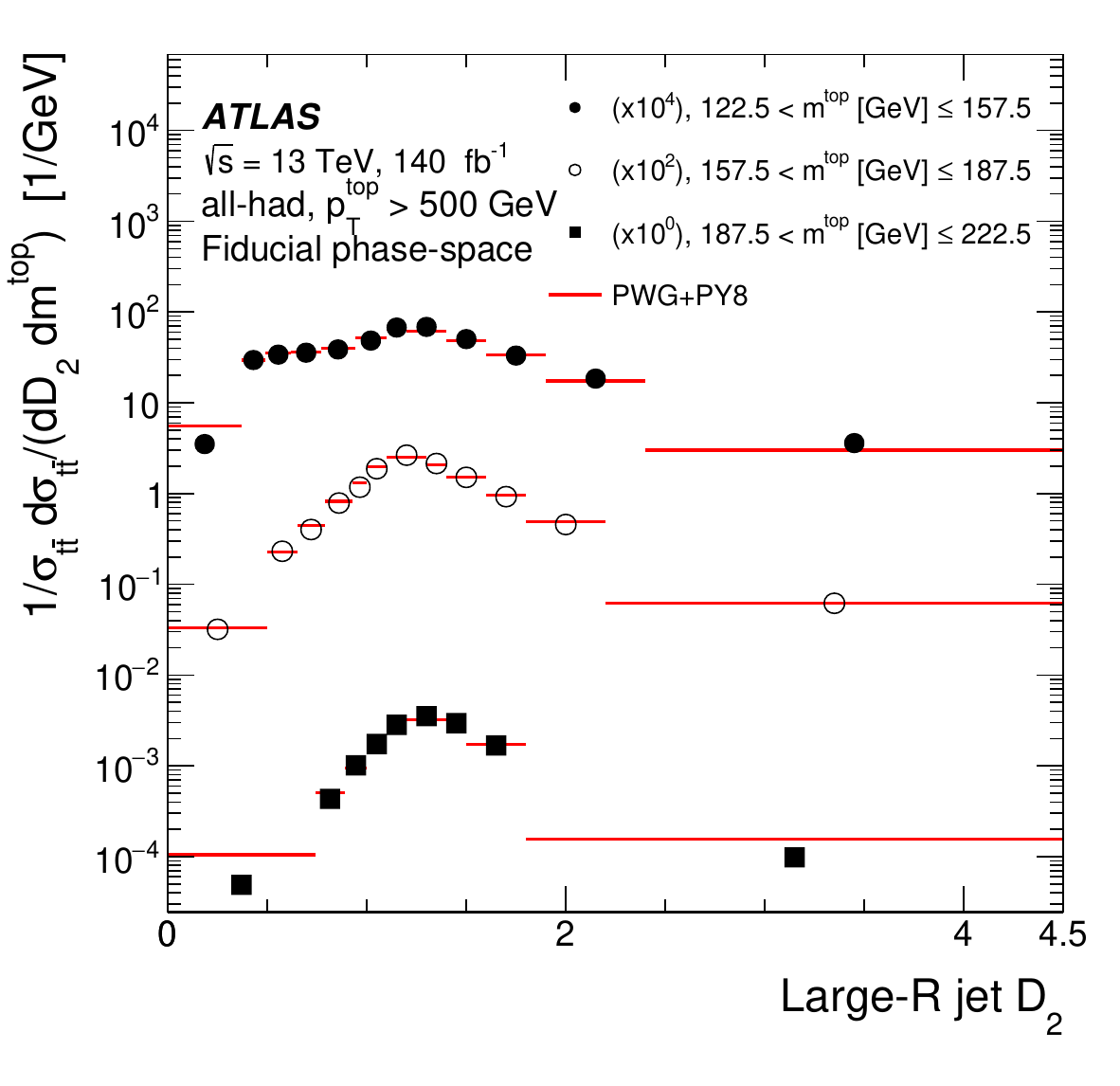}}\\
\subfloat[]{\label{fig:particle_level_results_allhad_dd_2:2}\includegraphics[width=\textwidth]{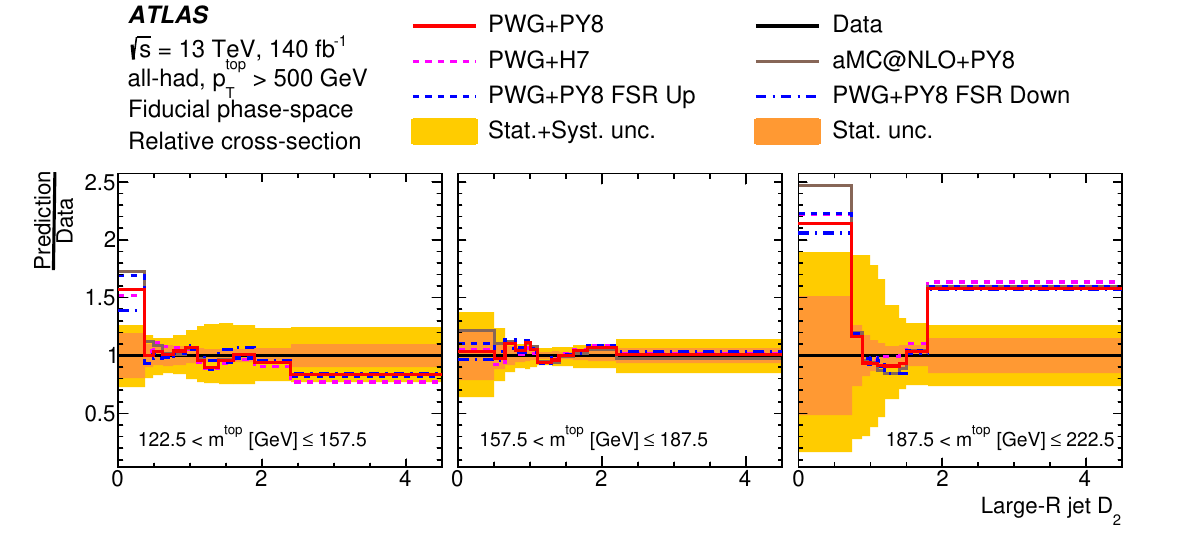}}
\caption{
Particle-level differential $t\bar{t}$ production cross-section as a function of $D_{2}$\ and jet mass,
for several NLOME+PS predictions of $t\bar{t}$ signal and the data.
The events shown are from the all-hadronic channel:
\subref{fig:particle_level_results_allhad_dd_2:1} shows the data and the nominal prediction, while
\subref{fig:particle_level_results_allhad_dd_2:2} shows the ratios of the predictions to the data.
The lighter band represents the total uncertainty of the measured differential cross-section,
while the darker band shows the statistical component.
}
\label{fig:particle_level_results_allhad_dd_2}
\end{figure}
 
\begin{figure}[t]
\centering
\subfloat[]{\label{fig:particle_level_results_allhad_dd_3:1}\includegraphics[width=0.6\textwidth]{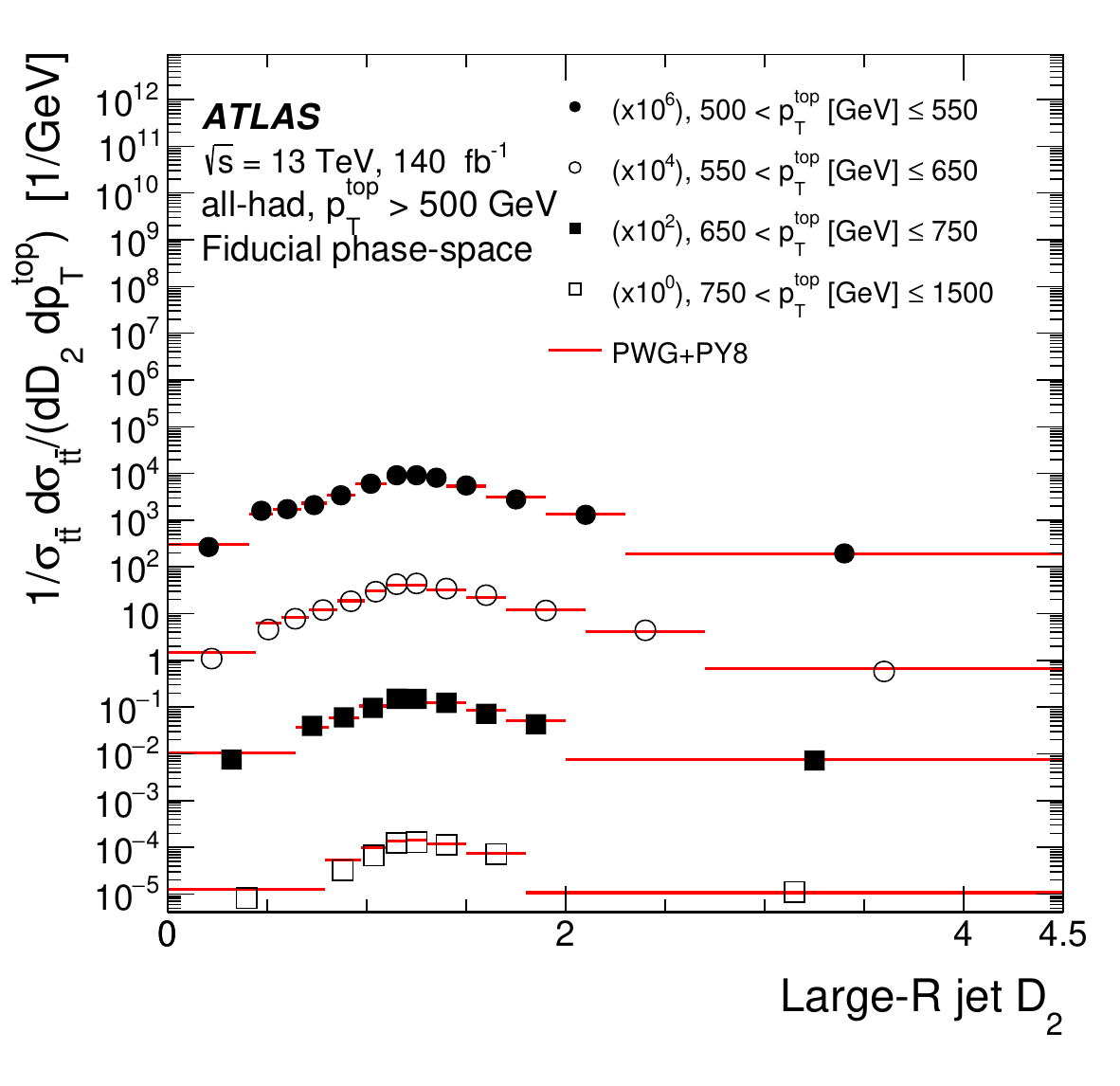}}\\
\subfloat[]{\label{fig:particle_level_results_allhad_dd_3:2}\includegraphics[width=\textwidth]{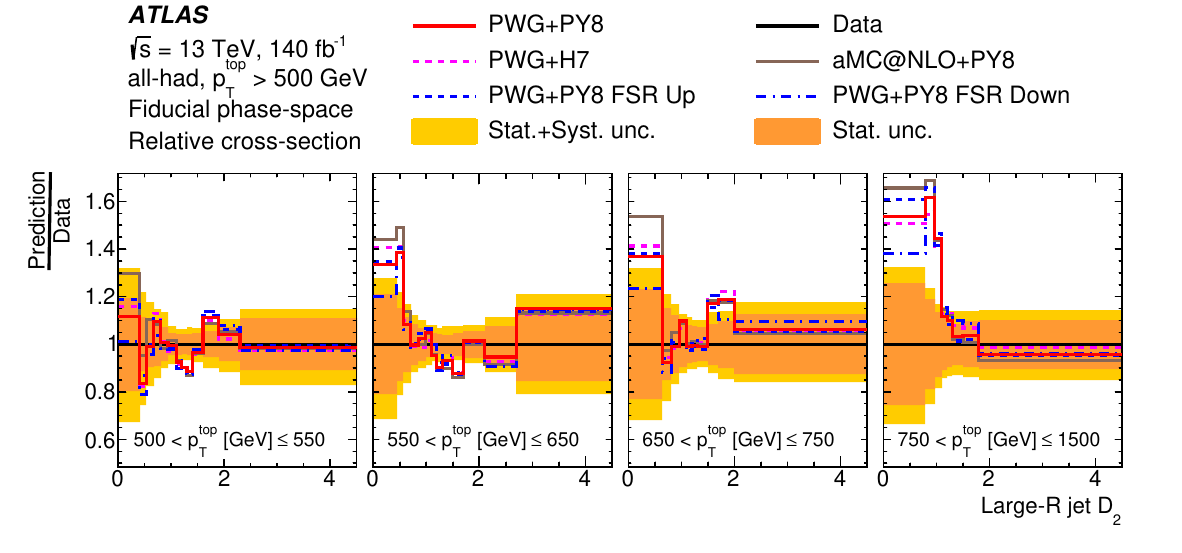}}
\caption{
Particle-level differential $t\bar{t}$ production cross-section as a function of $D_{2}$\ and jet \pT,
for several NLOME+PS predictions of $t\bar{t}$ signal and the data.
The events shown are from the all-hadronic channel:
\subref{fig:particle_level_results_allhad_dd_3:1} shows the data and the nominal prediction, while
\subref{fig:particle_level_results_allhad_dd_3:2} shows the ratios of the predictions to the data.
The lighter band represents the total uncertainty of the measured differential cross-section,
while the darker band shows the statistical component.
}
\label{fig:particle_level_results_allhad_dd_3}
\end{figure}


\FloatBarrier
\subsection{Comparisons of predicted and observed substructure variables}
\label{sec:results_covariance}
 
The level of agreement between the measured differential cross-sections and the SM predictions is quantified by a $\chi^2$ test-statistic and the related $p$-value extracted from the reduced $\chi^2$ distribution.
The $\chi^2$ calculation for the vector $V$, which contains the residuals between data and the theory predictions, is:
\begin{equation*}
\chi^2 = V^{T} \times C^{-1} \times V,
\end{equation*}
where the covariance matrix $C$\ incorporates the statistical uncertainties, the
bin-by-bin correlations induced by the
presence of the systematic uncertainties, and the regularization used to unfold the
data to  the particle-level differential cross-sections.
The covariance matrix is evaluated assuming no correlations between the different uncertainty
sources.
This method has been used in several \ttbar\ differential cross-section measurements~\cite{TOPQ-2015-06}.
 
The covariances evaluated for normalized spectra are not invertible since the number of rows and
columns is larger than the rank.
Consequently, one column and the corresponding row are removed from
the covariance matrix, as is the relevant element of the residuals vector.
The resulting $\chi^2$ does not depend on the row or column removed.
No theoretical uncertainties are included in the predictions used in the $\chi^2$ calculation
because each $\chi^2$\ value reflects the comparison of data with a specific model.
 
The $\chi^2$\ comparisons between the predicted and unfolded distributions are shown in
Tables~\ref{table:chi2_table_relative_ljets} and \ref{table:chi2_table_relative_allhad}.
The unfolded differential cross-sections are compared with
the five different MC predictions, and the resulting $\chi^2/\mathrm{NDF}$\ and
$p$-values are shown for each distribution.
 
\begin{table}[t]

\centering
\caption{$\chi^2$ and $p$-values quantifying the level of agreement between the unfolded spectra in the $\ell$+jets channel and several normalized NLO+PS predictions. PWG+PY8 corresponds to the \POWPY\ sample and PWG+H7 to the \POWHER\ sample.}
\resizebox{\textwidth}{!}{
\begin{tabular}[t]{c| r @{/} l r  | r @{/} l r  | r @{/} l r  | r @{/} l r  | r @{/} l r  } \toprule
\multirow{2}{*}{Observable} & \multicolumn{3}{c|}{\textsc{PWG+PY8}}& \multicolumn{3}{c|}{\textsc{PWG+H7}}& \multicolumn{3}{c|}{\textsc{aMC@NLO+PY8}}& \multicolumn{3}{c|}{\textsc{PWG+PY8(FSR Up)}}& \multicolumn{3}{c}{\textsc{PWG+PY8(FSR Down)}}\\
& \multicolumn{2}{c}{$\chi^{2}$/NDF} &  ~$p$-value& \multicolumn{2}{c}{$\chi^{2}$/NDF} &  ~$p$-value& \multicolumn{2}{c}{$\chi^{2}$/NDF} &  ~$p$-value& \multicolumn{2}{c}{$\chi^{2}$/NDF} &  ~$p$-value& \multicolumn{2}{c}{$\chi^{2}$/NDF} &  ~$p$-value\\
\hline
$\tau_{32}$ &{\ } 54 & 12 &<0.01 &{\ } 19 & 12 &0.09 &{\ } 15 & 12 &0.24 &{\ } 165 & 12 &<0.01 &{\ } 40 & 12 &<0.01\\
$\tau_{21}$ &{\ } 14 & 14 &0.41 &{\ } 7 & 14 &0.92 &{\ } 16 & 14 &0.32 &{\ } 42 & 14 &<0.01 &{\ } 8 & 14 &0.91\\
$\tau_{3}$ &{\ } 36 & 11 &<0.01 &{\ } 42 & 11 &<0.01 &{\ } 14 & 11 &0.23 &{\ } 130 & 11 &<0.01 &{\ } 23 & 11 &0.02\\
$ECF2$ &{\ } 25 & 18 &0.13 &{\ } 13 & 18 &0.78 &{\ } 15 & 18 &0.69 &{\ } 31 & 18 &0.03 &{\ } 24 & 18 &0.14\\
$D_{2}$ &{\ } 20 & 16 &0.20 &{\ } 17 & 16 &0.39 &{\ } 20 & 16 &0.20 &{\ } 37 & 16 &<0.01 &{\ } 15 & 16 &0.49\\
$C_{3}$ &{\ } 11 & 14 &0.65 &{\ } 6 & 14 &0.97 &{\ } 3 & 14 &1.00 &{\ } 35 & 14 &<0.01 &{\ } 3 & 14 &1.00\\
$p_\mathrm{T}^\mathrm{d,*}$ &{\ } 27 & 12 &<0.01 &{\ } 10 & 12 &0.58 &{\ } 11 & 12 &0.53 &{\ } 56 & 12 &<0.01 &{\ } 24 & 12 &0.02\\
$LHA$ &{\ } 14 & 17 &0.65 &{\ } 9 & 17 &0.92 &{\ } 20 & 17 &0.29 &{\ } 14 & 17 &0.69 &{\ } 19 & 17 &0.32\\
$D_{2}$ vs.\ $m^{\mathrm{top}}$ &{\ } 61 & 42 &0.03 &{\ } 62 & 42 &0.02 &{\ } 59 & 42 &0.05 &{\ } 118 & 42 &<0.01 &{\ } 44 & 42 &0.37\\
$D_{2}$ vs.\ $p^{\mathrm{top}}_\mathrm{T}$ &{\ } 71 & 56 &0.08 &{\ } 68 & 56 &0.13 &{\ } 70 & 56 &0.11 &{\ } 107 & 56 &<0.01 &{\ } 93 & 56 &<0.01\\
$\tau_{32}$ vs.\ $m^{\mathrm{top}}$ &{\ } 153 & 42 &<0.01 &{\ } 72 & 42 &<0.01 &{\ } 56 & 42 &0.07 &{\ } 413 & 42 &<0.01 &{\ } 77 & 42 &<0.01\\
$\tau_{32}$ vs.\ $p^{\mathrm{top}}_\mathrm{T}$ &{\ } 153 & 50 &<0.01 &{\ } 103 & 50 &<0.01 &{\ } 57 & 50 &0.23 &{\ } 360 & 50 &<0.01 &{\ } 114 & 50 &<0.01\\
\bottomrule
\end{tabular}
}
\label{table:chi2_table_relative_ljets}


\end{table}
 
\begin{table}[t]

\centering
\caption{$\chi^2$ and $p$-values quantifying the level of agreement between the unfolded spectra in the all-hadronic channel and several suitably normalized NLO+PS predictions. PWG+PY8 corresponds to the \POWPY\ sample and PWG+H7 to the \POWHER\ sample.}
\resizebox{\textwidth}{!}{
\begin{tabular}[t]{ c| r @{/} l r  | r @{/} l r  | r @{/} l r  | r @{/} l r  | r @{/} l r  } \toprule
\multirow{2}{*}{Observable} & \multicolumn{3}{c|}{\textsc{PWG+PY8}}& \multicolumn{3}{c|}{\textsc{PWG+H7}}& \multicolumn{3}{c|}{\textsc{aMC@NLO+PY8}}& \multicolumn{3}{c|}{\textsc{PWG+PY8(FSR Up)}}& \multicolumn{3}{c}{\textsc{PWG+PY8(FSR Down)}}\\
& \multicolumn{2}{c}{$\chi^{2}$/NDF} &  ~$p$-value& \multicolumn{2}{c}{$\chi^{2}$/NDF} &  ~$p$-value& \multicolumn{2}{c}{$\chi^{2}$/NDF} &  ~$p$-value& \multicolumn{2}{c}{$\chi^{2}$/NDF} &  ~$p$-value& \multicolumn{2}{c}{$\chi^{2}$/NDF} &  ~$p$-value\\
\hline
$\tau_{32}$ &{\ } 24 & 10 &<0.01 &{\ } 14 & 10 &0.20 &{\ } 9 & 10 &0.52 &{\ } 61 & 10 &<0.01 &{\ } 6 & 10 &0.82\\
$\tau_{21}$ &{\ } 7 & 10 &0.75 &{\ } 6 & 10 &0.80 &{\ } 6 & 10 &0.80 &{\ } 11 & 10 &0.36 &{\ } 6 & 10 &0.84\\
$\tau_{3}$ &{\ } 29 & 7 &<0.01 &{\ } 17 & 7 &0.02 &{\ } 10 & 7 &0.17 &{\ } 58 & 7 &<0.01 &{\ } 8 & 7 &0.29\\
$ECF2$ &{\ } 17 & 11 &0.10 &{\ } 12 & 11 &0.39 &{\ } 14 & 11 &0.26 &{\ } 20 & 11 &0.05 &{\ } 15 & 11 &0.19\\
$D_{2}$ &{\ } 11 & 12 &0.55 &{\ } 8 & 12 &0.82 &{\ } 8 & 12 &0.76 &{\ } 14 & 12 &0.27 &{\ } 7 & 12 &0.88\\
$C_{3}$ &{\ } 29 & 8 &<0.01 &{\ } 21 & 8 &<0.01 &{\ } 13 & 8 &0.13 &{\ } 57 & 8 &<0.01 &{\ } 10 & 8 &0.28\\
$p_\mathrm{T}^\mathrm{d,*}$ &{\ } 21 & 9 &0.01 &{\ } 6 & 9 &0.78 &{\ } 10 & 9 &0.35 &{\ } 35 & 9 &<0.01 &{\ } 8 & 9 &0.54\\
$LHA$ &{\ } 12 & 12 &0.49 &{\ } 9 & 12 &0.74 &{\ } 12 & 12 &0.46 &{\ } 12 & 12 &0.43 &{\ } 11 & 12 &0.53\\
$D_{2}$ vs.\ \mtop &{\ } 22 & 32 &0.91 &{\ } 27 & 32 &0.73 &{\ } 20 & 32 &0.95 &{\ } 28 & 32 &0.67 &{\ } 19 & 32 &0.96\\
$D_{2}$ vs.\ \pTtop &{\ } 29 & 43 &0.96 &{\ } 26 & 43 &0.98 &{\ } 28 & 43 &0.96 &{\ } 32 & 43 &0.88 &{\ } 26 & 43 &0.98\\
$\tau_{32}$ vs.\ \mtop &{\ } 30 & 27 &0.31 &{\ } 21 & 27 &0.79 &{\ } 15 & 27 &0.97 &{\ } 69 & 27 &<0.01 &{\ } 11 & 27 &1.00\\
$\tau_{32}$ vs.\ \pTtop &{\ } 49 & 37 &0.08 &{\ } 36 & 37 &0.53 &{\ } 34 & 37 &0.63 &{\ } 94 & 37 &<0.01 &{\ } 30 & 37 &0.79\\
\bottomrule
\end{tabular}
}
\label{table:chi2_table_relative_allhad}


\end{table}
 
These quantitative comparisons reflect the qualitative comparisons made earlier,
but also illustrate that overall agreement with the data is better for some MC predictions than others.
In particular, the predictions from \POWPY[8]\ (FSR Up) are in quite
poor agreement with the observations, with $p$-values less than 0.01 for all but two
\ellplusjets\ variables.
There is similarly poor agreement between the \POWPY[8]\ (FSR Up) predictions
and data in the all-hadronic
channel, although in that case the predictions for the
$\tau_{21}$, $D_{2}$\ and \LHA\ variables are in reasonable agreement with data, with a $p$-value larger than 0.3. Even for these variables the FSR Up predictions give the worst description of the data among the tested MC predictions.
In contrast, the \POWPY[8]\ (FSR Down) prediction provides a good description, with a $p$-value larger than 0.01 for all
single-differential observables except $\tau_{32}$\ in the \ellplusjets\ channel.
Overall, the measured distributions  favor a lower FSR scale, corresponding to a higher \alphaFSR value. This is consistent with recent tuning studies performed using ATLAS data~\cite{Kar:2021fqs} and with the recent CMS top-quark mass measurement using boosted top quarks, which employed the $\tau_{32}$ distribution to tune \alphaFSR~\cite{CMS:2022kqg}.
 
The predictions of the other NLOME+PS calculations are generally in good agreement with the data.
It is noteworthy that most predictions for the $\tau_{32}$\ variable are generally in poor agreement with the data, with $p$-values lower than 0.1 for several generator setups and with
the \POWPY[8]\ predictions having
$\chi^2/\mathrm{NDF}=54/12$\ and $24/10$\ for the \ellplusjets\ and
all-hadronic channels, respectively.
This shows that the NLOME+PS predictions are unable to  accurately model the three-body nature of boosted top-quark jets. Notably, the $\tau_{3}$\ variable is also generally described poorly by the predictions, at the same level as $\tau_{32}$ or worse,  which correlates with poor $\tau_{32}$\ agreement and further implicates the three-body modeling. One exception comes from \AMCatNLO, which provides a fair description of $\tau_{32}$ and $\tau_{3}$ in both channels, with $p$-values larger than 0.2.
The variables that are sensitive to the modeling of one-body and two-body substructure ($\tau_{21}$, \ECF\ and $D_2$) show good agreement with data, with $p$-values ranging from 0.2 to 0.82, depending on the generator and the channel.

The $C_{3}$ variable, a probe of three-body structure that is interpreted as a measure of the higher-order radiation relative to the lowest-order prediction,
is reasonably well modeled by most of the \ellplusjets\ calculations, with three $p$-values exceeding 0.9, but interestingly is not well modeled by most predictions in the all-hadronic channel.
This could be due to mismodeling of the internal radiation in the top-quark jet at higher \pT\
or perhaps the modeling of the energy flow between these higher-\pT\ jets in this all-hadronic final state.
 
The \ptdispersion\ variable is a measure of the distribution of the momentum between particles within the jet, and is therefore sensitive to the hadronization model. It is not well described by the \POWPY[8]\ model, but agreement improves with a higher value of \alphaFSR, especially in the all-hadronic channel, where the $p$-value reaches 0.54.
 
The \LHA\ variable, which is more sensitive to the modeling of the energy of the high-energy constituents, is well described by the predictions. This variable, and also \ECF, was selected because it was not expected to be highly sensitive to the FSR variations.  \POWHER[7]\ gives the best description of these variables in both the \ellplusjets\ and all-hadronic channels, with $p$-values of 0.92 and 0.74 respectively for \LHA.
 
The double-differential cross-sections provide more information about the correlations of the
two-body and three-body substructure measures, $D_{2}$\ and $\tau_{32}$.
Although $D_{2}$\ appears to be well modeled overall by most predictions, the
modeling of its correlations with \mtop\ and \pTtop\ is relatively poor in the \ellplusjets\ channel.
This arises from the difficulty of modeling $D_{2}$\ at lower \mtop\ and higher \pTtop.
For $\tau_{32}$, the correlations are even more poorly modeled in the \ellplusjets\ channel, with $p$-values less than 0.01 for several NLOME+PS predictions. The trends observed in the double-differential distributions for the two channels are similar, although the discrimination power in the all-hadronic channel is more limited by the statistical uncertainty.
 
Overall, the picture that emerges is that the \POWPY[8]\ (FSR Up) model is
not successful in describing the substructure of these high-\pT\ jets.
Compared with the nominal \POWPY[8]\ prediction, all \POWHER[7], \AMCatNLOPY[8]\ and \POWPY[8]\ (FSR Down) predictions are in better agreement with the measurements for most observables, and the \POWPY[8]\ (FSR Down) predictions are in closest agreement with the measurements in both the \ellplusjets\ and all-hadronic channels.
The correlations with \mtop\ and \pTtop\ provide some diagnostic information, which suggests
that the most significant disagreement tends to be at lower \mtop\ and higher \pTtop.
 
These results highlight the tension between the data and the predictions for key tagging variables such as $\tau_{32}$\ and $D_{2}$, and better agreement will be essential in improving the tuning of the showering parameters. Better agreement with data for these observables would also reduce the modeling uncertainties affecting the top-quark taggers and lead to scaling factors closer to unity when correcting for tagger performance differences between data and simulation.
\FloatBarrier


\section{Conclusion}
\label{sec:conclusions}

 
Top-quark jet substructure variables are measured for top-quark jets with $\pT > 350$~\GeV\ in a fiducial phase-space defined by the \pT\ and $\eta$\ of the jets.
These measurements use 140~fb$^{-1}$ of 13~\TeV\ $pp$ collision data collected by the ATLAS detector at the LHC, and are performed in both the lepton+jets and all-hadronic final states of \ttbar\ events.
The inclusion of the all-hadronic final state enables measurements of the substructure of
top-quark jets with average \pT\ above 500~\GeV.
This analysis is the first to investigate the substructure of top-quark jets with
variables defined using only charged components of the jet.
Compared to previous ATLAS measurements of top-quark-jet substructure,
this approach, together with the larger dataset analyzed, significantly reduces the uncertainties in the measured quantities.
These smaller uncertainties provide greater sensitivity to the differences between the various MC predictions.
Both single-differential cross-section measurements and a selected set of double-differential cross-section measurements are presented.
 
The observed substructure distributions are compared with several MC generator predictions, which are
all based on NLO calculations for the matrix element, LO calculations for the decay of
the top quark, and phenomenological models for parton showering and hadronization.
A quantitative comparison between the observed and predicted distributions shows that the $\tau_{21}$, \ECF\ and $D_2$\ variables are generally well described in both channels by the \POWPY[8]\ MC predictions,
which were used as the nominal predictions in the analysis.
In contrast, the distributions of $\tau_{32}$, $\tau_3$\ and $C_3$\ are poorly described by the nominal MC predictions, with the predicted jet substructure being more three-body-like than that observed.
These observations are consistent between the two channels.
The distribution of \pT\ dispersion, \ptdispersion, is not well modeled by the nominal MC prediction,
with the observed distributions being more strongly peaked at lower values.
 
Most of these features are also reflected in the alternative MC predictions.
Generally speaking, the MC predictions from the \HERWIG[7]\ parton-shower and hadronization model are a better match to the observed distributions, although the distributions of $\tau_{32}$, $\tau_3$, and $C_3$\ are poorly described.
The alternative matrix-element predictions from \AMCatNLOPY[8]\ are in better agreement with the observed distributions in both channels.
The predictions made with increased FSR scales are universally poorer descriptions of the data,
while for most of the substructure variables the models with decreased FSR scales are generally in better agreement than the nominal predictions,
indicating that the data favor an increased value of \alphaFSR.
In general, the predictions are in better agreement with the measurements in the all-hadronic channel.

The $\tau_{32}$\ and $D_{2}$\ variables,
which are sensitive to three-body and two-body modeling, have correlations
with \mtop\ and \pTtop\ that are poorly modeled by several NLOME+PS calculations,
indicating systematic problems in these calculations at low \mtop\ and high \pTtop.
 
These observations illustrate that the \HERWIG[7]\ model for parton showering and hadronization is a better match to the observed substructure than the \PYTHIA[8] model.
In addition, the models with increased  \alphaFSR predict jet substructure that is in better agreement with the observations.
These observations are consistent with the results of comparisons with the top-quark-jet kinematics observed in the all-hadronic final state,
and reinforce the need to improve the models used to predict the substructure of boosted top-quark jets.


\section*{Acknowledgements}
 
We thank CERN for the very successful operation of the LHC, as well as the
support staff from our institutions without whom ATLAS could not be
operated efficiently.
 
We acknowledge the support of
ANPCyT, Argentina;
YerPhI, Armenia;
ARC, Australia;
BMWFW and FWF, Austria;
ANAS, Azerbaijan;
CNPq and FAPESP, Brazil;
NSERC, NRC and CFI, Canada;
CERN;
ANID, Chile;
CAS, MOST and NSFC, China;
Minciencias, Colombia;
MEYS CR, Czech Republic;
DNRF and DNSRC, Denmark;
IN2P3-CNRS and CEA-DRF/IRFU, France;
SRNSFG, Georgia;
BMBF, HGF and MPG, Germany;
GSRI, Greece;
RGC and Hong Kong SAR, China;
ISF and Benoziyo Center, Israel;
INFN, Italy;
MEXT and JSPS, Japan;
CNRST, Morocco;
NWO, Netherlands;
RCN, Norway;
MEiN, Poland;
FCT, Portugal;
MNE/IFA, Romania;
MESTD, Serbia;
MSSR, Slovakia;
ARRS and MIZ\v{S}, Slovenia;
DSI/NRF, South Africa;
MICINN, Spain;
SRC and Wallenberg Foundation, Sweden;
SERI, SNSF and Cantons of Bern and Geneva, Switzerland;
MOST, Taiwan;
TENMAK, T\"urkiye;
STFC, United Kingdom;
DOE and NSF, United States of America.
In addition, individual groups and members have received support from
BCKDF, CANARIE, Compute Canada and CRC, Canada;
PRIMUS 21/SCI/017 and UNCE SCI/013, Czech Republic;
COST, ERC, ERDF, Horizon 2020 and Marie Sk{\l}odowska-Curie Actions, European Union;
Investissements d'Avenir Labex, Investissements d'Avenir Idex and ANR, France;
DFG and AvH Foundation, Germany;
Herakleitos, Thales and Aristeia programmes co-financed by EU-ESF and the Greek NSRF, Greece;
BSF-NSF and MINERVA, Israel;
Norwegian Financial Mechanism 2014-2021, Norway;
NCN and NAWA, Poland;
La Caixa Banking Foundation, CERCA Programme Generalitat de Catalunya and PROMETEO and GenT Programmes Generalitat Valenciana, Spain;
G\"{o}ran Gustafssons Stiftelse, Sweden;
The Royal Society and Leverhulme Trust, United Kingdom.
 
The crucial computing support from all WLCG partners is acknowledged gratefully, in particular from CERN, the ATLAS Tier-1 facilities at TRIUMF (Canada), NDGF (Denmark, Norway, Sweden), CC-IN2P3 (France), KIT/GridKA (Germany), INFN-CNAF (Italy), NL-T1 (Netherlands), PIC (Spain), ASGC (Taiwan), RAL (UK) and BNL (USA), the Tier-2 facilities worldwide and large non-WLCG resource providers. Major contributors of computing resources are listed in Ref.~\cite{ATL-SOFT-PUB-2021-003}.

\clearpage
\printbibliography
 
\clearpage
 
\begin{flushleft}
\hypersetup{urlcolor=black}
{\Large The ATLAS Collaboration}

\bigskip

\AtlasOrcid[0000-0002-6665-4934]{G.~Aad}$^\textrm{\scriptsize 102}$,
\AtlasOrcid[0000-0002-5888-2734]{B.~Abbott}$^\textrm{\scriptsize 120}$,
\AtlasOrcid[0000-0002-1002-1652]{K.~Abeling}$^\textrm{\scriptsize 55}$,
\AtlasOrcid[0000-0001-5763-2760]{N.J.~Abicht}$^\textrm{\scriptsize 49}$,
\AtlasOrcid[0000-0002-8496-9294]{S.H.~Abidi}$^\textrm{\scriptsize 29}$,
\AtlasOrcid[0000-0002-9987-2292]{A.~Aboulhorma}$^\textrm{\scriptsize 35e}$,
\AtlasOrcid[0000-0001-5329-6640]{H.~Abramowicz}$^\textrm{\scriptsize 151}$,
\AtlasOrcid[0000-0002-1599-2896]{H.~Abreu}$^\textrm{\scriptsize 150}$,
\AtlasOrcid[0000-0003-0403-3697]{Y.~Abulaiti}$^\textrm{\scriptsize 117}$,
\AtlasOrcid[0000-0003-0762-7204]{A.C.~Abusleme~Hoffman}$^\textrm{\scriptsize 137a}$
% ...
\bigskip
\\

% ...
$^{29}$Physics Department, Brookhaven National Laboratory, Upton NY; United States of America.\\
% ...
$^{35}$$^{(a)}$Facult\'e des Sciences Ain Chock, R\'eseau Universitaire de Physique des Hautes Energies - Universit\'e Hassan II, Casablanca;$^{(b)}$Facult\'{e} des Sciences, Universit\'{e} Ibn-Tofail, K\'{e}nitra;$^{(c)}$Facult\'e des Sciences Semlalia, Universit\'e Cadi Ayyad, LPHEA-Marrakech;$^{(d)}$LPMR, Facult\'e des Sciences, Universit\'e Mohamed Premier, Oujda;$^{(e)}$Facult\'e des sciences, Universit\'e Mohammed V, Rabat;$^{(f)}$Institute of Applied Physics, Mohammed VI Polytechnic University, Ben Guerir; Morocco.\\
% ...
$^{49}$Fakult\"{a}t Physik , Technische Universit{\"a}t Dortmund, Dortmund; Germany.\\
% ...
$^{55}$II. Physikalisches Institut, Georg-August-Universit\"{a}t G\"ottingen, G\"ottingen; Germany.\\
% ...
$^{102}$CPPM, Aix-Marseille Universit\'e, CNRS/IN2P3, Marseille; France.\\
% ...
$^{117}$Department of Physics, New York University, New York NY; United States of America.\\
% ...
$^{120}$Homer L. Dodge Department of Physics and Astronomy, University of Oklahoma, Norman OK; United States of America.\\
% ...
$^{137}$$^{(a)}$Departamento de F\'isica, Pontificia Universidad Cat\'olica de Chile, Santiago;$^{(b)}$Millennium Institute for Subatomic physics at high energy frontier (SAPHIR), Santiago;$^{(c)}$Instituto de Investigaci\'on Multidisciplinario en Ciencia y Tecnolog\'ia, y Departamento de F\'isica, Universidad de La Serena;$^{(d)}$Universidad Andres Bello, Department of Physics, Santiago;$^{(e)}$Instituto de Alta Investigaci\'on, Universidad de Tarapac\'a, Arica;$^{(f)}$Departamento de F\'isica, Universidad T\'ecnica Federico Santa Mar\'ia, Valpara\'iso; Chile.\\
% ...
$^{150}$Department of Physics, Technion, Israel Institute of Technology, Haifa; Israel.\\
$^{151}$Raymond and Beverly Sackler School of Physics and Astronomy, Tel Aviv University, Tel Aviv; Israel.\\
% ...
$^{a}$ Also Affiliated with an institute covered by a cooperation agreement with CERN.\\
$^{b}$ Also at An-Najah National University, Nablus; Palestine.\\
$^{c}$ Also at Borough of Manhattan Community College, City University of New York, New York NY; United States of America.\\
$^{d}$ Also at Center for High Energy Physics, Peking University; China.\\
$^{e}$ Also at Center for Interdisciplinary Research and Innovation (CIRI-AUTH), Thessaloniki; Greece.\\
$^{f}$ Also at Centro Studi e Ricerche Enrico Fermi; Italy.\\
% ...

\end{flushleft}


\end{document}